\documentclass[11pt]{article}

\usepackage{mathrsfs}
\usepackage{amsmath,amssymb}
\usepackage{epsfig}
\usepackage{relsize}
\usepackage{graphicx}
\usepackage{bbm}
\usepackage[shortlabels]{enumitem}
\usepackage{xcolor}
\usepackage{bm}
\usepackage{float}
\usepackage{cite}
\usepackage{tcolorbox}
\usepackage{nicefrac}

\definecolor{lightred}{RGB}{255,127,127}
\definecolor{lightgreen}{RGB}{127,255,127}
\definecolor{lightblue}{RGB}{127,127,255}
\definecolor{linkcolor}{rgb}{0,0,0.6}
\usepackage[ pdftex,colorlinks=true,
pdfstartview=FitV,
linkcolor= linkcolor,
citecolor= linkcolor,
urlcolor= linkcolor,
hyperindex=true,
hyperfigures=false]
{hyperref}

\numberwithin{equation}{section}

\newcommand{\noi}{\noindent}
\newcommand{\s}{\sigma}
\newcommand{\dd}{\text{d}}
\newcommand{\p}{\partial}
\newcommand{\Ac}{\mathcal{A}}
\newcommand{\Oc}{\mathcal{O}}
\newcommand{\Acm}{\mathcal{A}^{\,\text{can}}_{T^\ast \mathcal{M}}}
\newcommand{\Acg}{\mathcal{A}^{\,\text{can}}_{T^\ast G}}
\newcommand{\Hc}{\mathcal{H}}
\newcommand{\g}{\mathfrak{g}}
\newcommand{\R}{\mathbb{R}}

\newcommand{\C}{\mathbb{C}}
\newcommand{\Z}{\mathbb{Z}}
\newcommand{\CP}{\mathbb{CP}^1}
\newcommand{\Lc}{\mathcal{L}}
\newcommand{\Mc}{\mathcal{M}}
\newcommand{\Tr}{\text{Tr}}

\newcommand{\vp}{\varphi}
\newcommand{\Zc}{\mathcal{Z}}

\newcommand{\ze}{\zeta}

\newcommand{\Id}{\text{Id}}
\newcommand{\Rc}{\mathcal{R}}

\newcommand{\ad}{\text{ad}}
\newcommand{\Q}{\mathcal{Q}}
\newcommand{\Pc}{\mathcal{P}}
\newcommand{\Tc}{\mathcal{T}}
\newcommand{\Pexp}{\text{P}\overleftarrow{\text{exp}}}
\newcommand{\ti}[1]{_{\bm{\underline{#1}}}}
\newcommand{\td}{{\tt t}}
\newcommand{\ps}[2]{\left\langle#1,#2\right\rangle}
\newcommand{\psb}[2]{\bigl\langle#1,#2\bigr\rangle}

\newcommand{\psd}{\langle\cdot,\cdot\rangle}

\newcommand{\hv}{h^{\!\vee}}
\newcommand{\f}[2]{f_{#1}^{{\color{white}#1}#2}}
\newcommand{\F}[2]{F_{#1}^{{\color{white}#1}#2}}
\newcommand{\Jc}{\mathcal{J}}
\newcommand{\Wc}{\mathcal{W}}
\newcommand{\Kc}{\mathcal{K}}
\newcommand{\oms}{\omega_{\,\text{sym}}}
\newcommand{\rk}{\text{rk}\,\mathfrak{g}}
\newcommand{\Cox}{\text{Cox}\,\mathfrak{g}}
\renewcommand{\sl}{\mathfrak{sl}}
\newcommand{\su}{\mathfrak{su}}
\newcommand{\Ic}{\mathcal{I}}
\newcommand{\TM}{T^\ast \mathcal{M}}
\newcommand{\Lt}{\widetilde{L}}
\newcommand{\W}[1]{I_{\text{W}\hspace{-1pt}\text{Z}}\left[#1\right]}

\DeclareFontEncoding{LS1}{}{}
\DeclareFontSubstitution{LS1}{stix}{m}{n}
\DeclareSymbolFont{stixsymbols}{LS1}{stixscr}{m}{n}
\SetSymbolFont{stixsymbols}{bold}{LS1}{stixscr}{b}{n}
\DeclareMathSymbol{\kay}{\mathalpha}{stixsymbols}{"6B}
\DeclareMathSymbol{\hay}{\mathalpha}{stixsymbols}{"68}

\def\res{\mathop{\text{res}\,}}

\makeatletter
\DeclareFontFamily{OMX}{MnSymbolE}{}
\DeclareSymbolFont{MnLargeSymbols}{OMX}{MnSymbolE}{m}{n}
\SetSymbolFont{MnLargeSymbols}{bold}{OMX}{MnSymbolE}{b}{n}
\DeclareFontShape{OMX}{MnSymbolE}{m}{n}{
    <-6>  MnSymbolE5
   <6-7>  MnSymbolE6
   <7-8>  MnSymbolE7
   <8-9>  MnSymbolE8
   <9-10> MnSymbolE9
  <10-12> MnSymbolE10
  <12->   MnSymbolE12
}{}
\DeclareFontShape{OMX}{MnSymbolE}{b}{n}{
    <-6>  MnSymbolE-Bold5
   <6-7>  MnSymbolE-Bold6
   <7-8>  MnSymbolE-Bold7
   <8-9>  MnSymbolE-Bold8
   <9-10> MnSymbolE-Bold9
  <10-12> MnSymbolE-Bold10
  <12->   MnSymbolE-Bold12
}{}

\let\llangle\@undefined
\let\rrangle\@undefined
\DeclareMathDelimiter{\llangle}{\mathopen}%
                     {MnLargeSymbols}{'164}{MnLargeSymbols}{'164}
\DeclareMathDelimiter{\rrangle}{\mathclose}%
                     {MnLargeSymbols}{'171}{MnLargeSymbols}{'171}
\makeatother

\begin{document}

\thispagestyle{empty}

\begin{center}
\huge \textbf{Lectures on classical Affine Gaudin models} \\~\\
{\Large Sylvain Lacroix} \vspace{12pt}\\
\normalsize \textit{Institute for Theoretical Studies, ETH Z\"urich \\ Clausiusstrasse 47, 8092 Z\"urich, Switzerland} \vspace{12pt}\\
\large \begingroup\ttfamily\href{mailto:sylvain.lacroix@eth-its.ethz.ch}{\color{black}sylvain.lacroix@eth-its.ethz.ch}\endgroup \vspace{25pt}
\end{center}

\noi \textsc{Abstract:} \vspace{7pt}

\noi These lecture notes present an introduction to classical Affine Gaudin models, which provide a general framework for the systematic construction and study of a large class of integrable two-dimensional field theories. A key role is played by Kac-Moody currents, which are fields satisfying a particular Poisson bracket. After reviewing this notion, we discuss in detail the construction of Affine Gaudin models in the language of Hamiltonian field theories. Special emphasis is placed on their symmetries and conserved quantities, including the construction of infinite families of local and non-local Poisson-commuting charges in terms of Kac-Moody currents. Moreover, we study explicit examples of affine Gaudin models, making the link with the realm of integrable sigma-models. Finally, we mention briefly various perspectives concerning these theories, including the question of their quantisation.\vspace{9pt}

\noi Minimal prerequisites on classical Hamiltonian field theories and integrability are required to follow the presentation and a brief reminder of these notions is given at the beginning of the notes. Moreover, various exercises are included throughout the document. These notes were prepared for the Young Researchers Integrability School and Workshop held in Durham from 17 to 21 July 2023.

\newpage

\renewcommand{\baselinestretch}{0.98}\normalsize
\tableofcontents
\renewcommand{\baselinestretch}{1.0}\normalsize

\newpage

\section{Introduction}

Integrable systems are characterised by their high number of symmetries and conserved charges, which in particular allow the development of various methods for the exact computation of their physical observables. Due to this, they play a key role in theoretical and mathematical physics, either to obtain an exact description of various physical phenomena or as toy models for deepening our understanding of non-perturbative effects. An important class of such systems is formed by integrable $\s$-models, which are 2-dimensional field theories with applications in string theory, condensed matter, high-energy physics and various other domains. The construction of explicit examples of integrable $\s$-models and the study of their classical and quantum properties have been growing research subjects since the 70s, which are still currently active. These lectures are meant as an introduction to the formalism of \textit{classical affine Gaudin models}, which is a unifying framework encompassing many integrable $\s$-models and which have been the subject of various developments in the recent years.\\

The notion of Gaudin models was initially introduced by M. Gaudin in~\cite{Gaudin:1976,Gaudin:1983}, in the form of quantum integrable systems of spins in interaction. In that context, these spins are operators forming the Lie algebra $\su(2)$ or more generally a finite-dimensional Lie algebra. The integrable systems under consideration are thus described by a finite number of degrees of freedom and are then sometimes referred to as ``finite'' Gaudin models. These theories have been studied in great detail and were the subjects of many developments in the literature.

In their seminal work~\cite{Feigin:2007mr}, B. Feigin and E. Frenkel considered the Gaudin models obtained by replacing these finite-dimensional Lie algebras by so-called affine Lie algebras, which in particular are infinite-dimensional. The resulting theories, referred to as \textit{affine Gaudin models} (AGMs), thus possess infinitely many degrees of freedom. In fact, the structure of affine Lie algebras is such that these degrees of freedom can be understood as the Fourier modes of certain fundamental fields called Kac-Moody currents, which satisfy specific commutation relations. Crucially, this allows an interpretation of the AGMs as field theories\footnote{The transition from finite to affine Gaudin models thus amounts to passing from a system of interacting spins to a system of interacting fields. We note that there exists another standard way to obtain field theories from spin systems, namely by considering a continuum limit where the number of spins become infinite, in such a way that these degrees of freedom can be repackaged into continuous fields. In that case, the spin chain can thus be seen as a spatial discretisation of the field theory. This is not to be confused with the passage from finite Gaudin models to affine ones, where each spin is essentially replaced by a field, without taking any limit.}. In these lectures, we will focus mostly on the classical version of this construction. We note that these classical models were also introduced independently by A. M. Levin, M. A. Olshanetsky and A. Zotov in~\cite{Levin:2001nm}, using a more geometrical approach and under the name of affine Higgs bundles. More recently, the formalism of classical AGMs and its applications were further developed in the work~\cite{Vicedo:2017cge} by B. Vicedo and many subsequent references. These lectures will present a self-contained review of various of these results (we refer to the main text for further bibliographic details).\\

Let us now sketch the main topics addressed in the lectures.
Classical AGMs are defined in the Hamiltonian formulation, in terms of (classical) Kac-Moody currents. The latter are fields satisfying a specific Poisson bracket called a current algebra, which will play a crucial role in these lectures. The main interest of the AGM construction is that it ensures the existence of an infinite number of commuting symmetries / conserved charges built from the Kac-Moody currents, therefore making the model automatically integrable. This will be the main subject of section \ref{Sec:ClassicalAGM}, following the preliminary chapter \ref{Sec:Reminder}, which contains a general reminder about integrable Hamiltonian field theories. In section \ref{Sec:SpaceTime}, we will further discuss the dynamical aspects and the space-time symmetries of AGMs.

It was proven in~\cite{Vicedo:2017cge} that many known classical integrable $\s$-models can be recast as AGMs, by finding appropriate realisations of the current algebras in terms of the $\s$-model fields. Conversely, it was proposed in~\cite{Delduc:2019bcl} (and further developed in later works) that the formalism of AGMs can be used to construct new examples of integrable $\s$-models in a systematic way, extending consequently the panorama of such theories. This notion of realisation and the applications of AGMs to the study of classical integrable $\s$-models will be the main subjects of section \ref{Sec:Sigma}.\\

Beyond these classical aspects, it is hoped that the formalism of AGMs can also find applications towards the quantisation of integrable $\s$-models. This includes for instance the study of their renormalisation, as well as the question of their quantum integrability and spectrum, following the programme initially proposed by B. Feigin and E. Frenkel in~\cite{Feigin:2007mr} and which was part of the motivation behind the introduction of AGMs. For conciseness, we will not enter into the detail of these quantum perspectives in these lectures and will only give a brief overview in the concluding section \ref{Sec:Conclusion} (in particular, see the discussion there for references).\\

Let us finally note that these lectures also include various exercises, whose difficulty level is indicated using stars, from \raisebox{0.5pt}{$\Large\star$} (easiest) to \raisebox{0.5pt}{$\Large\star\star\star$} (most difficult).

\newpage

\section{Reminder on integrable Hamiltonian field theories}
\label{Sec:Reminder}

The main language that we will use in these lectures is the Hamiltonian formulation of integrable 2-dimensional field theories. The goal of this section is to give a quick reminder of some of the main features of this formalism that we will need in the next sections. For a more complete introduction to the subject, see for instance the textbooks~\cite{Babelon:2003qtg,Faddeev:1987ph} or the lectures~\cite{Driezen:2021cpd}. The reader familiar with these notions is invited to pass to the next section.

\subsection{2-dimensional field theories in the Hamiltonian formulation}
\label{SubSec:HFT}

\paragraph{Space-time.} Let us consider a classical 2-dimensional field theory (not necessarily integrable). We will denote its space-time manifold as $\Sigma$, which we will call the \textit{worldsheet}, following the terminology of $\s$-models and string theory. In the rest of these lectures, we will take this 2-dimensional manifold to be the cylinder, described by a time coordinate $t\in\R$ and a spatial coordinate $x\in[0,2\pi]$, requiring that our fields are periodic under the identification $x\sim x+2\pi$. We note however that most of the discussion below can be adapted to the case where $\Sigma$ is a plane, with a spatial coordinate $x\in\R$, where we ask that the fields decrease sufficiently fast at infinity.

\paragraph{Algebra of observables and Poisson bracket.} We will treat our field theory in the \textit{Hamiltonian formulation}. The first consequence of this approach is that we initially consider fields on a given time-slice, \textit{i.e.} at a fixed value of time $t$. The time-dependence will be reinstated later, as discussed at the end of this subsection. In this context, we now describe the theory in terms of a finite number of fundamental fields $\rho_\alpha(x)$, labelled by an index $\alpha$, and depending only on the space variable $x\in[0,2\pi]$. The physical observables of the theory are then functionals built from these fields, which can take various forms (local combinations of the fields and their derivatives, integrals, ...). This set of observables forms a \textit{Poisson algebra} $\Ac$, equipped with a Poisson bracket $\lbrace \cdot, \cdot \rbrace$. In these lectures, we will always work with brackets taking a local form on the fundamental fields $ \rho_\alpha(x)$, for which we have
\begin{equation}\label{Eq:PbRho}
\bigl\lbrace \rho_\alpha(x), \rho_\beta(y) \bigr\rbrace = \sum_{k=0}^{n_{\alpha,\beta}} F_{\alpha,\beta}^{(k)}\bigl( \rho_\gamma(x),\, \p_x\rho_\gamma(x),\,\dots \bigr) \,\p_x^k \delta(x-y)\,.
\end{equation}
Here, $\p_x^k\delta(x-y)$ is the $k^{\text{th}}$-derivative of the Dirac-distribution, which satisfies
\begin{equation}
\int_0^{2\pi} \p_x^k\delta(x-y) f(x) \dd x = (-1)^k \p_y^k f(y)\,,
\end{equation}
for any test function $f$. Moreover, $n_{\alpha,\beta} \in \Z_{\geq 0}$ is a positive integer and the $F_{\alpha,\beta}^{(k)}$'s are combinations of the fundamental fields and (a finite number of) their derivatives evaluated at the point $x$. The bracket \eqref{Eq:PbRho} is said to be local since it is a distribution with support $x=y$, with coefficients which are local functionals of the fields (in particular, it vanishes when the space coordinates $x$ and $y$ are distinct).

The equation \eqref{Eq:PbRho} fixes the Poisson bracket between any two observables in $\Ac$ built from the fundamental fields $\rho_\alpha(x)$. The resulting bracket $\lbrace \cdot,\cdot \rbrace$ should be skew-symmetric and should obey the Leibniz rule and the Jacobi identity
\begin{equation}\label{Eq:Jac}
\bigl\lbrace \lbrace \Oc_1, \Oc_2 \rbrace, \Oc_3 \bigr\rbrace + \bigl\lbrace \lbrace \Oc_2, \Oc_3 \rbrace, \Oc_1 \bigr\rbrace + \bigl\lbrace \lbrace \Oc_3, \Oc_1 \rbrace, \Oc_2 \bigr\rbrace = 0\,, \qquad \forall\,\Oc_1, \Oc_2, \Oc_3\in \Ac\,.
\end{equation}
In particular, this enforces various constraints that the coefficients $F_{\alpha,\beta}^{(k)}$ in equation \eqref{Eq:PbRho} should satisfy.

\paragraph{Example: canonical fields.} The most famous example of such a Poisson algebra is the one generated by so-called \textit{canonical fields}, which we will have the opportunity to work with later in these lectures. In the simplest case, the fundamental building blocks of this algebra are given by $N$ coordinates fields $\bigl( \phi^{\hspace{0.5pt}i}(x) \bigr)_{i=1,\dots,d}$ and $d$ conjugate momentum fields $\bigl( \pi_i(x) \bigr)_{i=1,\dots,d}$, all valued in $\R$, so that together they form a field valued in $\R^{2d}$. These are the field theory equivalents of the canonical coordinates $(q^i,p_i)$ of classical mechanics and are equipped with the canonical Poisson bracket
\begin{equation}\label{Eq:PbCan}
\bigl\lbrace \pi_i(x), \phi^{\,j}(y) \bigr\rbrace = \delta_{i}^{\;j} \,\delta(x-y), \qquad \bigl\lbrace \pi_i(x), \pi_{j}(y) \bigr\rbrace = \bigl\lbrace \phi^{\hspace{0.5pt}i}(x), \phi^{\,j}(y) \bigr\rbrace = 0
\end{equation}
for $i,j\in\lbrace 1,\dots,d\rbrace$. This clearly defines a particular case of the general local Poisson bracket \eqref{Eq:PbRho}. One easily checks that it satisfies the Jacobi identity, as required.

This setup can be generalised to the case where the canonical fields do not take values in the vector space $\R^{2d}$ but rather form a field valued in a cotangent bundle $T^\ast \Mc$, where $\Mc$ is a smooth manifold of dimension $d$. In that case, one obtains a set of local coordinate fields $\bigl( \phi^{\hspace{0.5pt}i}(x) \bigr)_{i=1,\dots,d}$ describing the $\Mc$-valued component by choosing a chart on $\Mc$ and a set of conjugate momentum fields $\bigl( \pi_i(x) \bigr)_{i=1,\dots,d}$ describing the cotangent fibers. The full description of the fundamental $T^\ast \Mc$-valued field then requires gluing together these local descriptions on different charts, which we will not describe here for simplicity. In a local chart, the fields $\bigl( \phi^{\hspace{0.5pt}i}(x), \pi_i(x) \bigr)_{i=1,\dots,d}$ satisfy the Poisson bracket \eqref{Eq:PbCan}. We will denote the corresponding Poisson algebra as $\Acm$. The configuration above corresponded to the example where $\Mc$ is the vector space $\R^d$, in which case we have global coordinates on $T^\ast \Mc=\R^{2d}$.

Although the algebra of canonical fields $\Acm$ is the most typical example of observables of a Hamiltonian field theory, we will encounter other ones in these lectures, in particular the Poisson algebra generated by classical Kac-Moody currents (see Subsection \ref{SubSec:KM}).

\paragraph{Hamiltonian and dynamics.} As mentioned above, the Poisson algebra $\Ac$ describes the observables of the field theory on a fixed time-slice. To describe the dynamics of this model, \textit{i.e.} define the time dependence of the fields, we have to choose a \textit{Hamiltonian} $\Hc \in \Ac$. In these lectures, it will always take the form of a local charge, \textit{i.e.} the integral
\begin{equation}
\Hc = \int_0^{2\pi} h\bigl( \rho_\alpha(x),\, \p_x \rho_\alpha(x),\, \dots \bigr)\, \dd x
\end{equation}
of a local combination of the fundamental fields and (a finite number of) their derivatives.  Given the choice of this $\Hc$, we then define the time evolution of any observable $\Oc \in \Ac$ as the Hamiltonian flow
\begin{equation}\label{Eq:DynH}
\p_t \Oc = \bigl\lbrace \Hc, \Oc \bigr\rbrace\,.
\end{equation}
In this context, we can then see the fundamental fields $\rho_\alpha(x,t)$ as functions of both the space coordinate $x$ and the time coordinate $t$. In particular, this time dependence is governed by the flow \eqref{Eq:DynH}, which takes the form of a partial differential equation
\begin{equation}\label{Eq:EoMRho}
\p_t \rho_\alpha(x,t) = G_\alpha \bigl(\rho_\beta(x,t), \, \p_x \rho_\beta(x,t), \,\dots \bigr)
\end{equation}
in the $(x,t)$ variables (of first order in $t$), which we refer to as the \textit{equation of motion} of $\rho_\alpha(x,t)$.\\

Let us finally comment on the relation to the Lagrangian formulation of field theories. This occurs when we consider a model based on the Poisson algebra $\Acm$ of canonical fields $(\phi^i,\pi_i)$ valued in a cotangent bundle $T^\ast \Mc$, as introduced in the previous paragraph. In that case, one can write down the equations of motion of the coordinate fields, thus expressing $\p_t\phi^i$ as a local combination of $(\phi^j,\pi_j)$ and their spatial derivatives. We then suppose that we can invert this relation to express the momenta $\pi_j$ as certain combinations of the coordinate fields $\phi^i$ and their derivatives with respect to both $x$ and $t$. Through this relation, the equations of motion for $\p_t \pi_j$ then induce a set of partial differential equations which involve only $\phi^i(x,t)$, but which are now of second order in $t$. This second order dynamics can be equivalently obtained from the variational principle of the action functional\vspace{-2pt}
\begin{equation}
S\bigl[ \phi^i \bigr] = \int_{\R} \left( \int_0^{2\pi} \p_t \phi^{\hspace{0.5pt}i}\,\pi_i\,\dd x - \Hc \right) \dd t\,,\vspace{-2pt}
\end{equation}
whose density is defined as the inverse Legendre transform of the Hamiltonian $\Hc$ (here, a summation over the repeated index $i\in\lbrace 1,\dots,N\rbrace$ is implied). This then takes the form of a Lagrangian 2-dimensional theory, defining the dynamics of a field valued in $\Mc$ -- locally described by coordinate fields $\bigl( \phi^i(x,t) \bigr)_{i=1,\dots,N}$.

\subsection{Integrable field theories and Lax formalism}
\label{SubSec:Lax}

\paragraph{Integrable structures.} We now pass to the review of integrable structures in 2-dimensional field theories. Let us consider the Hamiltonian theory described in the previous subsection, with fundamental fields $\rho_\alpha(x,t)$ and Hamiltonian $\Hc$. The integrability of this theory requires that it admits an infinite number of independent conserved charges which pairwise Poisson-commute. More concretely, this means that there exist functionally independent observables $\lbrace \Q_n \rbrace_{n\in I}\subset \Ac$, built from the fields $\rho_\alpha(x,t)$ and labelled by an index $n$ in an infinite set $I$, such that
\begin{equation}
\p_t \Q_n = \bigl\lbrace \Hc, \Q_n \bigr\rbrace = 0 \qquad \text{ and } \qquad \bigl\lbrace \Q_n, \Q_m \bigr\rbrace = 0\,,
\end{equation}
for all $n,m\in I$. We call this family of charges $\lbrace \Q_n \rbrace_{n\in I}$ the \textit{integrable structure} of the theory.

\paragraph{Lax connection.} In practice, building this integrable structure is a quite complicated task. A significant simplification is provided by the so-called Lax formalism, which we now describe. We say that the theory admits a \textit{Lax connection} if there exists two matrices $\Lc_x(\lambda,x,t)$ and $\Lc_t(\lambda,x,t)$, valued in a complexified Lie algebra $\g^\C$, built from the fundamental fields $\rho_\alpha(x,t)$ and depending meromorphically on an auxiliary complex variable $\lambda\in\C$, such that the equations of motion \eqref{Eq:EoMRho} are equivalent to the zero curvature equation
\begin{equation}\label{Eq:ZCE}
\p_t \Lc_x(\lambda,x,t) - \p_x \Lc_t(\lambda,x,t) + \bigl[ \Lc_t(\lambda,x,t), \Lc_x(\lambda,x,t) \bigr] = 0\,, \qquad \forall \,\lambda\in\C\,.
\end{equation}
Crucially, we ask that this condition holds for all values of the complex variable $\lambda\in\C$, which we call the \textit{spectral parameter}. Geometrically, these two matrices form a connection $\p_\mu + \Lc_\mu(\lambda)$ on $\Sigma$: the property \eqref{Eq:ZCE} is then equivalent to the flatness of this connection.

We call the spatial component $\Lc_x(\lambda,x,t)$ the \textit{Lax matrix} and define its monodromy as the path-ordered exponential
\begin{equation}
\Mc(\lambda,t) = \Pexp\left( - \int_0^{2\pi} \Lc_x(\lambda,x,t)\, \dd x \right)\,,
\end{equation}
which is then valued in a Lie group $G^\C$ with Lie algebra $\g^\C$. This is a non-local observable, which can be defined as an infinite series of nested integrals of $\Lc_x(\lambda,x,t)$. Let $V$ be a finite dimensional representation of the group $G^\C$: the monodromy $\Mc(\lambda,t)$ can then be realised as a matrix acting on $V$. In particular, we can consider its trace
\begin{equation}
\Tc_V(\lambda,t) = \Tr_V \bigl( \Mc(\lambda,t) \bigr)\,.
\end{equation}
Our main interest in this quantity is that, as a consequence of the zero curvature equation \eqref{Eq:ZCE} of $\Lc_\mu(\lambda)$, it is conserved under time evolution, \textit{i.e.}
\begin{equation}\label{Eq:dT}
\p_t \Tc_V(\lambda,t) = 0\,, \qquad \forall \lambda\in\C\,.
\end{equation}
In particular, since $\Tc_V(\lambda,t)$ is a meromorphic function of $\lambda$, we can expand it in power series around any point $\lambda_0\in\C$: the coefficients of this expansion then yield an infinite family of observables which are by construction conserved. The quantity $\Tc_V(\lambda,t)$ then plays the role of generating function for these conserved charges. In particular, the fact that we get an infinite number of such charges relies crucially on the dependence of $\Tc_V(\lambda)$, and thus ultimately of $\Lc_\mu(\lambda)$, on the spectral parameter $\lambda$. This is the main essence of the Lax formalism.\footnote{Although the Lax formalism is the most standard tool used to build integrable structures, let us note that there can exist situations where conserved charges can be built from other considerations. We will in fact encounter such a case in the construction of local charges in affine Gaudin models, in Subsections \ref{SubSec:Quad} and \ref{SubSec:Local}.}

\paragraph{Maillet bracket.} So far, we have used the notion of Lax connection to build an infinite family of conserved charges, extracted from the generating function $\Tc_V(\lambda)$. For these charges to form an integrable structure, they would have to also Poisson-commute pairwise. Since $\Tc_V(\lambda)$ is built from the Lax matrix $\Lc_x(\lambda)$, the Poisson bracket of these charges is ultimately controlled by that of the different components of $\Lc_x(\lambda)$. This raises the following question: can we find a sufficient condition on the Poisson brackets of these components which ensures that $\Tc_V(\lambda)$ and $\Tc_W(\mu)$ Poisson-commute, for any choice of representations $V$ and $W$ and for all values of the spectral parameters $\lambda,\mu\in\C$? This would then prove the integrability of the theory. Such a condition is given by the \textit{Maillet bracket}~\cite{Maillet:1985fn,Maillet:1985ek}, which will play an important role in these lectures and which we now review.\\

For that, it will be useful to first introduce the so-called tensorial notations. Let us suppose that we have two observables $X$ and $Y$ valued in the Lie algebra $\g^\C$. We will denote by $X\ti{1}$ and $X\ti{2}$ the two different embeddings of $X$ in the tensor product\footnote{Here, $U(\g^\C)$ is the universal enveloping algebra of $\g^\C$, whose complete formal definition we will not need. Essentially, it can be seen as the associative algebra generated by an identity element $\Id$ and the generators $\lbrace \td_a \rbrace_{a=1,\dots,\dim\g}$ of $\g^\C$, subject to the commutation relations $\td_a \td_b - \td_b \td_a = \f{ab}c\,\td_c$ (where $\f{ab}c$ are the structure constants of $\g^\C$).} $U(\g^\C) \otimes U(\g^\C)$, defined by
\begin{equation}
X\ti{1} = X \otimes \Id \qquad \text{ and } \qquad X\ti{2} = \Id \otimes X\,.
\end{equation}
Let us also introduce a basis $\lbrace \td_a \rbrace_{a=1,\dots,\dim\g}$ of the Lie algebra $\g$. We decompose $X$ and $Y$ along this basis as $X=X^a\,\td_a$ and $Y=Y^a\,\td_a$, where the components $X^a$ and $Y^a$ are then observables in $\Ac$. We will encode all the Poisson brackets between these components in a unique object
\begin{equation}\label{Eq:PbTens}
\bigl\lbrace X\ti{1} , Y\ti{2} \bigr\rbrace = \bigl\lbrace X^a, Y^b \bigr\rbrace\,\td_a \otimes \td_b\,,
\end{equation}
which is then valued in $\g^\C \otimes \g^\C$. One easily checks that this object is in fact independent of the choice of basis $\lbrace \td_a \rbrace_{a=1,\dots,\dim\g}$. We will use these notations extensively throughout the lectures.

We now have enough material to formulate the Maillet bracket. Let us consider the Lax matrix $\Lc_x(\lambda,x)$, where we omitted the time dependence since we want to consider Poisson brackets and are thus working on a fixed time-slice. It is an observable valued in $\g^\C$ and depending on the space coordinate $x$ and the spectral parameter $\lambda$. In particular, we can encode the Poisson bracket between the different components of this matrix, at different values of these variables, in a unique object $\bigl\lbrace \Lc_x(\lambda,x)\ti{1}, \Lc_x(\mu,y)\ti{2} \bigr\rbrace$ valued in $\g^\C \otimes \g^\C$, using the tensorial notation introduce above. We say that it takes the form of a \textit{Maillet bracket}~\cite{Maillet:1985fn,Maillet:1985ek} if
\begin{align}\label{Eq:Maillet}
\hspace{-10pt}\bigl\lbrace \Lc_x(\lambda,x)\ti{1}, \Lc_x(\mu,y)\ti{2} \bigr\rbrace
&= \bigl[ \Rc(\lambda,\mu)\ti{12}, \Lc_x(\lambda,x)\ti{1} \bigr]\, \delta(x-y) - \bigl[ \Rc(\mu,\lambda)\ti{21}, \Lc_x(\mu,y)\ti{2} \bigr] \, \delta(x-y)\\
& \hspace{40pt} - \left( \Rc(\lambda,\mu)\ti{12} + \Rc(\mu,\lambda)\ti{21} \right) \p_x\delta(x-y)\,, \notag
\end{align}
where $\Rc(\lambda,\mu)$ is a meromorphic function of $(\lambda,\mu)\in\C^2$ valued in $\g^\C \otimes \g^\C$, which we call the \textit{$\Rc$-matrix}. In general, this $\Rc$-matrix could also depend on the fields of the theory and thus on the space coordinates $x,y$ (in which case we would call it dynamical), but for simplicity we will restrict to the case where it is simply a fixed non-dynamical quantity. As mentioned above, this form for the bracket of the Lax matrix implies that
\begin{equation}
\bigl\lbrace \Tc_V(\lambda), \Tc_W(\mu) \bigr\rbrace = 0\,, \qquad \forall\,\lambda,\mu\in\C\,,
\end{equation}
and for all representations $V$ and $W$, hence ensuring the Poisson-commutation of the conserved charges produced by the Lax formalism and thus the integrability of the theory.

An important role in the Maillet bracket \eqref{Eq:Maillet} is played by the $\Rc$-matrix $\Rc(\lambda,\mu) \in \g^\C\otimes\g^\C$. Let us note that it has to satisfy certain specific properties to ensure the consistency of this bracket. In particular, the latter should obey the Jacobi identity \eqref{Eq:Jac}, which in the present case translates to a set of algebraic constraints on $\Rc(\lambda,\mu)$ and $\Lc_x(\lambda)$. A sufficient (but not necessary) condition for these constraints to be satisfied is the \textit{classical Yang-Baxter equation}
\begin{equation}\label{Eq:CYBE}
\left[ \Rc(\lambda_1,\lambda_2)\ti{12}, \Rc(\lambda_1,\lambda_3)\ti{13} \right] + \left[ \Rc(\lambda_1,\lambda_2)\ti{12}, \Rc(\lambda_2,\lambda_3)\ti{23} \right] + \left[ \Rc(\lambda_3,\lambda_2)\ti{32}, \Rc(\lambda_1,\lambda_3)\ti{13} \right] = 0\,,
\end{equation}
in which the $\Rc$-matrix is embedded in various ways in the triple tensor product $U(\g^\C)\otimes U(\g^\C) \otimes U(\g^\C)$.

\section{Classical affine Gaudin models and their integrable structure}
\label{Sec:ClassicalAGM}

In this section, we define the main protagonist of these lectures: a class of integrable classical field theories called affine Gaudin models (AGM)~\cite{Levin:2001nm,Feigin:2007mr,Vicedo:2017cge}. Our plan will be the following. We will start by describing the fundamental fields of the theory, called Kac-Moody currents, which in particular satisfy a specific Poisson bracket. We will then go on with the construction of the integrable structure of this model, which takes the form of an infinite family of Poisson-commuting observables built from the Kac-Moody currents (either as non-local charges, using the Lax formalism, or as higher-spin local charges, following another approach). Finally, we will discuss some additional internal symmetries of the model, its reality conditions and various generalisations of the construction.

Let us note that these results are for the moment purely ``kinematical'', in the sense that they do not require a notion of time evolution. They form a purely algebraic construction, which starts with a specific Poisson algebra generated by fields depending on a spatial coordinate $x$ and which produces an infinite number of Poisson-commuting charges built from these fields. In the following section \ref{Sec:SpaceTime}, we will define the dynamics of the theory by choosing an Hamiltonian $\Hc$ among these Poisson-commuting charges. By construction, this will specify a time evolution $\p_t = \lbrace \Hc,\cdot \rbrace$ with respect to which all these charges will be conserved. We will then obtain a proper integrable 2-dimensional field theory with space-time coordinates $(x,t)$. We postpone the discussion of all these dynamical aspects to Section \ref{Sec:SpaceTime}.

\subsection{Fundamental fields and Poisson algebra: the Kac-Moody currents}
\label{SubSec:KM}

\paragraph{Setup.} Let $\g$ be a real simple Lie algebra, with basis $\lbrace \td_a \rbrace_{a=1,\dots,\dim\g}$. We define its structure constants $\f{ab}c$ by the commutation relations
\begin{equation}
\bigl[ \td_a, \td_b \bigr] = \f{ab}c \, \td_c\,,
\end{equation}
which satisfy the Jacobi identity
\begin{equation}\label{Eq:JacF}
\f{ab}d\,\f{dc}e + \f{bc}d\,\f{da}e + \f{ca}d\,\f{db}e = 0\,.
\end{equation}
We will denote by $\psd$ the ad-invariant non-degenerate bilinear form\footnote{\label{FootForm}We fix the normalisation of $\psd$ so that it coincides with the opposite of the ``minimal'' bilinear form on $\g$, with respect to which the longest roots of $\g$ have length $2$. For instance, $\psd$ coincides with the opposite of the trace form in the fundamental representation for $\g=\mathfrak{su}(n)$. The minus sign was introduced so that the bilinear form is positive definite if $\g$ is compact. The main property of this normalisation is its relation with the Killing form, which reads
\begin{equation}
\ps{X}{Y} = - \frac{1}{2 \hv} \Tr\bigl(\ad_X \circ \ad_Y \bigr)\,, \qquad \forall\,X,Y\in\g\,,
\end{equation}
where $\hv$ is the dual Coxeter number of $\g$.} on $\g$, which then satisfies
\begin{equation}\label{Eq:AdInv}
\ps{[X,Y]}{Z} = \ps{X}{[Y,Z]}\,, \qquad \forall\,X,Y,Z\in\g\,.
\end{equation}
We let $\kappa_{ab} = \ps{\td_a}{\td_b}$ be the entries of this bilinear form in the basis, which then define a symmetric invertible $\dim\g\times\dim\g$ matrix. We will denote its inverse by $\kappa^{ab}$ and will use $\kappa_{ab}$ and $\kappa^{ab}$ to lower and raise Lie algebra indices, as commonly done. In particular, we define the dual basis $\lbrace \td^a \rbrace_{a=1,\dots,\dim\g}$ by $\td^a = \kappa^{ab}\td_b$, which then satisfies $\ps{\td^a}{\td_b}=\delta^a_{\;\,b}$. In components, the invariance \eqref{Eq:AdInv} of $\psd$ induces the following identities:
\begin{equation}\label{Eq:EtaInv}
\f{ab}d\,\kappa_{cd} + \f{ac}d\,\kappa_{bd} = 0 \qquad \text{ and } \qquad  \f{ab}c \, \kappa^{bd} + \f{ab}d \, \kappa^{cb} = 0\,.
\end{equation}

\paragraph{Kac-Moody current.} Let us now define one of the crucial notion used to construct affine Gaudin models, namely that of a \textit{classical Kac-Moody current}. This is a field $\Jc(x)$, depending on a periodic spatial coordinate $x\in[0,2\pi]$ and valued in the complexified\footnote{For most part of this section, we will consider complex fields and observables. We postpone the discussion of reality conditions to Subsection \ref{SubSec:Gen}: these will ensure that all physically relevant quantities are real.} algebra $\g^\C$. The main property of this field is its Poisson bracket, which, in terms of the basis decomposition $\Jc(x) = \Jc_a(x)\,\td^a$, reads
\begin{equation}\label{Eq:PbKM}
\bigl\lbrace \Jc_a(x), \Jc_b(y) \bigr\rbrace = \f{ab}c\,\Jc_c(x)\,\delta(x-y) - \ell\, \kappa_{ab}\,\p_x\delta(x-y)\,.
\end{equation}
Here $\ell\in\C$ is a complex parameter called the \textit{level} of the current. One checks that this bracket is skew-symmetric and satisfies the Jacobi identity, using the one \eqref{Eq:JacF} for the structure constants $\f{ab}c$ and the ad-invariance property \eqref{Eq:EtaInv} of $\kappa_{ab}$: we leave this as an exercise to the reader. The Kac-Moody bracket \eqref{Eq:PbKM} belongs to the general class of local Poisson brackets \eqref{Eq:PbRho} considered in the previous section and provides an example which does not take the form of a canonical bracket \eqref{Eq:PbCan}. The Poisson algebra generated by the fields $\Jc_a(x)$ can in fact be interpreted as the classical version of an infinite dimensional Lie algebra called an affine Kac-Moody algebra, giving their name to the Kac-Moody currents and to the affine Gaudin models.\vspace{10pt}

\begin{tcolorbox} \textit{\underline{Exercise 1:} The Kac-Moody bracket.} \raisebox{0.5pt}{\Large$\;\;\star$} \vspace{5pt}\\
Show that the Kac-Moody bracket \eqref{Eq:PbKM} is skew-symmetric and satisfies the Jacobi identity \eqref{Eq:Jac}, using the properties \eqref{Eq:JacF} and \eqref{Eq:EtaInv} of the structure constants $\f{ab}c$ and bilinear form $\kappa_{ab}$.
\end{tcolorbox}~\vspace{-12pt}

\paragraph{Tensorial notations and split Casimir.} The Poisson bracket \eqref{Eq:PbKM} is defined in terms of the components of $\Jc(x)$ in a basis. It is natural to wonder if one can reformulate this bracket in a basis-independent way. This is done using the tensorial notations, as introduced in equation \eqref{Eq:PbTens} and the surrounding paragraph. We introduce the so-called \textit{split quadratic Casimir}
\begin{equation}\label{Eq:Cas}
C\ti{12} = \kappa^{ab}\, \td_a \otimes \td_b = \td_a \otimes \td^a\,,
\end{equation}
which is valued in the tensor product $\g\otimes\g$. One easily checks that it does not depend on the choice of basis $\lbrace \td_a \rbrace_{a=1,\dots,\dim\g}$, is symmetric (\textit{i.e.} $C\ti{12}=C\ti{21}$) and satisfies the completeness relation
\begin{equation}\label{Eq:CasComp}
\ps{C\ti{12}}{X\ti{2}\hspace{1pt}}\null\hspace{-1pt}\ti{2} = X\,, \qquad \forall\,X\in\g\,,
\end{equation}
where $\psd\ti{2}$ means that we are taking the bilinear form on the second tensor factor. Moreover, the ad-invariance property \eqref{Eq:EtaInv} of the bilinear form implies the identity
\begin{equation}\label{Eq:CasId}
\bigl[ C\ti{12}, X\ti{1} + X\ti{2} \bigr] = 0 \,, \qquad \forall\,X\in\g\,.
\end{equation}
Using this language, one rewrites the Kac-Moody bracket \eqref{Eq:PbKM} in the following basis-independent way:
\begin{equation}\label{Eq:PbKMtens}
\bigl\lbrace \Jc(x)\ti{1}, \Jc(y)\ti{2} \bigr\rbrace = \bigl[ C\ti{12}, \Jc(x)\ti{1} \bigr]\, \delta(x-y) - \ell\,C\ti{12}\,\p_x\delta(x-y)\,.
\end{equation}
We leave the proof of this reformulation as an exercise to the reader. The skew-symmetry of this bracket is made manifest by the Casimir identity \eqref{Eq:CasId}.\vspace{10pt}

\begin{tcolorbox} \textit{\underline{Exercise 2:} The Kac-Moody bracket in tensorial notations.} \raisebox{0.5pt}{\Large$\;\;\star\star$} \vspace{5pt}\\
We strongly encourage the reader to do this short exercise, to gain some familiarity and intuition with the manipulation of tensorial notations, since the later will be used extensively throughout these lectures.\vspace{4pt}\\
1. Show that the Kac-Moody bracket \eqref{Eq:PbKM} can be reformulated as \eqref{Eq:PbKMtens} using tensorial notations.\vspace{4pt}\\
2. Establish the skew-symmetry and Jacobi identity of this bracket in tensorial notations, using the Jacobi identity for the Lie bracket $[\cdot,\cdot]$ and the property \eqref{Eq:CasId} of the Casimir.\vspace{2pt}\\
\textit{Hint:} first convince yourself that, in tensorial notations, the Jacobi identity amounts to the vanishing of the $(\g^\C\otimes\g^\C\otimes\g^\C)$--valued quantity
\begin{equation*}
\Bigl\lbrace \bigl\lbrace \Jc(x_1)\ti{1}, \Jc(x_2)\ti{2} \bigr\rbrace, \Jc(x_3)\ti{3} \Bigr\rbrace + \Bigl\lbrace \bigl\lbrace \Jc(x_2)\ti{2}, \Jc(x_3)\ti{3} \bigr\rbrace , \Jc(x_1)\ti{1} \Bigr\rbrace + \Bigl\lbrace \bigl\lbrace \Jc(x_3)\ti{3},  \Jc(x_1)\ti{1}\bigr\rbrace , \Jc(x_2)\ti{2} \Bigr\rbrace\,.
\end{equation*}
\end{tcolorbox}~\vspace{-12pt}

\paragraph{Momentum.} Let us consider the quantity
\begin{equation}\label{Eq:PJ}
\Pc_{\Jc} = \frac{1}{2\ell} \int_0^{2\pi} \psb{\Jc(x)}{\Jc(x)}\,\dd x\, = \frac{\kappa^{ab}}{2\ell} \int_0^{2\pi} \Jc_a(x)\,\Jc_b(x)\,\dd x\,.
\end{equation}
One shows (see exercise below) that
\begin{equation}\label{Eq:PbPj}
\bigl\lbrace \Pc_{\Jc}, \Jc(x) \bigr\rbrace = \p_x \Jc(x)\,.
\end{equation}
The observable $\Pc_{\Jc}$ is then the momentum of the Kac-Moody current $\Jc(x)$, \textit{i.e.} the generator of infinitesimal spatial translations $\p_x = \lbrace \Pc_{\Jc},\cdot\rbrace$. The reader familiar with 2-dimensional CFTs might recognise in the definition \eqref{Eq:PJ} a classical version of the Segal-Sugawara construction -- see \textit{e.g.}~\cite{DiFrancesco:1997nk}.$\!$ \vspace{10pt}

\begin{tcolorbox} \textit{\underline{Exercise 3:} Momentum of the Kac-Moody current.} \raisebox{0.5pt}{\Large$\;\;\star$} \vspace{5pt}\\
Show equation \eqref{Eq:PbPj} from the Poisson bracket of the Kac-Moody current $\Jc(x)$, either in components \eqref{Eq:PbKM} or in tensorial notations \eqref{Eq:PbKMtens}.
\end{tcolorbox}~\vspace{-12pt}

\paragraph{Poisson algebra of the AGM.} The algebra of observables $\Ac$ of an affine Gaudin model is defined as the Poisson algebra of $N$ independent Kac-Moody currents $\Jc_1(x),\dots,\Jc_N(x)$, where $N\in\Z_{\geq 1}$ is a positive integer. Concretely, this means that $\Ac$ is generated by $N\times\dim\g$ scalar fields $\bigl( \Jc_{r,a}(x) \bigr)_{r=1,\dots,N}^{a=1,\dots,\dim\g}$ -- the components of the currents -- satisfying the Poisson bracket
\begin{equation}\label{Eq:PbJa}
\bigl\lbrace \Jc_{r,a}(x), \Jc_{s,b}(y) \bigr\rbrace = \delta_{rs}\Bigl( \f{ab}c\,\Jc^c_r(x)\,\delta(x-y) - \ell_r\, \kappa_{ab}\,\p_x\delta(x-y)\Bigr)\,.
\end{equation}
This bracket vanishes for $r\neq s$ and depends on $N$ continuous parameters, the levels $\ell_r\in\C$ of the currents. Equivalently, in tensorial notations, we have
\begin{equation}\label{Eq:PbJ}
\bigl\lbrace \Jc_r(x)\ti{1}, \Jc_s(y)\ti{2} \bigr\rbrace = \delta_{rs} \Bigl( \bigl[ C\ti{12}, \Jc_r(x)\ti{1} \bigr]\, \delta(x-y) - \ell_r\,C\ti{12}\,\p_x\delta(x-y) \Bigr)\,.
\end{equation}
In ``physical language'', the phase space of the AGM is the (infinite-dimensional) space of configurations of the fields $\Jc_{r,a}(x)$, equipped with the bracket \eqref{Eq:PbJa}, which is thus determined by the choice of the Lie algebra $\g$, the number of currents $N$ and the collection of levels $\ell_1,\dots,\ell_N$. 

\subsection{Lax matrix and non-local charges}
\label{SubSec:LaxG}

We now come to the construction of Poisson-commuting non-local charges in AGMs, using the Lax formalism reviewed in Subsection \ref{SubSec:Lax}.\vspace{-4pt}

\paragraph{Gaudin Lax matrix, twist function and sites.} Let us pick $N$ complex numbers $\sigma_1,\dots,\sigma_N\in\C$, which we call the \textit{sites} of the AGM. We think of these as points $\sigma_r$ in the complex plane, to which we formally attach the Kac-Moody currents $\Jc_r(x)$ introduced earlier. From this data, we define the \textit{Gaudin Lax matrix} as the following meromorphic function of the spectral parameter $\lambda\in\C$:
\begin{equation}\label{Eq:Gaudin}
\Gamma(\lambda,x) = \sum_{r=1}^N \frac{\Jc_r(x)}{\lambda-\sigma_r}\,.
\end{equation}
Note that $\Gamma(\lambda,x)$ will not be the Lax matrix of the theory, in the precise sense of Subsection \ref{SubSec:Lax}, but will be closely related to it (this terminology comes from a more abstract construction of AGMs, which we do not develop here). In particular, it allows to encode all the fields $\Jc_r(x)$ of the theory in a unique object, using the spectral parameter dependence as the fundamental organising principle.\\

Similarly, we define the so-called \textit{twist function} of the model as
\begin{equation}\label{Eq:Twist}
\vp(\lambda) = \sum_{r=1}^N \frac{\ell_r}{\lambda-\sigma_r} - \ell_\infty\,,
\end{equation}
in terms of the levels $\ell_r$ and an additional constant term\footnote{\label{Footnote:Inf}This constant term can be seen as a double pole at $\lambda=\infty$ in the 1-form $\vp(\lambda)\,\dd \lambda$ and can thus be formally interpreted as coming from a site at infinity in the Riemann sphere $\CP=\C \sqcup \lbrace \infty \rbrace$. Such a site is treated in a slightly different way as the finite ones $\sigma_r\in\C$ (in particular, there is no current attached to it): we will not develop this formulation further here and will simply treat $\ell_\infty$ as encoding the freedom to add a constant term in the twist function.} $\ell_\infty$, which we suppose are all non-vanishing. This function encodes in a unique object all the continuous parameters defining the AGM, namely:\vspace{-3pt}
\begin{itemize}\setlength\itemsep{0.2pt}
\item the sites $\bm{\sigma}=(\sigma_1,\dots,\sigma_N) \in \C^N$ ;
\item the levels $\bm{\ell}=(\ell_1,\dots,\ell_N,\ell_\infty) \in (\C\setminus\lbrace 0\rbrace)^{N+1}$.
\end{itemize}

\paragraph{The key formula.} The Gaudin Lax matrix \eqref{Eq:Gaudin} is defined in terms of the Kac-Moody currents $\Jc_r(x)$, which satisfy the Poisson bracket \eqref{Eq:PbJ}. One can thus determine the bracket of this matrix with itself. This turns out to take a quite simple form, which will serve as a key formula for our construction. Due to the importance of this result, we detail its proof below. From the definition \eqref{Eq:Gaudin} and using tensorial notations, we have
\begin{equation*}
\bigl\lbrace \Gamma(\lambda,x)\ti{1}, \Gamma(\mu,y)\ti{2} \bigr\rbrace = \sum_{r,s=1}^N \frac{1}{(\lambda-\sigma_r)(\mu-\sigma_s)} \bigl\lbrace \Jc_r(x)\ti{1}, \Jc_s(y)\ti{2} \bigr\rbrace\,.
\end{equation*}
Reinserting the fundamental Kac-Moody bracket \eqref{Eq:PbJ} and using the Kronecker delta $\delta_{rs}$ to perform the sum over $s$, we get
\begin{equation*}
\bigl\lbrace \Gamma(\lambda,x)\ti{1}, \Gamma(\mu,y)\ti{2} \bigr\rbrace = \sum_{r=1}^N \frac{1}{(\lambda-\sigma_r)(\mu-\sigma_r)} \Bigl( \bigl[ C\ti{12}, \Jc_r(x)\ti{1} \bigr]\, \delta(x-y) - \ell_r\,C\ti{12}\,\p_x\delta(x-y) \Bigr)\,.
\end{equation*}
We now use the circle lemma
\begin{equation}
\frac{1}{(\lambda-\sigma_r)(\mu-\sigma_r)} = \frac{1}{\mu-\lambda} \left( \frac{1}{\lambda-\sigma_r} - \frac{1}{\mu-\sigma_r}  \right)\,,
\end{equation}
which is simply proven by putting all fractions on the same denominator. We then rewrite the above bracket as
\begin{align*}
\hspace{-10pt}\bigl\lbrace \Gamma(\lambda,x)\ti{1}, \Gamma(\mu,y)\ti{2} \bigr\rbrace &= \frac{1}{\mu-\lambda} \sum_{r=1}^N \left( \left[ C\ti{12}, \frac{\Jc_r(x)\ti{1}}{\lambda-\sigma_r} - \frac{\Jc_r(x)\ti{1}}{\mu-\sigma_r} \right]\, \delta(x-y)  \right. \\
& \hspace{90pt} \left. -\, C\ti{12} \left( \frac{\ell_r}{\lambda-\sigma_r} - \frac{\ell_r}{\mu-\sigma_r} \right) \p_x\delta(x-y) \right)\,.
\end{align*}
Noting that the sums over $r$ form the expressions for the Gaudin Lax matrix \eqref{Eq:Gaudin} and the twist function \eqref{Eq:Twist}, we finally put the bracket in the form\footnote{\label{Footnote:Linf}Note that the constant term $\ell_\infty$ in the twist function \eqref{Eq:Twist} disappears from the combination $\vp(\lambda)-\vp(\mu)$. The key formula \eqref{Eq:PbGaudin} then holds for any value of this term.}
\begin{equation}\label{Eq:PbGaudin}
\bigl\lbrace \Gamma(\lambda,x)\ti{1}, \Gamma(\mu,y)\ti{2} \bigr\rbrace = -\left[ C\ti{12},\frac{\Gamma(\lambda,x)\ti{1} - \Gamma(\mu,x)\ti{1}}{\lambda-\mu} \right]\, \delta(x-y) + C\ti{12}\,\frac{\vp(\lambda)-\vp(\mu)}{\lambda-\mu} \, \p_x\delta(x-y)\,.
\end{equation}
In particular, this key formula follows a quite simple structure, where the dependence on the number and the positions of the sites $\sigma_1,\dots,\sigma_N$ is completely reabsorbed in the functions $\Gamma(\lambda)$ and $\vp(\lambda)$. This will allow us to treat all AGMs in a uniform way.

\paragraph{Lax matrix and Maillet bracket.} Using the Casimir identity \eqref{Eq:CasId} and the fact that $C\ti{12}=C\ti{21}$, the key formula \eqref{Eq:PbGaudin} can be rewritten in the following suggestive form
\begin{align}\label{Eq:PbGaudinR}
\hspace{-40pt}\bigl\lbrace \Gamma(\lambda,x)\ti{1}, \Gamma(\mu,y)\ti{2} \bigr\rbrace
&= \bigl[ \Rc^0(\lambda,\mu)\ti{12}, \Gamma(\lambda,x)\ti{1} \bigr] \delta(x-y) - \bigl[ \Rc^0(\mu,\lambda)\ti{21}, \Gamma(\mu,y)\ti{2} \bigr] \delta(x-y)\\
& \hspace{40pt} - \left( \Rc^0(\lambda,\mu)\ti{12}\,\vp(\lambda) + \Rc^0(\mu,\lambda)\ti{21}\vp(\mu) \right) \p_x\delta(x-y)\,, \notag
\end{align}
where we defined
\begin{equation}\label{Eq:R0}
\Rc^0(\lambda,\mu)\ti{12} = \frac{C\ti{12}}{\mu-\lambda} \quad \in \quad \g^\C\otimes\g^\C\,.
\end{equation}
From the Casimir identity \eqref{Eq:CasId} and with a bit of algebra, one easily checks that $\Rc^0(\lambda,\mu)$ satisfies the classical Yang-Baxter equation \eqref{Eq:CYBE}: we will call it the \textit{standard $\Rc$-matrix}.

The bracket \eqref{Eq:PbGaudinR} takes a form quite similar to that of a Maillet bracket \eqref{Eq:Maillet}, except for the presence of $\vp(\lambda)$ and $\vp(\mu)$ in the $\p_x\delta(x-y)$ term. We can remove those by considering
\begin{equation}\label{Eq:Lax}
\Lc_x(\lambda,x) = \frac{\Gamma(\lambda,x)}{\vp(\lambda)}\,,
\end{equation}
which, as the notation suggests, will be identified as the \textit{Lax matrix} of the model. The equation \eqref{Eq:PbGaudinR} then implies that $\Lc_x(\lambda)$ satisfies a Maillet bracket \eqref{Eq:Maillet}, with $\Rc$-matrix
\begin{equation}\label{Eq:RTwist}
\Rc(\lambda,\mu)\ti{12} = \Rc^0(\lambda,\mu)\ti{12} \; \vp(\mu)^{-1}\,.
\end{equation}
The latter obeys the classical Yang-Baxter equation \eqref{Eq:CYBE}, since $\Rc^0$ does: indeed, passing from $\Rc^0$ to $\Rc$ on the left-hand side of equation \eqref{Eq:CYBE} simply amounts to a multiplication by a global factor $\vp(\lambda_2)^{-1}\vp(\lambda_3)^{-1}$. We will call $\Rc(\lambda,\mu)$ an $\Rc$-matrix with twist function\footnote{Such $\Rc$-matrices, built by ``twisting'' the standard $\Rc^0(\lambda,\mu)$ by $\vp(\mu)^{-1}$, were studied \textit{e.g.} in~\cite{Maillet:1985ec,Reyman:1988sf,Sevostyanov:1995hd,Vicedo:2010qd}.}. We note that, contrarily to $\Rc^0(\lambda,\mu)$, it is not skew-symmetric: in particular, this implies the presence of a term proportional to $\p_x\delta(x-y)$ in the Maillet bracket \eqref{Eq:Maillet}, which is called the non-ultralocal term.

\paragraph{Poisson-commuting non-local charges.} Now that we identified a Lax matrix satisfying a Maillet bracket, we can apply the general procedure reviewed in Subsection \ref{SubSec:Lax} to construct an infinite family of (generally non-local) charges which are pairwise Poisson-commuting. Following this procedure, these charges are encoded in the generating function
\begin{equation}\label{Eq:T}
\Tc_V(\lambda) = \Tr_V \left[ \Pexp\left( - \int_0^{2\pi} \Lc_x(\lambda,x)\,\dd x \right) \right]\,,
\end{equation}
where the path-orderered exponential is valued in a group $G^\C$ with Lie algebra $\g^\C$ and $\Tr_V$ denotes the trace in a representation $V$ of $G^\C$. We recall that, as a consequence of the Maillet bracket of $\Lc_x(\lambda)$, one has\footnote{Although we derived the equation \eqref{Eq:PbT} from the Maillet bracket here, let us note that this result was initially proposed in~\cite{Feigin:2007mr} using a different argument, based on the abstract construction of AGMs in terms of affine Lie algebras.}
\begin{equation}\label{Eq:PbT}
\bigl\lbrace \Tc_V(\lambda), \Tc_W(\mu) \bigr\rbrace = 0\,, \qquad \forall \,\lambda,\mu\in\C\,,
\end{equation}
for any two representations $V$ and $W$, ensuring that we indeed get Poisson-commuting charges by expanding $\Tc_V(\lambda)$ as a power series in $\lambda$. These charges are in general non-local and are built from the fundamental fields $\Jc_r(x)$ as complicated linear combinations of nested integrals
\begin{equation}
\int_0^{2\pi} \dd x_1 \int_0^{x_1} \dd x_2\, \cdots \int_0^{x_{k-1}} \dd x_k \; \Tr_V\bigl( \Jc_{r_1}(x_1) \cdots \Jc_{r_k}(x_k)\bigr)\,,
\end{equation}
with coefficients depending on the sites $\bm{\s}$ and the levels $\bm{\ell}$, in a way which ensures their Poisson-commutativity. They form half of the integrable structure of the theory (the other half being composed by higher-degree local charges, as we will see in the next subsections) and are thus an important outcome of the AGM construction.

\paragraph{Analytical structure of the Lax matrix.} Let us now give a brief description of the analytical structure of $\Lc_x(\lambda)$ as a function of the spectral parameter $\lambda$. From its definition \eqref{Eq:Lax} in terms of $\Gamma(\lambda)$ and $\vp(\lambda)$ and the expressions \eqref{Eq:Gaudin} and \eqref{Eq:Twist} of the latter, it is clear that $\Lc_x(\lambda)$ has poles at the zeroes of $\vp(\lambda)$. We will suppose that these zeroes are all simple. We denote them by $\bm{\ze}=(\ze_1,\dots,\ze_N)\in \C^N$. A direct analysis of the behaviour of $\Lc_x(\lambda)$ around $\lambda=\ze_i$ and $\lambda=\infty$ shows that
\begin{equation}\label{Eq:LPoles}
\Lc_x(\lambda) = \sum_{i=1}^N \frac{1}{\vp'(\ze_i)} \frac{\Gamma(\ze_i)}{\lambda-\ze_i}\,.
\end{equation}
Note that the quantities $\vp'(\ze_i)$ are non-vanishing since we supposed that $\ze_i$ is a simple zero of $\vp(\lambda)$.

\paragraph{Remarks on the parameters.} So far, we have seen the parameters of the AGM as the sites $\bm{\sigma}=(\sigma_1,\dots,\sigma_N) \in \C^N$ and the levels $\bm{\ell}=(\ell_1,\dots,\ell_N,\ell_\infty) \in (\C\setminus\lbrace 0\rbrace)^{N+1}$. In the previous paragraph, we have shown that the analytical structure of the Lax matrix $\Lc_x(\lambda)$ is directly related to the zeroes $\bm{\ze}=(\ze_1,\dots,\ze_N)\in \C^N$ of the twist function. These zeroes are the roots of a polynomial of degree $N$, with coefficients depending on the defining parameters $\bm{\sigma}$ and $\bm{\ell}$: in practice, it is thus a difficult, if not impossible, task to express them in terms of these parameters. Therefore, it is  often useful to think of the defining data of the model as the the sites $\bm{\sigma}$, the zeroes $\bm{\ze}$ and the constant term $\ell_\infty$ instead. Indeed, this data completely specifies the twist function in its factorised form
\begin{equation}
\vp(\lambda) = -\ell_\infty \frac{\prod_{i=1}^N (\lambda-\ze_i)}{\prod_{r=1}^N (\lambda-\sigma_r)}\,.
\end{equation}
In this formulation, the levels $\ell_r = \res_{\lambda=\sigma_r} \vp(\lambda)\,\dd \lambda$ are seen as specific rational combinations of the defining parameters $(\sigma_r,\ze_i,\ell_\infty)$.\\

Let us finally observe that there is a certain redundancy in the data of the $2N+1$ parameters $(\sigma_r,\ze_i,\ell_\infty)$. Indeed, let us consider the transformation
\begin{equation}\label{Eq:ChangeSpec}
\sigma_r \longmapsto a\sigma_r + b\,, \qquad \ze_i \longmapsto a\ze_i + b\,, \qquad \ell_\infty \longmapsto a^{-1} \ell_\infty\,,
\end{equation}
where $a$ and $b$ are complex numbers, with $a$ non-zero. This leaves the levels $\ell_r$ invariant and thus does not change the algebra of observables of the model. Moreover, one shows that it leaves the Lax matrix $\Lc_x(\lambda)$ invariant if we also transform the spectral parameter according to $\lambda\mapsto a\lambda+b$.\footnote{We note that the Gaudin Lax matrix and the twist function are also left invariant under this transformation if we see them as 1-forms $\Gamma(\lambda)\,\dd\lambda$ and $\vp(\lambda)\,\dd\lambda$ on $\C$, rather than functions (with $\dd\lambda$ transforming as $a\,\dd\lambda$).} Thus, the family of Poisson-commuting charges extracted from the generating function \eqref{Eq:T}, which is the main outcome of the above construction, is left unaffected under this change of parameters. This redundancy can be fixed by setting two of the parameters to specific values: the model is then described by the $2N-1$ remaining ones. 

\subsection{Quadratic local charges}
\label{SubSec:Quad}

\paragraph{Motivation.} In the previous subsection, we have shown how to build Poisson-commuting charges from the generating function $\Tc_V(\lambda)$ defined in equation \eqref{Eq:T}. As mentioned earlier, these quantities are generally non-local when expressed in terms of the Kac-Moody currents. However, it is known that the charges obtained from the series expansion around a pole of the Lax matrix $\Lc_x(\lambda)$ are in fact local: we refer to the textbook~\cite{Babelon:2003qtg} for details about this construction, which goes under the name of abelianisation procedure. In the present case, the poles of $\Lc_x(\lambda)$ are located at the zeroes $\ze_i$ of $\vp(\lambda)$: one can thus extract local charges from the expansion at these points. Doing so, one finds that their densities, despite being local, are in fact quite complicated objects, which for instance involve square roots of fields. Moreover, these charges are all dimension-less: thus, they cannot encode the momentum and Hamiltonian of a relativistic field theory, which we would want to belong to the integrable structure. This motivates the search for other types of local charges in involution, which are not contained in the generating function $\Tc_V(\lambda)$ and have non-trivial scaling dimensions. As we shall now explained, this turns out to be possible. Here, we will focus on the simplest examples of such charges, whose densities are built as quadratic combinations of the Kac-Moody currents. We will discuss their higher-degree generalisations in the next subsection.

\paragraph{Main result.} Recall that $\bm{\ze}=(\ze_1,\dots,\ze_N) \in \C^N$ are the zeroes of the twist function $\vp(\lambda)$, which we supposed are simple. We associate a local charge with each of these zeroes, defined as:
\begin{equation}\label{Eq:Qi}
\Q_i = -\frac{1}{2\vp'(\ze_i)} \int_{0}^{2\pi} \psb{\Gamma(\ze_i,x)}{\Gamma(\ze_i,x)} \, \dd x\,,
\end{equation}
where the prefactor has been introduced for future convenience. By construction, it is clear that the density of this charge is a quadratic polynomial in the fundamental fields $\Jc_r^a(x)$. Moreover, one finds that $\Q_i$ is of scaling dimension $1$, \textit{i.e.} is expressed in units of inverse length. For completeness, we also note that it can be interpreted more formally as the residue
\begin{equation}\label{Eq:QiRes}
\Q_i = - \begin{array}c \text{res}\\[-3pt] \lambda=\ze_i \end{array} \int_0^{2\pi} \frac{\psb{\Gamma(\lambda,x)}{\Gamma(\lambda,x)}}{2\,\vp(\lambda)} \,\dd x\,\dd\lambda\,.
\end{equation}
Our main claim, that we shall prove below, is that these charges Poisson-commute between themselves and with the non-local ones. More precisely, we have
\begin{equation}\label{Eq:PbQ}
\bigl\lbrace \Q_i, \Q_j \bigr\rbrace = 0 \qquad \text{ and } \qquad \bigl\lbrace \Q_i, \Tc_V(\lambda) \bigr\rbrace = 0  \,,
\end{equation}
for any $i,j\in\lbrace 1,\dots, N\rbrace$, $\lambda\in\C$ and representation $V$.

\paragraph{Some useful Poisson brackets.} In order to prove the result \eqref{Eq:PbQ}, let us derive some useful intermediate Poisson brackets. We introduce
\begin{equation}
\Wc(\lambda,x) = \frac{1}{2} \psb{\Gamma(\lambda,x)}{\Gamma(\lambda,x)}\,.
\end{equation}
We now want to determine the Poisson bracket of this quantity with the Gaudin Lax matrix. This computation uses some of the standard techniques that will be employed throughout the lectures: we thus detail it as a simple illustration of these ideas and will then allow ourselves to skip some of the details in later computations. Using tensorial notations and the Leibniz rule, one finds
\begin{equation*}
\bigl\lbrace \Gamma(\lambda,x), \Wc(\mu,y) \bigr\rbrace = \Bigl\langle \bigl\lbrace \Gamma(\lambda,x)\ti{1}, \Gamma(\mu,y)\ti{2} \bigr\rbrace , \Gamma(\mu,y)\ti{2} \Bigr\rangle\ti{2}\,,
\end{equation*}
where we recall that the $\psd\ti{2}$ means we apply the bilinear form on the second tensor factor, so that the end result is a $\g^\C$-valued quantity. We can now use our key formula \eqref{Eq:PbGaudin} for the Poisson bracket of $\Gamma$ with itself. We then rewrite the above bracket as
\begin{equation*}
-\left[ \psb{C\ti{12}}{\Gamma(\mu,x)\ti{2}},\frac{\Gamma(\lambda,x)\ti{1} - \Gamma(\mu,x)\ti{1}}{\lambda-\mu} \right]\, \delta(x-y) + \psb{C\ti{12}}{\Gamma(\mu,y)\ti{2}}\,\frac{\vp(\lambda)-\vp(\mu)}{\lambda-\mu} \, \p_x\delta(x-y)\,.
\end{equation*}
Note that we have exchanged the order of application of the bilinear form $\psd\ti{2}$ and of the commutator in the first term, which we are allowed to do since they act on different tensor spaces. Applying the completeness identity \eqref{Eq:CasComp}, we thus get
\begin{equation}\label{Eq:PbGaudinW}
\bigl\lbrace \Gamma(\lambda,x), \Wc(\mu,y) \bigr\rbrace = \frac{\bigl[\Gamma(\lambda,x),\Gamma(\mu,x) \bigr]}{\lambda-\mu}\, \delta(x-y) + \Gamma(\mu,y)\,\frac{\vp(\lambda)-\vp(\mu)}{\lambda-\mu} \, \p_x\delta(x-y)\,,
\end{equation}
where we also used $[\Gamma(\mu,x),\Gamma(\mu,x)]=0$.\\

This is the first result that we will need. To get the second one, we have to reinsert equation \eqref{Eq:PbGaudinW} into the bracket
\begin{equation*}
\bigl\lbrace \Wc(\lambda,x), \Wc(\mu,y) \bigr\rbrace = \Bigl\langle \bigl\lbrace \Gamma(\lambda,x), \Wc(\mu,y) \bigr\rbrace, \Gamma(\lambda,x) \Bigr\rangle\,.
\end{equation*}
The contribution of the first term of \eqref{Eq:PbGaudinW} is proportional to $\bigl\langle [\Gamma(\lambda),\Gamma(\mu)], \Gamma(\lambda)\bigr\rangle$, which vanishes by ad-invariance of $\psd$. We thus get a unique term
\begin{equation}\label{Eq:PbWW}
\bigl\lbrace \Wc(\lambda,x), \Wc(\mu,y) \bigr\rbrace = \bigl\langle \Gamma(\lambda,x),\Gamma(\mu,y)\bigr\rangle\,\frac{\vp(\lambda)-\vp(\mu)}{\lambda-\mu} \, \p_x\delta(x-y)\,.
\end{equation}

\paragraph{Poisson-commutation of the $\bm{\Q_i}$'s.} We now have enough material to prove the first equation in \eqref{Eq:PbQ}, \textit{i.e.} $\lbrace \Q_i,\Q_j \rbrace=0$. It is clear that this is true for $i=j$ by skew-symmetry. For the case $i\neq j$, we evaluate the Poisson bracket \eqref{Eq:PbWW} at $\lambda=\ze_i$ and $\mu=\ze_j$. Using the fact that $\vp(\ze_i)=\vp(\ze_j)=0$, we simply find that the bracket vanishes: thus, not only do we have the Poisson-commutation of $\Q_i$ and $\Q_j$ but in fact, we see that their densities themselves Poisson-commute.\\

Although it is not necessary at this point, it will be useful later to compute the Poisson algebra of the densities
\begin{equation}\label{Eq:q}
q_i(x) = -\frac{\psb{\Gamma(\ze_i,x)}{\Gamma(\ze_i,x)}}{2\vp'(\ze_i)} =  -\frac{\Wc(\ze_i,x)}{\vp'(\ze_i)}
\end{equation}
of the charges $\Q_i$. As observed above, they Poisson-commute for $i\neq j$. For $i=j$, the evaluation of \eqref{Eq:PbWW} at $\lambda=\mu=\ze_i$ produces a derivative of the twist function, hence
\begin{equation*}
\bigl\lbrace q_i(x), q_i(y) \bigr\rbrace = \frac{ \bigl\langle \Gamma(\ze_i,x),\Gamma(\ze_i,y)\bigr\rangle}{\vp'(\ze_i) }\, \p_x\delta(x-y)\,.
\end{equation*}
To conclude, we need the identity
\begin{equation}\label{Eq:DerDelta}
f(x,y)\,\p_x\delta(x-y) = \p_y f(x,y)\bigl|_{y=x} \,\delta(x-y) + f(x,y)\bigl|_{y=x}\,\p_x\delta(x-y)\,.
\end{equation}
Using it and recognising $q_i(x)$ and its derivative in the computation, we finally get
\begin{equation}\label{Eq:Pbq}
\bigl\lbrace q_i(x), q_j(y) \bigr\rbrace = -\delta_{ij} \Bigl( \p_x q_i(x)\,\delta(x-y) + 2 q_i(x)\, \p_x\delta(x-y)\Bigr) \,.
\end{equation}
This is in fact $N$ independent copies of the classical Virasoro algebra (with vanishing central charge). This gives another example of a local Poisson algebra of the form \eqref{Eq:PbRho}.

\paragraph{Zero curvature equation.} We now want to prove the second equation in \eqref{Eq:PbQ}. Recall that the generating function $\Tc_V(\lambda)$ of the non-local charges is built from the Lax matrix $\Lc_x(\lambda)$ as in equation \eqref{Eq:T}. It will thus be useful to compute the Poisson bracket of $\Q_i$ with this matrix. For that, recall the expression \eqref{Eq:Lax} of $\Lc_x(\lambda)$ and that \eqref{Eq:q} of the density $q_i$ of $\Q_i$. We thus get
\begin{equation*}
\bigl\lbrace q_i(y), \Lc_x(\lambda,x) \bigr\rbrace = \frac{\bigl\lbrace \Gamma(\lambda,x), \Wc(\ze_i,y) \bigr\rbrace}{\vp(\lambda)\,\vp'(\ze_i)}\,.
\end{equation*}
We can now use the intermediate bracket \eqref{Eq:PbGaudinW} computed earlier, evaluating it at $\mu=\ze_i$ and using $\vp(\ze_i)=0$. We get
\begin{equation*}
\bigl\lbrace q_i(y), \Lc_x(\lambda,x) \bigr\rbrace = \bigl[\Lc_x(\lambda,x), \Kc_i(\lambda,x) \bigr]\, \delta(x-y) + \Kc_i(\lambda,y) \, \p_x\delta(x-y)\,,
\end{equation*}
where
\begin{equation}
\Kc_i(\lambda,x) = \frac{1}{\vp'(\ze_i)}\frac{\Gamma(\ze_i,x)}{\lambda-\ze_i}\,.
\end{equation}
We finally integrate over $y$ using the $\delta$-distribution and its derivative, yielding
\begin{equation}\label{Eq:ZCEi}
\bigl\lbrace \Q_i, \Lc_x(\lambda,x) \bigr\rbrace - \p_x \Kc_i(\lambda,x) + \bigl[\Kc_i(\lambda,x), \Lc_x(\lambda,x) \bigr] = 0 \,.
\end{equation}
This is the fundamental result of this paragraph. Indeed, it shows that the evolution of the Lax matrix $\Lc_x(\lambda,x)$ under the Hamiltonian flow $\lbrace \Q_i,\cdot\rbrace$ generated by the local charge takes the form of a zero curvature equation. This is the analogue of the Lax equation \eqref{Eq:ZCE} for a 2d integrable field theory where the time evolution $\p_t$ is replaced by the Hamiltonian flow $\lbrace \Q_i,\cdot\rbrace$ and the temporal Lax component $\Lc_t(\lambda)$ is replaced by $\Kc_i(\lambda)$. In particular, through this parallel, the equivalent of the conservation equation \eqref{Eq:dT} becomes $\bigl\lbrace \Q_i, \Tc_V(\lambda) \bigr\rbrace = 0$, which is what we wanted to prove.\vspace{10pt}

\begin{tcolorbox} \textit{\underline{Exercise 4:} Poisson brackets, Poisson brackets everywhere.} \raisebox{0.5pt}{\Large$\;\;\star\star$} \vspace{5pt}\\
We encourage the reader to rederive on their own the (numerous) Poisson brackets computed above. These are rather simple illustrations of the standard techniques employed throughout these lectures, including for instance the use of tensorial notations.
\end{tcolorbox}~\vspace{-13pt}

\paragraph{Momentum and Hamiltonian.} In the following section \ref{Sec:SpaceTime}, we will show that the momentum $\Pc$ of the theory, defined as the generator of the spatial derivative $\p_x=\lbrace \Pc,\cdot\rbrace$, is in fact equal to the sum of all the quadratic local charges $\Q_i$, $i\in\lbrace 1,\dots,N\rbrace$. In analogy with this result, we will define the Hamiltonian $\Hc$, \textit{i.e.} the generator of the time derivative $\p_t = \lbrace \Hc,\dot \rbrace$, as a linear combination of the $\Q_i$'s, therefore specifying a dynamics for the theory. By construction, all the Poisson-commuting charges $\Q_i$ and $\Tc_V(\lambda)$ will then be conserved under this time evolution. We postpone the discussion of these dynamical aspects and of the space-time symmetries of the resulting theory to Section \ref{Sec:SpaceTime}.

\subsection{Higher degree local charges}
\label{SubSec:Local}

In the previous subsection, we have shown the existence of quadratic local charges that Poisson-commute between themselves and with the trace of the monodromy. It is natural to ask whether the integrability of the theory also manifests itself through the existence of \textit{higher degree local charges}, which would generalise the quadratic ones. This is indeed the case, following a method proposed in~\cite{Lacroix:2017isl}, which itself was inspired by the construction~\cite{Evans:1999mj,Evans:2000hx} of such charges in the so-called Principal Chiral Model\footnote{As we will explain in Subsection \ref{SubSec:PCM}, the Principal Chiral Model is in fact a particular example of an affine Gaudin model. The results of~\cite{Lacroix:2017isl} can thus be seen as a generalisation of~\cite{Evans:1999mj,Evans:2000hx} to all affine Gaudin models, which include many other examples of integrable $\s$-models.}. In the present subsection and Appendix \ref{App:Local}, we revisit these results, presenting them in a slightly different, but essentially equivalent, way than~\cite{Lacroix:2017isl}.

\paragraph{Main result.} The higher degree local charges are naturally associated with the zeroes $\bm{\ze}=(\ze_1,\dots,\ze_N)$ of the twist function. More precisely, for each zero $\ze_i$, we define an infinite tower of local charges
\begin{equation}\label{Eq:Qip}
\Q_i^{(p)} = \frac{1}{(p+1)\,\vp'(\ze_i)^{(p+1)/2}} \int_0^{2\pi} \kappa_{p+1} \bigl( \Gamma(\ze_i,x), \dots, \Gamma(\ze_i,x) \bigr)\,\dd x\,,
\end{equation}
where the prefactor has been introduced for future convenience and where
\begin{equation}
\begin{array}{lccc}
\kappa_{p+1} : & \g^\C \oplus \dots \oplus \g^\C       & \longrightarrow & \C \\
               & (X_1,\dots,X_{p+1}) & \longmapsto & \kappa_{p+1}^{a_1\,\cdots\, a_{p+1}}\,X_{1,a_1} \cdots X_{p+1,a_{p+1}}
\end{array}
\end{equation}
is a symmetric $(p+1)$--multilinear form on $\g^\C$, equivalently characterised by a symmetric $(p+1)$--tensor $\kappa_{p+1}^{a_1\,\cdots \,a_{p+1}}$ in components (using a basis decomposition $X_k = X_{k,a}\,\td^a$). For a fixed zero $\ze_i$, the associated charges $\Q_i^{(p)}$ are indexed by the integer $p\in E$, taking value in an infinite set $E \subset \Z_{\geq 1}$ called the \textit{affine exponents of $\g$}, which essentially labels their degree. More precisely, the density of $\Q_i^{(p)}$ is a polynomial of degree $p+1$ in the Kac-Moody currents. In particular, $\Q_i^{(p)}$ is of dimension $p$, \textit{i.e.} is expressed in units of length to the power $-p$. The set $E$ depends on the Lie algebra $\g$ but always starts with $p=1$: the corresponding quadratic tensor $\kappa^{(2)}$ is simply the bilinear form $-\psd$ and the associated charge $\Q_i^{(1)}$ thus coincides with the quadratic charge defined in equation \eqref{Eq:Qi}.

Generalising the equation \eqref{Eq:PbQ} found in the quadratic case, the main claim of this subsection is that the local charges $\Q_i^{(p)}$ Poisson-commute between themselves and with the non-local charges:
\begin{equation}\label{Eq:PbQp}
\bigl\lbrace \Q_i^{(p)}, \Q_j^{(q)} \bigr\rbrace = 0 \qquad \text{ and } \qquad \bigl\lbrace \Q_i^{(p)}, \Tc_V(\lambda) \bigr\rbrace = 0  \,,
\end{equation}
for any $i,j\in\lbrace 1,\dots, N\rbrace$, any $p,q\in E$, any $\lambda\in\C$ and any representation $V$.

\paragraph{The EHMM-tensors.} The tensors $\kappa_{p+1}$, $p\in E$, satisfy certain fundamental properties, which in the end ensure that the main result \eqref{Eq:PbQp} holds.  Here, we will express them in components for concreteness: we refer to Appendix \ref{App:EHMM-Tensors} for a basis-independent formulation of these properties. The first one is that they are ad-invariant, in the sense that
\begin{equation}\label{Eq:adKappa}
\f{bc}{a_1}\,\kappa_{p+1}^{c\,a_2\,\cdots\, a_{p+1}} + \f{bc}{a_2}\,\kappa_{p+1}^{a_1\,c\,a_3\,\cdots\, a_{p+1}} + \f{bc}{a_{p+1}}\,\kappa_{p+1}^{a_1\,\cdots\, a_{p}\,c} = 0\,,
\end{equation}
for all $a_1,\dots,a_{p+1},b\in\lbrace 1,\dots,\dim\g\rbrace$.\\

Following the standard convention, we will use round brackets $(a_1\cdots a_k)$ to indicate symmetrisation over the indices $a_1,\dots,a_k$, meaning that we are averaging over permutations of these indices. Using this notation, the main property satisfied by the tensors $\kappa_{p+1}$ can be written as follows:
\begin{equation}\label{Eq:KappaKappa}
\kappa_{p+1}^{\,(a_1\, \cdots\, a_p}\null_{\,c} \; \kappa_{q+1}^{\,b_1\, \cdots\, b_{q-1})\, b_q\,c} = \kappa_{p+1}^{(a_1\, \cdots\, a_p}\null_{\,c} \; \kappa_{q+1}^{b_1\, \cdots\,b_{q-1}\, b_q)\,c}\,.
\end{equation}
Grasping the content of this identity requires a bit of ``index gymnastic'': let us then decode the notation explicitly. Both sides of the equation contain an implicit sum over the repeated indices $c\in\lbrace 1,\dots,\dim\g\rbrace$, one of which being lowered using the bilinear form. We are thus considering the contraction of $\kappa_{p+1}$ and $\kappa_{q+1}$, which forms a tensor depending on $p+q$ indices $a_1,\dots,a_p,b_1,\dots,b_{q}$. This tensor is clearly symmetric with respect to the first $p$ indices and with respect to the last $q$ indices. However, it is in general not invariant under permutations mixing these two sets. On the left-hand side of equation \eqref{Eq:KappaKappa}, we consider the symmetrisation over the first $p+q-1$ indices $a_1,\dots,a_p,b_1,\dots,b_{q-1}$: the main content of the identity is then that this is enough to make the tensor completely symmetric, since on the right-hand side we symmetrise over all $p+q$ indices.\\

The properties \eqref{Eq:adKappa} and \eqref{Eq:KappaKappa} pose quite non-trivial constraints on the tensors $\kappa_{p+1}$. That solutions to these constraints exist is non-obvious and was proven in the works~\cite{Evans:1999mj,Evans:2000hx} of Evans, Hassan, MacKay and Mountain -- see in particular~\cite[Equation (2.39)]{Evans:2000hx}: we will then refer to these solutions $\kappa_{p+1}$ as EHMM-tensors. In particular, this is where the restriction on the degrees of these tensors makes its apparition: indeed, the specific solutions found in~\cite{Evans:1999mj,Evans:2000hx} have degrees $p+1$ with $p$ running through the set $E$ of affine exponents of $\g$. For conciseness, we will not describe the full construction of these tensors in the lectures but gather a few more details as well as some low-degree examples in the Appendix \ref{App:EHMM-Tensors}.

\paragraph{Poisson-commutation of the local charges.} Let us now discuss the proof of the main result \eqref{Eq:PbQp}. We will focus here on the main steps, relegating the detailed computations to the Appendix \ref{App:LocalPb}. For $p\in E$, we define
\begin{equation}\label{Eq:Wp}
\Wc^{(p+1)}(\lambda,x) = \frac{1}{p+1}\; \kappa_{p+1} \bigl( \Gamma(\lambda,x), \dots, \Gamma(\lambda,x) \bigr)\,,
\end{equation}
so that the density of the charge $\Q_i^{(p)}$ is given by
\begin{equation}\label{Eq:Density}
q_i^{(p)}(x) = \frac{\Wc^{(p+1)}(\ze_i,x)}{\vp'(\ze_i)^{(p+1)/2}}\,.
\end{equation}
Starting with our key formula \eqref{Eq:PbGaudin}, namely the Poisson algebra of the Gaudin Lax matrix, one can compute the Poisson bracket between $\Wc^{(p+1)}(\lambda,x)$ and $\Wc^{(q+1)}(\mu,y)$, for $p,q\in E$ and $\lambda,\mu\in\C$. Using the ad-invariance \eqref{Eq:adKappa} of the EHMM-tensors as well as the property \eqref{Eq:KappaKappa}, one can put this bracket in the following form -- see equation \eqref{Eq:PbWpWqApp}:
\begin{eqnarray}\label{Eq:PbWpWq}
\bigl\lbrace \Wc^{(p+1)}(\lambda,x), \Wc^{(q+1)}(\mu,y) \bigr\rbrace \!\!\!\!&=&\!\!\!\! \Bigl( \vp(\lambda)\,\Ac^{(p,q)}_1(\lambda,\mu,x) - \vp(\mu)\,\Ac^{(p,q)}_1(\lambda,\mu,x) + \p_x \Ac^{(p,q)}_2(\lambda,\mu,x) \Bigr) \delta(x-y) \notag \\
&& \hspace{30pt} + \Ac^{(p,q)}_3(\lambda,\mu,x)\,\p_x \delta(x-y)\,.
\end{eqnarray}
Here, the fields $\Ac^{(p,q)}_k(\lambda,\mu,x)$ are local combinations of $\Gamma(\lambda,x)$ and $\Gamma(\mu,x)$, which are regular at $\lambda=\mu$. Up to a global factor, the Poisson bracket of the densities $q_i^{(p)}(x)$ and $q_j^{(q)}(y)$ is obtained by evaluating \eqref{Eq:PbWpWq} at $\lambda=\ze_i$ and $\mu=\ze_j$: using $\vp(\ze_i)=\vp(\ze_j)=0$, it is clear that the terms involving $\Ac^{(p,q)}_1$ vanish under this evaluation. Moreover, the two remaining terms, involving $\Ac^{(p,q)}_2$ and $\Ac^{(p,q)}_3$, are spatial derivatives and hence vanish once integrated over $x$ and $y$. Thus, we simply get
\begin{equation}
\bigl\lbrace \Q_i^{(p)}, \Q_j^{(q)} \bigr\rbrace = 0\,,
\end{equation}
proving the first part of equation \eqref{Eq:PbQp}. In fact, looking more closely at the intermediate steps in the derivation of the bracket \eqref{Eq:PbWpWq}, one finds a slightly stronger result \eqref{Eq:PbWWpij}, namely that the densities associated with different zeroes Poisson-commute:
\begin{equation}\label{Eq:Pbqiqj}
\bigl\lbrace q_i^{(p)}(x), q_j^{(q)}(y) \bigr\rbrace = 0\,, \qquad \text{ if } \; i\neq j\,.
\end{equation}
For $i=j$, the right-hand side of this equation does not vanish but is formed by spatial derivatives.

\paragraph{Zero curvature equations.} The second part of the main result \eqref{Eq:PbQp} is proven in Appendix \ref{App:LocalPb}, using a similar approach as for the quadratic case. More precisely, we compute the evolution of the Lax matrix $\Lc(\lambda,x)$ under the Hamiltonian flow $\bigl\lbrace \Q_i^{(p)}, \cdot \bigr\rbrace$ and find that it takes the form of a zero curvatuve equation. Explicitely, there exists a matrix $\Kc_i^{(p)}(\lambda,x)$ such that
\begin{equation}\label{Eq:ZCEip}
\bigl\lbrace \Q_i^{(p)}, \Lc_x(\lambda,x) \bigr\rbrace - \p_x \Kc_i^{(p)}(\lambda,x) + \bigl[\Kc_i^{(p)}(\lambda,x), \Lc_x(\lambda,x) \bigr] = 0 \,.
\end{equation}
This implies that $\Q_i^{(p)}$ Poisson-commutes with the trace of the monodromy of $\Lc_x(\lambda,x)$, \textit{i.e.} $\Tc_V(\lambda)$, hence proving the second equation in \eqref{Eq:PbQp}.

The zero curvature equation \eqref{Eq:ZCEip} can be interpretred as the commutation of the connections $\p_x + \Lc_x(\lambda)$ and $\bigl\lbrace \Q_i^{(p)}, \cdot \bigr\rbrace + \Kc_i^{(p)}(\lambda)$.
It was argued in~\cite{Lacroix:2017isl} (at least in some cases) that these connections in fact all commute one with another, \textit{i.e.}
\begin{equation}
\bigl\lbrace \Q_i^{(p)}, \Kc_j^{(q)}(\lambda) \bigr\rbrace - \bigl\lbrace \Q_j^{(q)},\Kc_i^{(p)}(\lambda) \bigr\rbrace + \bigl[\Kc_i^{(p)}(\lambda), \Kc_j^{(q)}(\lambda) \bigr] = 0 \,.
\end{equation}
This forms what is sometimes called an \textit{integrable hierarchy}, \textit{i.e.} an infinite number of commuting Hamiltonian flows which admit compatible zero curvature representations.\vspace{10pt}

\begin{tcolorbox} \textit{\underline{Exercise 5:} Wow that's a lot of Poisson brackets!} \raisebox{0.5pt}{\Large$\;\;\star\star\star$} \vspace{5pt}\\
As an exercise, the motivated reader can rederive on their own the main results of this subsection, in particular the Poisson brackets \eqref{Eq:PbWpWq} and \eqref{Eq:ZCEip}. This can be done following the Appendix \ref{App:LocalPb}, filling up the technical blanks and convincing oneself of the various steps.\vspace{4pt}\\
For the readers more versed in components manipulations, one can also rederive some of these results using the projection of \eqref{Eq:PbGaudin} along a basis rather than tensorial notations.
\end{tcolorbox}

\paragraph{Integrable structure of the AGM.} Let us quickly summarise what we have proven so far. In Subsection \ref{SubSec:LaxG}, we argued that the AGM admits an infinite number of Poisson-commuting non-local charges, extracted from the generating functions $\Tc_V(\lambda)$. In this subsection and the previous one, we further constructed an infinite number of higher-degree local charges, which Poisson-commute one with another and with the non-local charges as well. Together, all these quantities form what we will call the \textit{integrable structure} of the AGM. More formally, we define it as the subalgebra $\Zc \subset \Ac$ generated by the non-local and local charges: this is an infinitely-generated subalgebra of $\Ac$ which is Poisson-commutative, in the sense that the Poisson bracket $\lbrace \cdot,\cdot\rbrace$ of $\Ac$ vanishes when restricted on $\Zc$.

\paragraph{Change of spectral parameter.} Recall the transformation \eqref{Eq:ChangeSpec} of the defining parameters of the AGM: in Subsection \ref{SubSec:LaxG}, we have explained that it can be interpreted as a change of spectral parameter $\lambda \mapsto a\,\lambda+b$. In particular, we used this idea to prove the invariance of the family of non-local charges of the AGM under this transformation. Similarly, one finds that the local charges $\Q_i^{(p)}$ built in the current subsection are also left unaffected. This transformation of the parameters thus leaves invariant the full integrable structure $\Zc$ of the AGM.\vspace{10pt}

\begin{tcolorbox} \textit{\underline{Exercise 6:} Invariance under change of spectral parameter} \raisebox{0.5pt}{\Large$\;\;\star$} \vspace{5pt}\\
Compute the transformation of the quantities $\Gamma(\ze_i)$ and $\vp'(\ze_i)$ under the change of parameters \eqref{Eq:ChangeSpec}. Deduce from this result that the density \eqref{Eq:Density} of the local charge $\Q_i^{(p)}$ is left invariant under this transformation.
\end{tcolorbox}~\vspace{-8pt}

\subsection{Global diagonal symmetry}
\label{SubSec:Diag}

Now that we have defined the integrable structure of the AGM, let us describe some of its additional properties, including its so-called diagonal symmetry.

\paragraph{The symmetry.} Let $G^\C$ be a connected Lie group with Lie algebra $\g^\C$. For $h\in G^\C$ a constant element in this group, we consider the \textit{diagonal action}
\begin{equation}\label{Eq:Diag}
\Jc_r \longmapsto h^{-1}\,\Jc_r\,h\,, \qquad \forall\,r\in\lbrace 1,\dots,N\rbrace\,,
\end{equation}
which acts simultaneously on all the Kac-Moody currents $\Jc_r$ of the AGM. From their definitions \eqref{Eq:Gaudin} and \eqref{Eq:Lax}, it is clear that the Gaudin Lax matrix $\Gamma(\lambda)$ and Lax matrix $\Lc_x(\lambda)$ transform in the following way under this action:
\begin{equation}
\Gamma(\lambda) \longmapsto h^{-1}\,\Gamma(\lambda)\,h\qquad \text{ and } \qquad \Lc_x(\lambda) \longmapsto h^{-1}\,\Lc_x(\lambda)\,h\,.
\end{equation}
In particular, the monodromy of $\Lc_x(\lambda)$ also transforms by the adjoint action of $h$, so that the generating function \eqref{Eq:T} is invariant. Similarly, thanks to the ad-invariance of the EHMM-tensor $\kappa_{p+1}$, the local charge \eqref{Eq:Qip} is also unaffected by this transformation. In conclusion, we found
\begin{equation}\label{Eq:DiagTQ}
\Tc_V(\lambda) \longmapsto \Tc_V(\lambda) \qquad \text{ and } \qquad \Q_i^{(p)} \longmapsto \Q_i^{(p)}\,.
\end{equation}
Thus the diagonal action \eqref{Eq:Diag} is a symmetry of the integrable structure of the AGM. We note that this is a global symmetry, in the sense that the parameter $h$ is a constant element of the group $G^\C$.

\paragraph{Generator.} The diagonal symmetry \eqref{Eq:Diag} is generated by the Noether charge
\begin{equation}\label{Eq:Mu}
\mu = -\begin{array}c \text{res}\\[-3pt] \lambda=\infty \end{array} \; \int_0^{2\pi} \Gamma(\lambda,x)\,\dd x \, \dd\lambda = \sum_{r=1}^N \int_0^{2\pi} \Jc_r(x)\,\dd x\,.
\end{equation}
More precisely, the infinitesimal transformation
\begin{equation}\label{Eq:DiagInf}
\delta_\epsilon \Jc_r = \bigl[ \Jc_r, \epsilon \bigr]
\end{equation}
induced by the action \eqref{Eq:Diag}, where $\epsilon\in\g^\C$ is an infinitesimal Lie algebra-valued symmetry parameter, coincides with the Hamiltonian flow
\begin{equation}
\delta_\epsilon = \lbrace \ps{\mu}{\epsilon}, \cdot \rbrace\,.
\end{equation}
This is easily proven from the Poisson bracket \eqref{Eq:PbJ} of the Kac-Moody currents and is left as an exercise to the reader.

The symmetry property \eqref{Eq:DiagTQ} means that the generator $\mu$ Poisson-commutes with the charges $\Tc_V(\lambda)$ and $\Q_i$: this can be verified explicitly. We finally note that the components of $\mu = \mu_a\,\td^a$ in a basis decomposition are not pairwise Poisson-commuting. In fact they satisfy a classical version of the commutation relations of the Lie algebra $\g^\C$, \textit{i.e.}
\begin{equation}\label{Eq:PbMu}
\bigl\lbrace \mu_a, \mu_b \bigr\rbrace = \f{ab}c\,\mu_c\,.
\end{equation}
This is to be expected since $\mu$ generates a $\g^\C$--action. \vspace{10pt}

\begin{tcolorbox} \textit{\underline{Exercise 7:} Generator of the diagonal symmetry} \raisebox{0.5pt}{\Large$\;\;\star$} \vspace{5pt}\\
Use the Kac-Moody bracket \eqref{Eq:PbJ} to prove that the charge \eqref{Eq:Mu} generates the infinitesimal transformation \eqref{Eq:DiagInf}. Check as well the Poisson bracket \eqref{Eq:PbMu}, or its tensorial equivalent
\begin{equation}
\bigl\lbrace \mu\ti{1},\mu\ti{2} \bigr\rbrace = \bigl[ C\ti{12}, \mu\ti{1} \bigr]\,.
\end{equation}
\end{tcolorbox}~\vspace{-4pt}

\subsection{Further developments and generalisations}
\label{SubSec:Gen}

\paragraph{Reality conditions.} So far, we have considered all the parameters and observables of the AGM  as complex quantities. We now quickly discuss how to extract a real theory from this construction. For that, we will impose certain reality conditions on the parameters of the AGM, which can be summarised compactly as the following requirement on the twist function:
\begin{equation}
\overline{\vp(\lambda)} = \vp(\overline{\lambda})\,.
\end{equation}
Here and in what follows, $\overline{z}$ denotes the conjugate of the complex number $z\in\C$. In terms of the sites $\bm{\sigma}=(\sigma_1,\dots,\sigma_N)$ and levels $\bm{\ell}=(\ell_1,\dots,\ell_N,\ell_\infty)$ appearing in the partial fraction decomposition \eqref{Eq:Twist} of the twist function, this requirement is equivalent to the following. Each site is either a real number $\s_r \in \R$, in which case the corresponding level $\ell_r$ is also real, or belongs to a pair of complex conjugate sites $(\s_r,\s_{\bar r})\in\C^2$ with $\s_{\bar r}=\overline{\s_r}$ (for some $\bar r \in \lbrace 1,\dots,N\rbrace$), in which case the corresponding levels $\ell_r$ and $\ell_{\bar r}$ are also complex conjugate one of another, \textit{i.e.} $\ell_{\bar r}=\overline{\ell_r}$. Moreover, the constant term $\ell_\infty$ is real. Similarly, the zeroes $\bm{\ze}=(\ze_1,\dots,\ze_N)$ of the twist function are either real or coming in pairs of complex conjugates: here we will suppose for simplicity that they are all real.\\

Now that we have imposed reality conditions on parameters, we have to discuss the ones on the fundamental fields of the model, \textit{i.e.} the Kac-Moody currents $\Jc_r$. For that, recall that we initially started from the data of a real simple Lie algebra $\g$ and later considered its complexification $\g^\C$. This $\g$ is thus a real form of $\g^\C$, which can be seen as the set of fixed points $\g=\lbrace X\in\g^\C\,|\,\tau(X)=X \bigr\rbrace$ of an involutive antilinear automorphism $\tau$ of $\g^\C$. More explicitly, the latter is an operator $\tau : \g^\C \to \g^\C$, with $\g^\C$ seen as a real vector space of dimension $2\dim\g$, which satisfies $\tau^2=\Id$, $\tau(\lambda X+\mu Y) = \overline{\lambda}\,\tau(X) + \overline{\mu}\,\tau(Y)$ and $\tau([X,Y])=\bigl[\tau(X),\tau(Y)\bigr]$, where $\lambda,\mu\in\C$ and $X,Y\in\g^\C$. For instance, the real form $\g=\su(k)$ of $\g^\C=\sl(k,\C)$ corresponds to $\tau : X \mapsto -X^\dagger$. The real algebra $\g=\sl(k,\R)$ shares the same complexification $\g^\C=\sl(k,\C)$ but corresponds to $\tau: X \mapsto \overline{X}$.

If $\s_r\in\R$ is a real site, we will restrict the corresponding Kac-Moody current $\Jc_r$ to be valued in the real form $\g$ (in other words to be fixed by $\tau$). In the case of a pair of complex conjugate sites $(\s_r,\s_{\bar r})$, the associated currents $(\Jc_r,\Jc_{\bar r})$ are valued in the complexification $\g^\C$ but we impose that they are conjugate one to another by the antilinear automorphism $\tau$, \textit{i.e.} $\Jc_{\bar r} = \tau\bigl( \Jc_r \bigr)$. One checks that these restrictions are compatible with the Poisson brackets \eqref{Eq:PbJ} and more generally with the construction of the AGM described in the previous subsections. We also note that under these requirements, the Kac-moody currents encode a total of $N\,\dim\g$ real fields.

The above reality conditions on the parameters and the fields translate to the following property of the Gaudin Lax matrix \eqref{Eq:Gaudin} and Lax matrix \eqref{Eq:Lax}:
\begin{equation}\label{Eq:RealityGaudinLax}
\tau\bigl( \Gamma(\lambda) \bigr) = \Gamma(\overline\lambda) \qquad \text{ and } \qquad \tau\bigl( \Lc_x(\lambda) \bigr) = \Lc_x(\overline\lambda) \,.
\end{equation}
In particular, one shows that the trace of the monodromy of $\Lc_x(\lambda)$ satisfies
\begin{equation}
\overline{\Tc_V(\lambda)} = \Tc_V(\overline\lambda)\,.
\end{equation}
Thus the non-local charges extracted from the expansion of $\Tc_V(\lambda)$ around a real point $\lambda_0\in\R$ are all real. Similarly, the higher-degree local charges $\Q_i^{(p)}$ are also real, due to our assumption above that the zeroes $\ze_i$ belong to $\R$. Thus, the integrable structure of the model is generated entirely by real observables. Finally, we note that, under these reality conditions, the diagonal symmetry discussed in Subsection \ref{SubSec:Diag} becomes an action of the real Lie group $G$, rather than its complefixication $G^\C$: more precisely, the symmetry parameter $h$ appearing in the transformation \eqref{Eq:Diag} should now be taken in the real group $G$.

\paragraph{Higher-order poles.} We now discuss certain generalisations of the AGM construction. The first one, which we treat in the present paragraph, will be useful in Subsection \ref{SubSec:PCM} to recover the Principal Chiral Model. Recall that the Gaudin Lax matrix \eqref{Eq:Gaudin} and the twist function \eqref{Eq:Twist} have been introduced as functions of the spectral parameter $\lambda$ with simple poles at the sites $\s_r$, $r\in\lbrace 1,\dots,N\rbrace$. One can extend this setup to include higher-order poles~\cite{Vicedo:2017cge}. We then consider the following generalisations of equations \eqref{Eq:Gaudin} and \eqref{Eq:Twist}:
\begin{equation}\label{Eq:Mult}
\Gamma(\lambda,x) = \sum_{r=1}^N \sum_{p=0}^{m_r-1} \frac{\Jc_r^{[p]}(x)}{(\lambda-\s_r)^{p+1}} \qquad \text{ and } \qquad \vp(\lambda) = \sum_{r=1}^N \sum_{p=0}^{m_r-1} \frac{\ell_r^{[p]}}{(\lambda-\s_r)^{p+1}} - \ell_\infty\,,
\end{equation}
where $m_r\in\Z_{\geq 1}$ is the order of the pole at $\lambda=\s_r$, which we also call the \textit{multiplicity} of the site $\s_r$. We will suppose that the highest-order coefficients $\ell_r^{[m_r-1]}$ in the twist function are non-vanishing, as well as the constant term $\ell_\infty$.

The $\g^\C$--valued fields $\bigl( \Jc_r^{[p]}(x) \bigr)_{r=1,\dots,N}^{p=0,\dots,m_r-1}$ appearing in the definition of $\Gamma(\lambda,x)$ are the higher-order equivalents of the Kac-Moody currents, which are called \textit{Takiff currents}. Their main characteristic is their Poisson bracket, which reads
\begin{equation}\label{Eq:PbTakiff}
\bigl\lbrace \Jc_r^{[p]}(x)\ti{1}, \Jc_s^{[q]}(y)\ti{2} \bigr\rbrace = \delta_{rs} \left\lbrace \begin{array}{lcl}
\bigl[ C\ti{12}, \Jc_r^{[p+q]}(x)\ti{1} \bigr]\, \delta(x-y) - \ell_r^{[p+q]}\,C\ti{12}\,\p_x\delta(x-y) & & \text{ if }\, p+q < m_r \\[5pt]
0 & & \text{ if }\, p+q \geq m_r
\end{array} \right.\,.
\end{equation}
This is a clear generalisation of the Kac-Moody bracket \eqref{Eq:PbJ}, which we recover in the case $m_r=1$.\footnote{For a non-exhaustive list of references on these Takiff currents and their quantum analogues, see~\cite{Babichenko:2012uq,Vicedo:2017cge,Rasmussen:2019zfu,Quella:2020uhk}.} The main result that we will need (see exercise below) is that this choice of bracket ensures that the Poisson algebra of the Gaudin Lax matrix stays the same as in the simple pole case, \textit{i.e.} our key formula \eqref{Eq:PbGaudin} still holds. As this was the starting point for the construction of the AGM integrable structure, all the results of the previous subsections essentially apply in the exact same way in this more general case. In particular, we build non-local charges from the monodromy of the Lax matrix \eqref{Eq:Lax} and local ones using the formula \eqref{Eq:Qip}, in terms of the zeroes $\ze_i$ of the twist function. Note that we still suppose that these zeroes are simple: there are then
\begin{equation}\label{Eq:M}
M = \sum_{r=1}^N m_r
\end{equation}
of them. These non-local and local charges are all pairwise Poisson-commuting, forming the integrable structure of the AGM with multiplicities. As in the simple poles case, a convenient choice of parameters for the AGM are the sites $\bm\s=(\s_1,\dots,\s_N)$, the zeroes $\bm\ze = (\ze_1,\dots,\ze_M)$ and the constant term $\ell_\infty$, in terms of which the twist function is defined in the factorised form
\begin{equation}
\vp(\lambda) = - \ell_\infty \frac{\prod_{i=1}^M (\lambda-\ze_i)}{\prod_{r=1}^N (\lambda-\s_r)^{m_r}}\,.
\end{equation}
The levels $\bigl( \ell_r^{[p]} \bigr)_{r=1,\dots,N}^{p=0,\dots,m_r-1}$ are then obtained as rational combinations of these parameters by performing the partial fraction decomposition of $\vp(\lambda)$.  Similarly to what was done in the previous paragraph for simple poles, one can impose reality conditions on these parameters and the Takiff currents to ensure that the relevant physical observables of the model are real.

\vspace{10pt}

\begin{tcolorbox} \textit{\underline{Exercise 8:} AGM with multiplicities} \raisebox{0.5pt}{\Large$\;\;\star\star$} \vspace{5pt}\\
1. Show that the Takiff bracket \eqref{Eq:PbTakiff} ensures that the Gaudin Lax matrix satisfies the key formula \eqref{Eq:PbGaudin}, where $\Gamma(\lambda,x)$ and $\vp(\lambda)$ are defined as in equation \eqref{Eq:Mult}.\vspace{2pt} \\
\textit{Hint:} the following identity, valid for any distinct $\lambda,\mu,\s\in\C$, will be useful
\begin{equation*}
\sum_{k=0}^p \frac{1}{(\lambda-\s)^{k+1}(\mu-\s)^{p-k+1}} = \frac{1}{\mu-\lambda} \left( \frac{1}{(\lambda-\s)^{p+1}} - \frac{1}{(\mu-\s)^{p+1}} \right) \,.
\end{equation*}
2. {\Large$\,\star\star\star\,$}  Consider the transformation \eqref{Eq:ChangeSpec} of the parameters $(\s_1,\dots,\s_N,\ze_1,\dots,\ze_M,\ell_\infty)$. How does it act on the levels $\bigl( \ell_r^{[p]} \bigr)_{r=1,\dots,N}^{p=0,\dots,m_r-1}$? Following the same type of reasoning as in the simple poles case, show that the integrable structure of the AGM with multiplicities is left invariant under this transformation if one also considers an appropriate transformation of the higher-order Takiff currents $\Jc_r^{[p]}(x)$ for $p>0$. \vspace{4pt}\\
3. Generalise the discussion of the global diagonal symmetry of Subsection \ref{SubSec:Diag} to the higher-order poles case.
\end{tcolorbox}~\vspace{-12pt}

\paragraph{Vanishing $\bm{\ell_\infty}$ and local diagonal symmetry.} So far, we have always considered twist functions $\vp(\lambda)$ containing a non-vanishing constant term $\ell_\infty$ -- see \eqref{Eq:Twist} or \eqref{Eq:Mult}. This coefficient is part of the defining parameters of the AGM and in particular appears in the expression of the charges forming its integrable structure. We now discuss what happens when we take $\ell_\infty=0$. As argued in the footnote \ref{Footnote:Linf}, the key formula \eqref{Eq:PbGaudin} for the Poisson algebra of the Gaudin Lax matrix is independent of the choice of $\ell_\infty$. This result thus still holds in the case $\ell_\infty=0$. Most of the formal constructions presented in this section, which are based on this key formula, therefore still apply: for instance, one can construct Poisson-commuting non-local and local charges, forming the integrable structure of the model.\\

The main difference will concern the diagonal symmetry discussed in Subsection \ref{SubSec:Diag}. In fact, taking $\ell_\infty=0$ will promote this symmetry from a global one to a local one. More precisely, let us consider the following transformation of the currents:
\begin{equation}\label{Eq:DiagLoc}
\Jc_r^{[p]}(x) \longmapsto h(x)^{-1}\,\Jc_r^{[p]}(x)\,h(x) + \ell_r^{[p]}\, h(x)^{-1} \p_x h(x)\,,
\end{equation}
where $h(x)$ is an arbitrary group-valued function of the coordinate $x$ (which we take to be periodic in $x\sim x+2\pi$ when the spatial direction is a circle). In the case where $h$ is a constant group element, independent of $x$, we recover the diagonal symmetry considered in equation \eqref{Eq:Diag} -- restricting also to the case of simple poles. We will refer to \eqref{Eq:DiagLoc} as the \textit{local diagonal transformation}.

Strating from equation \eqref{Eq:Mult}, one easily shows that it acts on the Gaudin Lax matrix as
\begin{equation}\label{Eq:DiagGaudin}
\Gamma(\lambda,x) \longmapsto h(x)^{-1}\,\Gamma(\lambda,x)\,h(x) + \vp(\lambda)\, h(x)^{-1} \p_x h(x)\,.
\end{equation}
We stress that this equation holds only in the case $\ell_\infty=0$ considered in the present paragraph: for non-vanishing $\ell_\infty$, one would have to replace $\vp(\lambda)$ by $\vp(\lambda)+\ell_\infty$ in the second term of the transformation. In terms of the Lax matrix \eqref{Eq:Lax}, we then simply get
\begin{equation}
\Lc_x(\lambda,x) \longmapsto h(x)^{-1}\,\Lc_x(\lambda,x)\,h(x) + h(x)^{-1} \p_x h(x)\,.
\end{equation}
The reader might recognise in this expression the gauge transformation of the connection $\p_x + \Lc_x(\lambda,x)$ by the local parameter $h(x)$. It is well known that the generating function $\Tc_V(\lambda)$, defined in \eqref{Eq:T} as the trace of the monodromy of $\Lc_x(\lambda,x)$ along the spatial circle, is not modified by such a gauge transformation. The non-local charges extracted from $\Tc_V(\lambda)$ are thus invariant under the local diagonal transformation \eqref{Eq:DiagLoc}.

Evaluating the equation \eqref{Eq:DiagGaudin} at a zero $\lambda=\ze_i$ of the twist function, we find that the specific linear combination $\Gamma(\ze_i,x)$ of the currents is a covariant field, \textit{i.e.} it transforms as
\begin{equation}
\Gamma(\ze_i,x) \longmapsto h(x)^{-1}\,\Gamma(\ze_i,x)\,h(x)\,,
\end{equation}
without any derivative term. By ad-invariance of the EHMM-tensors $\kappa_{p+1}$, the local charges defined in equation \eqref{Eq:Qip} are thus also unaffected by the transformation. We thus conclude that \eqref{Eq:DiagLoc} is a local symmetry of the whole AGM integrable structure. In particular, it means that the fields contained in the currents $\Jc_r^{[p]}(x)$ are not all physical: some of these degrees of freedom can be ``gauged away'' by the local symmetry. We stress that all this reasoning fails when we take $\ell_\infty$ to be non-zero, in which case only the global part of the diagonal symmetry survives.\\

As mentioned in footnote \ref{Footnote:Inf}, the coefficient $\ell_\infty$ contributes to a second order pole at $\lambda=\infty$ in the 1-form $\vp(\lambda)\,\dd\lambda$. In the present case, where we assume $\ell_\infty=0$, the infinity becomes either a simple pole or a regular point of $\vp(\lambda)\,\dd\lambda$, depending on whether the sum of the levels $\ell_r^{[0]}$ is non-zero or vanishing, respectively. These two cases are treated in a slightly different way: the latter requires the additional notion of first-class constraints, which has to do with Dirac's treatment of gauge symmetries in Hamiltonian systems~\cite{dirac1964lectures,Henneaux:1992ig}. We will not develop these aspects further in the lectures and refer for instance to~\cite{Vicedo:2017cge,Lacroix:2019xeh} for detailed discussions. Let us mention however that these considerations eventually allow one to define AGMs associated to arbitrary 1-forms $\vp(\lambda)\,\dd\lambda$ on $\CP$, with any type of pole structures including at $\lambda=\infty$. In this more general family of AGMs, the redundancy \eqref{Eq:ChangeSpec} in the parameters of the model, which corresponds to the freedom of performing dilations and translations of the spectral parameter $\lambda\in\C$, gets upgraded to a 3-parameter symmetry group, corresponding to all biholomorphic actions on $\lambda\in\CP$, \textit{i.e.} Möbius transformations $\lambda \mapsto \frac{a\lambda+b}{c\lambda+d}$.

\paragraph{Cyclotomic/dihedral AGMs.} Our last generalisation will be the so-called cyclotomic and dihedral AGMs introduced in~\cite{Vicedo:2017cge}.\footnote{For a non-exhaustive list of references concerning cyclotomic Gaudin models based on finite algebras, see~\cite{Skrypnyk:2006,Vicedo:2014zza}.} Compared to the standard AGM described in the rest of this section, these models depend on an additional ingredient, namely a Lie algebra automorphism $\Omega : \g^\C \rightarrow \g^\C$ of finite order $T\in\Z_{\geq 1}$, \textit{i.e.} such that $\Omega^T=\Id$. This automorphism defines an action of the cyclic group $\mathbb{Z}_T = \mathbb{Z} / T\mathbb{Z}$ on the Lie algebra $\g^\C$. We also consider an action of $\mathbb{Z}_T$ on the complex plane of the spectral parameter, generated by the multiplication $\lambda \mapsto \omega\,\lambda$, where $\omega=\exp(\frac{2\mathrm{i}\pi}{T})$ is a $T^{\text{th}}$ root of unity. The main property of the cyclotomic AGM is that its Gaudin Lax matrix $\Gamma(\lambda,x)$ and twist function $\vp(\lambda)$ are equivariant 1-forms with respect to these $\mathbb{Z}_T$ actions, in the sense that
\begin{equation}\label{Eq:Equiv}
\Gamma(\omega\lambda,x)\, \dd(\omega \lambda) = \Omega\bigl( \Gamma(\lambda,x) \bigr)\,\dd\lambda \qquad \text{ and } \qquad \vp(\omega\lambda)\, \dd(\omega \lambda) = \vp(\lambda)\, \dd\lambda\,.
\end{equation}
In particular, if $\Gamma(\lambda)$ and $\vp(\lambda)$ have a pole at a site $\s_r \in\C$, then they also have a pole at all the points $\omega^k\s_r$ in the $\mathbb{Z}_T$--orbit of $\s_r$, with coefficients directly related to the ones at $\s_r$ by appropriate multiplications by $\omega$ and actions of the automorphism $\Omega$. This equivariance property is ensured by changing the definition $\Gamma(\lambda,x)$ and $\vp(\lambda)$ in terms of the Takiff currents $\Jc_r^{[p]}(x)$ and their levels $\ell_r^{[p]}$. For instance, in the case of simple poles, we take the Gaudin Lax matrix to be
\begin{equation}
\Gamma(\lambda,x) = \frac{1}{T} \sum_{k=0}^{T-1} \sum_{r=1}^N \frac{\Omega^k\Jc_r(x)}{\lambda-\omega^k\s_r}\,,
\end{equation}
where $\s_1,\dots,\s_N \in \C$ are the sites of the models and $\Jc_1(x),\dots,\Jc_N(x)\in\g^\C$ are the associated Kac-Moody currents. One easily checks that this quantity satisfies the equivariance property \eqref{Eq:Equiv}. The above definition can also be extended to the case of higher-order poles.

The other main property of this cyclotomic Gaudin Lax matrix is that it satisfies a modified version of the key formula \eqref{Eq:PbGaudin}, describing its Poisson bracket. In fact, this modified formula still takes the form \eqref{Eq:PbGaudinR} in terms of an $\Rc$-matrix $\Rc^0(\lambda,\mu) \in \g^\C\otimes\g^\C$ but the latter is different from the standard one \eqref{Eq:R0}. More precisely, one finds a non-skew-symmetric $\Rc^0(\lambda,\mu)$, expressed in terms of the split Casimir $C\ti{12}$ with appropriate actions of the automorphisms $\Omega$ and multiplications of $\lambda,\mu$ by powers of $\omega$, in a way that we will not detail here. Suffice to say that this matrix satisfies the classical Yang-Baxter equation \eqref{Eq:CYBE}. This ensures that we can construct a Lax matrix with Maillet bracket and thus that we can build Poisson-commuting non-local charges, as in Subsection \ref{SubSec:LaxG} for the non-cyclotomic case. Similarly, the construction of higher-degree local charges of Subsections \ref{SubSec:Quad} and \ref{SubSec:Local} also generalises to this case, following~\cite{Lacroix:2017isl,Evans:2000qx,Evans:2005zd}. One obtains this way the integrable structure of cyclotomic AGMs.

One of the important property of this integrable structure is that it possesses a gauge symmetry, similar to the local diagonal symmetry \eqref{Eq:DiagLoc} discussed earlier. The main difference in the cyclotomic case is that the symmetry group is not the full group $G^\C$ but rather the subgroup $H^\C$ formed by fixed points of the automorphism $\Omega$ (or more precisely its lift to the group).

Finally, let us mention that dihedral AGMs are essentially cyclotomic ones together with reality conditions. In particular, the latter are phrased in terms of an involutive antilinear automorphism $\tau$ of $\g^\C$, characterising the real form $\g=\lbrace X\in\g^\C\,|\,\tau(X)=X \bigr\rbrace$ (see beginning of this subsection for details). To ensure the compatibility of these reality conditions with the cyclotomy, the $\mathbb{Z}_T$--automorphism $\Omega$ has to satisfy the property $\tau \circ \Omega = \Omega^{-1} \circ \tau$~\cite{Vicedo:2017cge}. This compatibility condition can be reformulated as the fact that the automorphisms $\Omega$ and $\tau$ generate the dihedral group $\mathbb{Z}_T \rtimes \mathbb{Z}_2$, giving their name to the models. For a brief overview of the applications of dihedral AGMs to the study of integrable $\s$-models, we refer to the discussion in Subsection \ref{SubSec:Pano}.

\section{Hamiltonian, dynamics and space-time symmetries}
\label{Sec:SpaceTime}

The formalism developed so far in Section \ref{Sec:ClassicalAGM} never makes any reference to a time coordinate $t$. It is based on a Poisson algebra $\Ac$ generated by fields $\Jc_r^{[p]}(x)$ on a fixed-time slice and results in the construction of a large integrable structure in $\Ac$. In this section, we address the question of how to define a 2-dimensional integrable field theory from this algebraic construction, by specifying a well-chosen Hamiltonian dynamics $\p_t = \lbrace \Hc, \cdot \rbrace$. The description of the momentum and Hamiltonian of this theory is the subject of the first subsection, while the second one discusses its space-time symmetries, following mostly the approach developed in~\cite{Delduc:2019bcl}.

\subsection{Momentum and Hamiltonian}
\label{SubSec:MomHam}

\paragraph{Momentum.} Recall the quadratic local charges $\Q_i$ defined by equation \eqref{Eq:Qi} and consider the following particular linear combination:
\begin{equation}\label{Eq:P}
\Pc = \sum_{i=1}^M \Q_i\,.
\end{equation}
One checks that it can be rewritten as\footnote{Here, we assume for simplicity that the sites $\s_r$ of the AGM under consideration all have multiplicity $m_r=1$, \textit{i.e.} the twist function has only simple poles. The results of this paragraph still hold in the case with arbitrary multiplicities, replacing the Kac-Moody currents $\Jc_r$ by the Takiff currents $\Jc_r^{[p]}$ and with an appropriate change~\cite{Vicedo:2017cge,Delduc:2019bcl} of the definition of $\Pc_r$ in equation \eqref{Eq:Pr}. This is the subject of the question 2, in the exercise 9 below.}
\begin{equation}\label{Eq:Pr}
\Pc = \sum_{r=1}^N \Pc_r\,, \qquad \text{ where } \qquad \Pc_r = \frac{1}{2\ell_r} \int_0^{2\pi}\psb{\Jc_r(x)}{\Jc_r(x)}\,\dd x\,.
\end{equation}
We will prove this formula in an exercise below. Comparing to equation \eqref{Eq:PJ}, we recognise in $\Pc_r$ the momentum of the Kac-Moody current $\Jc_r(x)$. Adapting equation \eqref{Eq:PbPj} to the case of $N$ independent currents, we thus have
\begin{equation}\label{Eq:PbPJ}
\bigl\lbrace \Pc_s, \Jc_r(x) \bigr\rbrace = \delta_{rs}\,\p_x \Jc_r(x)\,.
\end{equation}
In particular, one then has
\begin{equation}
\bigl\lbrace \Pc, \Jc_r(x) \bigr\rbrace = \p_x \Jc_r(x)
\end{equation}
for all $r\in\lbrace 1,\dots,N\rbrace$. Since the Kac-Moody currents $\Jc_r(x)$ are the fundamental fields of the AGM, we thus identified the observable $\Pc$ as the momentum of this Hamiltonian theory, \textit{i.e.} the generator of the spatial derivative $\p_x=\lbrace \Pc,\cdot\rbrace$. \vspace{10pt}

\begin{tcolorbox} \textit{\underline{Exercise 9:} Momentum.} \raisebox{0.5pt}{\Large$\;\;\star\star\star$} \vspace{5pt}\\
1. Prove equation \eqref{Eq:Pr} from the definition \eqref{Eq:P} of $\Pc$.\vspace{2pt} \\ \textit{Hint:} one possible approach is to use the representation \eqref{Eq:QiRes} of $\Q_i$ as a residue and the fact that the sum of the residues of a 1-form in $\CP$ vanish.\vspace{4pt}\\
2. Generalise the results of this paragraph to the case of AGMs with multiplicities (see equation \eqref{Eq:Mult} and the surrounding paragraph), where the fundamental fields are the Takiff currents $\Jc_r^{[p]}(x)$, starting with the same definition \eqref{Eq:P} of $\Pc$.\vspace{3pt}\\
\textit{Hint:} You should find that the quantity $\Pc_r$ in equation \eqref{Eq:Pr} is replaced by~\cite{Vicedo:2017cge,Delduc:2019bcl}
\begin{equation}\label{Eq:PrTak}
\Pc_r = \frac{1}{2} \sum_{p,q=0}^{m_r-1} \xi_r^{[p+q]}\, \int_0^{2\pi} \psb{\Jc_r^{[p]}(x)}{\Jc_r^{[q]}(x)}\,\dd x\,,
\end{equation}
where the coefficients $\xi_r^{[p]}$ are defined by the power series expansion of $\vp(\lambda)^{-1}$ around $\lambda=\s_r$:
\begin{equation}
\frac{1}{\vp(\lambda)} = \sum_{p=0}^{2m_r-2} \xi_r^{[p]} (\lambda-\s_r)^{p+1} + O\bigl( (\lambda-\s_r)^{2m_r-1} \bigr)\,.
\end{equation}
Note that $\xi_r^{[p]}=0$ for $0 \leq p \leq m_r - 2$ since $\vp(\lambda)^{-1}$ has a zero of order $m_r$ at $\lambda=\s_r$.
\end{tcolorbox}~\vspace{-12pt}

\paragraph{Hamiltonian and dynamics.} We now want to choose a Hamiltonian $\Hc$, to define the dynamics $\p_t = \lbrace \Hc,\cdot\rbrace$ and obtain a 2-dimensional field theory. It is natural to take $\Hc$ as part of the integrable structure of the AGM. Indeed, in that case, the charges $\Q_i^{(p)}$ and $\Tc_V(\lambda)$ generating this structure are all conserved and Poisson-commuting by construction, thus proving automatically the integrability of the theory in the sense of Subsection \ref{SubSec:Lax}. In order to obtain a local 2-dimensional field theory, we restrict our choice of $\Hc$ to be a local charge: at this point, we could then take it to be any of the $\Q_i^{(p)}$'s or more generally a linear combination thereof. We will ultimately be interested in relativistic theories, for which the Hamiltonian has dimension 1. This further restricts $\Hc$ to be built as a linear combination of the quadratic charges $\Q_1,\dots,\Q_M$.\footnote{Here and for the rest of this section, we are back to the general case of an AGM with arbitrary multiplicities $m_r\in\Z_{\geq 1}$. Recall that we then have $M=\sum_{r=1}^N m_r$ quadratic local charges $\Q_i$.} More concretely, we then define
\begin{equation}\label{Eq:H}
\Hc = \sum_{i=1}^M \epsilon_i\,\Q_i\,,
\end{equation}
where $\epsilon_i \in \R$ are real parameters, which for the moment are free to take any values. At this point, we note the analogy with the momentum \eqref{Eq:P}, which generates the spatial derivative $\p_x = \lbrace \Pc,\cdot\rbrace$ and which corresponds to all the coefficients $\epsilon_i$ being equal to 1. Having defined the dynamics $\p_t = \lbrace \Hc,\cdot \rbrace$, we will now consider observables in $\Ac$ to be time-dependent: in particular, we now see the currents $\Jc_r^{[p]}(x,t)$ as 2-dimensional fields depending on the space-time coordinates $(x,t)$.

Recall that with the reality conditions imposed in Subsection \ref{SubSec:Gen}, the charges $\Q_i$ are real observables: this ensures that the Hamiltonian itself is real. Let us now suppose that the real Lie algebra $\g$ considered to define these reality conditions is the compact real form of $\g^\C$, in which case the bilinear form $\psd$ is definite positive on $\g$. One checks from equation \eqref{Eq:RealityGaudinLax} that the fields $\Gamma(\ze_i)$'s are all valued in $\g$ as we suppose that the zeroes $\ze_i$ belong to $\R$. Thus, one finds that the local charge $\Q_i$, given by equation \eqref{Eq:Qi}, is always of the same sign as $-\vp'(\ze_i)$, for all fields configurations. Choosing the coefficient $\epsilon_i$ to also be of this sign for each $i$, we find that the Hamiltonian $\Hc$ is positive and in particular bounded below. This is a property often asked to define physical field theories.

\paragraph{Equations of motion and Lax connection.} The Hamiltonian \eqref{Eq:H} can be expressed as a quadratic combination of the Takiff currents
\begin{equation}\label{Eq:HJ}
\Hc = \frac{1}{2} \sum_{r,s=1}^N \sum_{p=0}^{m_r-1}\, \sum_{q=0}^{m_s-1} a_{rs}^{[pq]} \; \int_0^{2\pi} \psb{\Jc_r^{[p]}(x,t)}{\Jc_s^{[q]}(x,t)}\,\dd x\,,
\end{equation}
for some symmetric coefficients $a_{rs}^{[pq]}=a_{sr}^{[qp]}$ defined in a rather complicated way in terms of the parameters $(\s_r,\ze_i,\ell_\infty,\epsilon_i)$ and whose explicit expression we will not need. From this formula and the Takiff bracket \eqref{Eq:PbTakiff}, one can derive the equations of motion of the currents $\Jc_r^{[p]}(x,t)$. This is the subject of the exercise 10 below, which yields
\begin{equation}\label{Eq:EoMJ}
\p_t \Jc_r^{[p]} =  \sum_{s=1}^{N} \sum_{q=0}^{m_r-1-p}\,\sum_{n=1}^{m_s-1} \, a_{rs}^{[qn]}\,\Bigl( \ell_r^{[p+q]}\,\p_x \Jc_s^{[n]} +  \bigl[ \Jc_r^{[p+q]}, \Jc_s^{[n]} \bigr] \Bigr)\,.\vspace{4pt}
\end{equation}

\begin{tcolorbox} \textit{\underline{Exercise 10:} Equations of motion of the currents.} \raisebox{0.5pt}{\Large$\;\;\star$} \vspace{5pt}\\
Using the expression \eqref{Eq:HJ} of the Hamiltonian and the Takiff bracket \eqref{Eq:PbTakiff}, compute the time derivative $\p_t\Jc_r^{[p]} = \bigl\lbrace \Hc, \Jc_r^{[p]} \bigr\rbrace$ of the Takiff currents and show that it takes the form \eqref{Eq:EoMJ}.
\end{tcolorbox}~\vspace{-10pt}

The equations of motion \eqref{Eq:EoMJ} are quite non-trivial, in particular since the coefficients $a_{rs}^{[pq]}$ are complicated combinations of the parameters of the theory. However, an important point here is that this dynamics can always be recast in a compact way in terms of a Lax connection. Indeed, it is clear from equations \eqref{Eq:ZCEi} and \eqref{Eq:H} that the time evolution of the Lax matrix $\Lc_x(\lambda)$ takes the form of a zero curvature equation
\begin{equation}\label{Eq:ZCEg}
\p_t \Lc_x(\lambda) - \p_x \Lc_t(\lambda) + \bigl[ \Lc_t(\lambda), \Lc_x(\lambda) \bigr] = 0\,, \qquad \forall \lambda\in\C\,,
\end{equation}
where the temporal component of the Lax connection is defined as
\begin{equation}
\Lc_t(\lambda) = \sum_{i=1}^M \epsilon_i \, \Kc_i(\lambda)\,.
\end{equation}
It is instructive to compare this expression with the one \eqref{Eq:LPoles} of the spatial component. Combined together, we find that the 2-dimensional Lax connection reads:
\begin{equation}\label{Eq:Ltx}
\Lc_x(\lambda) = \sum_{i=1}^M \frac{1}{\vp'(\ze_i)}\frac{\Gamma(\ze_i)}{\lambda-\ze_i} \qquad \text{ and } \qquad \Lc_t(\lambda) = \sum_{i=1}^M \frac{\epsilon_i}{\vp'(\ze_i)}\frac{\Gamma(\ze_i)}{\lambda-\ze_i}\,.
\end{equation}

In principle, one could check by hand that the equations of motion \eqref{Eq:EoMJ} can be recast as the zero curvature equation \eqref{Eq:ZCEg} for the Lax connection \eqref{Eq:Ltx}, after reexpressing the latter as a linear combination of the currents $\Jc_r^{[p]}$ and finding the explicit expression of the coefficients $a_{rs}^{[pq]}$. This would be a quite tedious endeavour, which we will not attempt here. We also note that it would be rather arduous to guess the expression \eqref{Eq:Ltx} of the flat Lax connection directly from the equations of motion \eqref{Eq:EoMJ} if we did not already know it from the AGM formalism. Even more so, it would very difficult to determine which choice of coefficients $a_{rs}^{[pq]}$ in an Hamiltonian of the form \eqref{Eq:HJ} would lead to an integrable theory if we were not guided by AGMs. This shows the power of the AGM construction: indeed, it produces a Hamiltonian which is automatically integrable and also provides an explicit expression for the Lax connection of the theory.\footnote{Note that, technically, although the AGM formalism produces integrable Hamiltonians of the form \eqref{Eq:HJ}, it might not give all of them. It is however quite tempting to conjecture that this is the case.}

\subsection{Space-time symmetries}

Now that we have defined a dynamical field theory from the AGM, with 2-dimensional coordinates $(x,t)$, we come to the discussion of its space-time symmetries.\vspace{-4pt}

\paragraph{Space-time translations and energy-momentum tensor.} By construction, the model is invariant under \textit{space-time translations} $(x,t)\mapsto(x+a,t+b)$. The corresponding conserved Noether charges are the momentum $\Pc$ and Hamiltonian $\Hc$, given by equations \eqref{Eq:P} and \eqref{Eq:H}. Since they are both expressed as linear combinations of the quadratic local charges $\Q_i$, it is clear that they Poisson-commute, \textit{i.e.} $\lbrace \Hc, \Pc \rbrace = 0$, as expected from the conservation of $\Pc$.

Following the standard conventions for field theories, we will use the labels $0$ and $1$ to denote respectively temporal and spatial components of tensors, as for instance in the derivatives $\p_0=\p_t$ and $\p_1 = \p_x$. We will also use greek letters $\mu,\nu,\dots \in \lbrace 0,1 \rbrace$ for unspecified space-time indices. In these notations, the translation symmetry of the theory is reflected in the local conservation equation
\begin{equation}\label{Eq:ConsEM}
\p_\mu T^\mu_{\;\,\nu} = \p_0 T^0_{\;\,\nu} + \p_1 T^1_{\;\,\nu} = 0
\end{equation}
of the \textit{energy-momentum tensor} $T^\mu_{\;\,\nu}$. The components $T^0_{\;\,0}$ and $T^0_{\;\,1}$ are defined as the densities of the Hamiltonian and momentum respectively:
\begin{equation}
\Hc = \int_0^{2\pi} T^0_{\;\,0}(x,t) \,\dd x \qquad \text{ and } \qquad \Pc = \int_0^{2\pi} T^0_{\;\,1}(x,t) \,\dd x \,.
\end{equation}
The local conservation equation \eqref{Eq:ConsEM} then ensures that the time derivative of these densities are also spatial derivatives, so that $\Hc$ and $\Pc$ are conserved charges.\\

In the present case, we can read the components $T^0_{\;\,0}$ and $T^0_{\;\,1}$ from the expressions \eqref{Eq:H} and \eqref{Eq:P} of $\Hc$ and $\Pc$, yielding
\begin{equation}\label{Eq:T0}
T^0_{\;\,0}(x,t) = \sum_{i=1}^M \epsilon_i\,q_i(x,t) \qquad \text{ and } \qquad T^0_{\;\,1}(x,t) = \sum_{i=1}^M q_i(x,t)\,,
\end{equation}
where $q_i(x,t)$ is the density \eqref{Eq:q} of the quadratic local charge $\Q_i$. The other two components of $T^\mu_{\;\,\nu}$ can be determined through the local conservation equation \eqref{Eq:ConsEM}. More precisely, one has to compute the time derivatives $\p_0 T^0_{\;\,0}$ and $\p_0 T^0_{\;\,1}$ explicitly and put them in the form of spatial derivatives to read off $T^1_{\;\,0}$ and $T^1_{\;\,1}$ . We leave that as an exercise to the reader (see below). In the end, one finds
\begin{equation}\label{Eq:T1}
T^1_{\;\,0}(x,t) = -\sum_{i=1}^M \epsilon^2_i\,q_i(x,t) \qquad \text{ and } \qquad T^1_{\;\,1}(x,t) = -\sum_{i=1}^M \epsilon_i\,q_i(x,t)\,.
\end{equation}

\begin{tcolorbox} \textit{\underline{Exercise 11:} Energy-momentum tensor} \raisebox{0.5pt}{\Large$\;\;\star$} \vspace{5pt}\\
Use the Poisson-algebra \eqref{Eq:Pbq} of the quadratic densities $q_i(x,t)$ to compute the time derivative
\begin{equation}
\p_0 T^0_{\;\,\mu}(x,t) = \bigl\lbrace \Hc, T^0_{\;\,\mu}(x,t) \bigr\rbrace
\end{equation}
for $\mu=0,1$. Put it in the form of a spatial derivative
\begin{equation}
\p_0 T^0_{\;\,\mu}(x,t) = -\p_1 T^1_{\;\,\mu}(x,t)
\end{equation}
and use this to determine the spatial component $T^1_{\;\,\mu}(x,t)$ of the conserved energy-momentum tensor. You should find equation \eqref{Eq:T1}.
\end{tcolorbox}~\vspace{-12pt}

\paragraph{Classical scale invariance.} Now that we have determined the components $T^\mu_{\;\,\nu}$ of the energy-momentum tensor in equations \eqref{Eq:T0} and \eqref{Eq:T1}, we can use this to derive additional results about the space-time symmetries of the AGM. For instance, it is well-known that the theory is scale-invariant if the energy-momentum tensor is traceless, \textit{i.e.} if
\begin{equation}
T^\mu_{\;\,\mu} = T^0_{\;\,0} + T^1_{\;\,1} = 0\,.
\end{equation}
From equations \eqref{Eq:T0} and \eqref{Eq:T1}, this condition clearly holds in the present case. The AGM, with the choice of Hamiltonian \eqref{Eq:H}, is thus invariant under dilations $(x,t)\mapsto(\lambda x,\lambda t)$, at least classically. In general, this property will be broken at the quantum level and the theory will develop a scale anomaly, reflected in particular in its non-trivial renormalisation group flow. We will briefly discuss some of these aspects in Subsection \ref{SubSec:RG}.

\paragraph{Relativistic invariance.} We now turn our attention to  relativistic invariance. To do so, we first have to define a Lorentzian metric $\eta=\bigl( \eta_{\mu\nu} \bigr)_{\mu,\nu=0,1}$ on the worldsheet, with respect to which we will consider Lorentz transformations. Here, we will take it to be the Minkowski metric $\eta = \text{diag}(+1,-1)$, so that the time direction $t$ has positive signature and the spatial one $x$ has negative signature. The Lorentz transformations that leave this metric invariant are the boosts\footnote{Let us note that these boosts are not well-defined on the cylinder $\mathbb{S}^1\times \R$, as they do not preserve the periodicity $x\sim x+2\pi$ of the spatial direction. For this choice of worldsheet, the relativistic symmetry of the theory is then mostly meant as the invariance under infinitesimal Lorentz boosts, which can still be considered locally.}
\begin{equation}
(x,t)\mapsto(\cosh(\theta)\,x+\sinh(\theta)\,t,\cosh(\theta)\,t+\sinh(\theta)\,x)\,.
\end{equation}
As usual we denote by $\bigl( \eta^{\mu\nu} \bigr)_{\mu,\nu=0,1}$ the inverse of $\bigl( \eta_{\mu\nu} \bigr)_{\mu,\nu=0,1}$ and use it to define the energy-momentum tensor with raised indices:
\begin{equation}
T^{\mu\nu} = \eta^{\nu\rho}\,T^\mu_{\;\;\rho}\,.
\end{equation}
In matrix terms, we then have
\begin{equation}\label{Eq:Tup}
\begin{pmatrix}
T^{00} & T^{01} \\
T^{10} & T^{11} 
\end{pmatrix} =
\begin{pmatrix}
T^0_{\;\,0} & -T^0_{\;\,1} \\
T^1_{\;\,0} & -T^1_{\;\,1}
\end{pmatrix} = \sum_{i=1}^M \, q_i \begin{pmatrix}
\epsilon_i & -1 \\
-\epsilon_i^2 & \epsilon_i
\end{pmatrix}\,,
\end{equation}
where we used equations \eqref{Eq:T0} and \eqref{Eq:T1} to obtain the last equality.

In these notations, the theory is relativistic invariant if and only if $T^{\mu\nu}$ is symmetric\footnote{There is a technical subtlety here, due to the non-uniqueness of the energy-momentum tensor. Indeed, one can consider its modification $T^{\mu\nu}_\Sigma=T^{\mu\nu} + \p_\rho\,\Sigma^{\rho\mu\nu}$, where $\Sigma$ is a tensor which is skew-symmetric in the first two indices $\rho$ and $\mu$: one easily checks that $T^{\mu\nu}_\Sigma$ is also conserved, \textit{i.e.} $\p_\mu T^{\mu\nu}_\Sigma=0$. Moreover, such a modification changes the conserved charges $\int_0^{2\pi} T^{0\mu}\,\dd x$ only by boundary terms, which vanish due to the periodic boundary conditions. In this context, the relativistic invariance of the theory is then more precisely equivalent to the existence of a choice of $\Sigma$ such that $T^{\mu\nu}_\Sigma$ is symmetric. In the present case, one can convince oneself that adding such a modification to the tensor \eqref{Eq:Tup} will create terms which are not expressible in terms of the densities $q_i$ only and that we do not have to consider such modifications in our treatment of relativistic invariance.}. Using equation \eqref{Eq:Tup}, it is clear that this happens if and only if
\begin{equation}
\epsilon_i^2 = 1\,, \qquad \forall\,i\in\lbrace 1,\dots,M \rbrace\,.
\end{equation}
Thus, relativistic invariance imposes that each of the coefficients $\epsilon_i$ in the definition \eqref{Eq:H} of the Hamiltonian is equal to either $+1$ or $-1$. We will restrict to this relativistic case in the rest of the lectures. We then introduce the sets
\begin{equation}\label{Eq:Ipm}
\Ic_\pm = \bigl\lbrace\, i\in\lbrace 1,\dots,M\rbrace\;\bigl|\; \epsilon_i = \pm 1 \bigr\rbrace\,,
\end{equation}
which form a partition $\lbrace 1,\dots,M \rbrace = \Ic_+ \sqcup \Ic_-$. Recall that these indices $i\in\lbrace 1,\dots,M\rbrace$ are naturally associated with the zeroes $\bm\ze=(\ze_i)_{i=1,\dots,M}$ of the twist function: we thus obtain a partition of the zeroes into two subsets $\bm{\ze_\pm}=(\ze_i)_{i\in\Ic_\pm}$.

\paragraph{Light-cone structure.} Now that we have discussed the relativistic invariance of the model, let us investigate its light-cone structure. We introduce the light-cone coordinates $x^\pm$ and their derivatives:
\begin{equation}
x^\pm = t \pm x \qquad \text{ and } \qquad \p_\pm = \frac{1}{2} \bigl( \p_t \pm \p_x \bigr)\,.
\end{equation}
By construction, these derivatives are generated by the charges
\begin{equation}\label{Eq:PLc}
\Pc_\pm = \frac{1}{2} \bigl( \Hc \pm \Pc \bigr) = \pm \sum_{i\in\Ic_\pm} \Q_i\,.
\end{equation}
In this equation, we have used the subsets $\Ic_\pm \subset \lbrace 1,\dots, M\rbrace$ introduced in \eqref{Eq:Ipm}. In particular, we see that the separation of the zeroes of the twist function into the two subsets $\bm{\ze_\pm}=(\ze_i)_{i\in\Ic_\pm}$ is naturally related to the light-cone directions of the worldsheet.\\

This is also visible in the integrable structure of the theory. For instance, the zero curvature equation of the Lax connection can be written as
\begin{equation}
\p_+ \Lc_-(\lambda) - \p_-\Lc_+(\lambda) + \bigl[ \Lc_+(\lambda),\Lc_-(\lambda) \bigr] = 0\,,
\end{equation}
where we have introduced the light-cone components $\Lc_\pm(\lambda) = \frac{1}{2} \bigl( \Lc_t(\lambda) \pm \Lc_x(\lambda) \bigr)$. From equation \eqref{Eq:Ltx}, one sees that these components read
\begin{equation}\label{Eq:Lpm}
\Lc_\pm(\lambda) = \pm \sum_{i\in\Ic_\pm} \frac{1}{\vp'(\ze_i)}\frac{\Gamma(\ze_i)}{\lambda-\ze_i}\,.
\end{equation}
We thus observe that the subsets of zeroes  $\bm{\ze_\pm}=(\ze_i)_{i\in\Ic_\pm}$ control the analytical structure of the light-cone Lax connection, \textit{i.e.} the position of the poles of $\Lc_\pm(\lambda)$. In particular, for a relativistic AGM, $\Lc_+(\lambda)$ and $\Lc_-(\lambda)$ do not share any common pole.

Let us finally comment on the densities $q_i^{(p)}(x,t)$ of the local charges $\Q_i^{(p)}$. Using the property \eqref{Eq:Pbqiqj} and the expression \eqref{Eq:PLc} of the light-cone Hamiltonians $\Pc_\pm$, one easily shows that
\begin{equation}
\p_\mp q_i^{(p)}(x,t) = 0\,, \qquad \text{ if } \quad i\in\Ic_\pm\,.
\end{equation}
In other words, the density of $\Q_i^{(p)}\hspace{-0.8pt}$ is left-moving if $i\hspace{-0.5pt}\in\hspace{-0.75pt}\Ic_+$ and right-moving if $i\hspace{-0.5pt}\in\hspace{-0.75pt}\Ic_-$, at least classically. In general, we expect this property to be broken at the quantum level by the conformal anomaly.

\section[Realisations of AGMs and integrable \texorpdfstring{$\s$}{sigma}-models]{Realisations of AGMs and integrable \texorpdfstring{$\bm\s$}{sigma}-models}
\label{Sec:Sigma}

We now want to explain the relation between AGMs and integrable $\s$-models. For that we will need to introduce the notion of Kac-Moody/Takiff realisations, which will be the main subject of the first subsection. In the second one, we will explain how such a realisation produces an integrable $\s$-model from an AGM. The third subsection will concern the detailed treatment of the simplest example of integrable $\s$-model that can be obtained this way, the so-called Principal Chiral Model. Finally, in the fourth and last subsection, we will discuss briefly some other examples and will give an overview of the panorama of integrable $\s$-models arising from realisations of AGMs.

\subsection{Takiff realisations from canonical fields}
\label{SubSec:Real}

\paragraph{Canonical fields and bracket.} The fundamental fields of the AGM are the Kac-Moody currents $\Jc_r(x)$ attached to its sites, or more generally the Takiff currents $\Jc_r^{[p]}(x)$ in the case of sites with multiplicities. To introduce the notion of realisation, we first have to consider another type of fields and of Poisson bracket, namely the canonical ones. We already discussed those in the reminder subsection \ref{SubSec:HFT}: to ease the reading, we briefly recall the main concepts that we will need here. Let us fix a real smooth manifold $\Mc$ of dimension $d$ and consider a field $\Phi(x)$ depending periodically on the spatial variable $x\in[0,2\pi]$ and valued in $\Mc$. In a local chart, $\Phi(x)$ can be equivalently described by $d$ \textit{coordinate fields} $\bigl( \phi^{\hspace{0.5pt}i}(x) \bigr)_{i=1,\dots,d}$. To obtain the full description of $\Phi(x)$, one has to consider all the local coordinate charts covering $\Mc$ and patch together these local descriptions: for simplicity, we will not enter into these global considerations here and will always work with the local formulation.

To each coordinate field $\phi^{\hspace{0.5pt}i}(x)$, we associate a \textit{conjugate momentum field} $\pi_i(x)$. Geometrically, the $2d$ fields $\bigl( \phi^{\hspace{0.5pt}i}(x), \pi_i(x) \bigr)_{i=1,\dots,d}$ form the local description of a field valued in the cotangent bundle $T^\ast \Mc$, with the coordinates $\phi^{\hspace{0.5pt}i}$ describing the basis $\Mc$ and the momenta $\pi_i$ describing the cotangent fibers. As above, we will not get into the details of this global geometric formulation. Concretely, the main result that we will need is that the cotangent bundle $T^\ast \Mc$ is equipped with a natural Poisson structure, which can be used to define the so-called \textit{canonical Poisson bracket} on the fields $\bigl( \phi^{\hspace{0.5pt}i}(x), \pi_i(x) \bigr)_{i=1,\dots,d}$.  The latter reads
\begin{equation}\label{Eq:PbCan2}
\bigl\lbrace \pi_i(x), \phi^{\,j}(y) \bigr\rbrace = \delta_{i}^{\;j} \,\delta(x-y), \qquad \bigl\lbrace \pi_i(x), \pi_{j}(y) \bigr\rbrace = \bigl\lbrace \phi^{\hspace{0.5pt}i}(x), \phi^{\,j}(y) \bigr\rbrace = 0\,,
\end{equation}
making $\bigl( \phi^{\hspace{0.5pt}i}(x), \pi_i(x) \bigr)$ the field theory equivalents of the canonical coordinates $(q^{\hspace{0.5pt}i},p_i)$ in classical mechanics. We will denote by $\Acm$ the Poisson algebra generated by these canonical fields.

\paragraph{Takiff realisation.} Let us now consider combinations of the canonical fields $\bigl( \phi^{\hspace{0.5pt}i}(x), \pi_i(x) \bigr)$ taking the specific form
\begin{equation}\label{Eq:RealTakiff}
J_{r,\,a}^{[p]} (x) = A_{r,p,a}^{\hspace{19pt}i}\bigl( \Phi(x) \bigr)\,\pi_i(x) + B_{r,p,a\,;\,i}\bigl( \Phi(x) \bigr)\,\p_x\phi^{\hspace{0.5pt}i}(x)
\end{equation}
and indexed by the same labels
\begin{equation*}
r\in\lbrace 1,\dots,N\rbrace\,, \qquad p\in\lbrace 0,\dots,m_r-1 \rbrace\qquad \text{ and } \qquad a\in\lbrace 1,\dots,\dim\g\bigr\rbrace
\end{equation*}
as the components of the Takiff currents $\Jc_r^{[p]}=\Jc_{r,\,a}^{[p]}\,\td^a$ underlying our AGM. Note that a sum over $i\in\lbrace 1,\dots,d \rbrace$ is implied in the equation \eqref{Eq:RealTakiff}. The $J_{r,\,a}^{[p]} (x)$'s are thus composite fields, built as linear combinations of the momenta $\pi_i(x)$ and the derivatives $\p_x\phi^{\hspace{0.5pt}i}(x)$, with coefficients $A_{r,p,a}^{\hspace{19pt}i}$ and $B_{r,p,a\,;\,i}$ given by functions of the coordinate fields $\phi^{\,j}(x)$ but not of their derivatives or the momenta. 

With respect to the canonical Poisson structure \eqref{Eq:PbCan2}, the Poison bracket of two of these fields $J_{r,\,a}^{[p]} (x)$ and $J_{s,\,b}^{[q]} (y)$ contains a term proportional to the Dirac distribution $\delta(x-y)$, with coefficient depending linearly on $\pi_i(x)$ and $\p_x\phi^{\hspace{0.5pt}i}(x)$ and non-linearly on the fields $\phi^{\hspace{0.5pt}i}(x)$, and a term proportional to $\p_x \delta(x-y)$, whose coefficient can depend on the fields $\phi^{\hspace{0.5pt}i}(x)$ but not on their derivatives or the momenta. This is reminiscent of the current-type brackets considered in the previous sections. Guided by this observation, we say that the fields $J_{r,\,a}^{[p]} (x)$ define a \textit{Takiff realisation} if their brackets take the very specific form
\begin{equation}\label{Eq:PbTakiffReal}
\bigl\lbrace J_{r,\,a}^{[p]}(x), J_{s,\,b}^{[q]}(y) \bigr\rbrace = \delta_{rs} \left\lbrace \begin{array}{lcl}
\f{ab}c \, J_{r,\,c}^{[p+q]}(x)\, \delta(x-y) - \ell_r^{[p+q]}\,\kappa_{ab}\,\p_x\delta(x-y) & & \text{ if }\, p+q < m_r \\[5pt]
0 & & \text{ if }\, p+q \geq m_r
\end{array} \right.\,.
\end{equation}
Written in terms of the Lie algebra-valued fields $J_r^{[p]}(x)=J_{r,\,a}^{[p]}(x)\,\td^a$, this becomes the Takiff bracket \eqref{Eq:PbTakiff}, with the abstract Takiff currents $\Jc_r^{[p]}(x)$ replaced by the composite fields $J_r^{[p]}(x)$. Obtaining such a bracket is quite non-trivial and requires choosing very fine-tuned coefficients $A_{r,p,a}^{\hspace{19pt}i}$ and $B_{r,p,a\,;\,i}$ in the definition \eqref{Eq:RealTakiff}. If we manage to do so, we then have realised the Takiff currents and their bracket in terms of the canonical fields $\bigl( \phi^{\hspace{0.5pt}i}(x), \pi_i(x) \bigr)_{i=1,\dots,d}$. More formally, this amounts to a Poisson map from the Takiff Poisson algebra $\Ac$ to the canonical one $\Acm$, sending the abstract currents $\Jc_r^{[p]}(x)$ in $\Ac$ to the composite fields $J_r^{[p]}(x)$ in $\Acm$.\\

As explained in Subsection \ref{SubSec:Gen}, we impose certain reality conditions on the Takiff currents of an AGM, which are either real or come in pairs of complex conjugates. If one considers a realisation $J_r^{[p]}(x)$ of these currents, as here through equation \eqref{Eq:RealTakiff}, we will suppose that the fields $J_r^{[p]}(x)$ satisfy the same reality conditions. Let us be a bit more explicit. Here, we took $\Mc$ to be a real manifold: the canonical fields $\bigl( \phi^{\hspace{0.5pt}i}(x), \pi_i(x) \bigr)_{i=1,\dots,d}$ are thus all real. Let us then consider the components $J_{r,\,a}^{[p]}(x)$ of the realised currents, with respect to a basis $\lbrace \td^a \rbrace_{a=1,\dots,\dim\g}$ of the real form $\g$, defined as the composite fields \eqref{Eq:RealTakiff} in terms of $\bigl( \phi^{\hspace{0.5pt}i}(x), \pi_i(x) \bigr)$. We then ask that each of these components is either a real field $J_{r,\,a}^{[p]}(x) \in \R$ or belongs to a pair of complex conjugate fields $\bigl(J_{r,\,a}^{[p]}(x),J_{\bar r,\,a}^{[p]}(x)\bigr)$. We will consider these requirements as part of the definition of a Takiff realisation.

\paragraph{An $\bm{\sl(2,\R)}$ example.} Before explaining the use of Takiff realisations, let us describe one simple example. For this paragraph, we consider the Lie algebra $\g=\sl(2,\R)$, with basis
\begin{equation}\label{Eq:sl2}
\td_1 = \begin{pmatrix}
0 & 1 \\ 0 & 0
\end{pmatrix}\,, \qquad
\td_2 = \begin{pmatrix}
1 & 0 \\ 0 & -1
\end{pmatrix}\,, \qquad
\td_3 = \begin{pmatrix}
0 & 0 \\ 1 & 0
\end{pmatrix}\,.
\end{equation}
With respect to this basis, the bilinear form $\kappa_{ab} = \ps{\td_a}{\td_b}$ considered in the previous sections reads
\begin{equation}\label{Eq:KappaSl2}
\bigl( \kappa_{ab} \bigr)_{a,b=1,2,3} = -\begin{pmatrix}
0 & 0 & 1 \\ 0 & 2 & 0 \\ 1 & 0 & 0
\end{pmatrix}\,.
\end{equation}
To build a Takiff realisation, we will work with the canonical Poisson algebra $\Ac_{T^\ast G}$, where $G=SL(2,\R)$ is the Lie group of $2\times 2$ matrices of determinant 1, whose Lie algebra is $\sl(2,\R)$. $\Ac_{T^\ast G}$ is generated by $\dim G = 3$ coordinate fields $\phi^{\hspace{0.5pt}i}(x)$, parametrising a $G$-valued field\footnote{The $G$-valued field $g(x)$ considered here is the equivalent of the $\Mc$-valued field $\Phi(x)$ used in the previous paragraph. Throughout the lectures, we will reserve the notation $g$ for group-valued fields.} $g(x)$ in a local chart, and the 3 conjugate momenta $\pi_i(x)$. Here, we choose the parametrisation
\begin{equation}\label{Eq:gSL2}
g(x) = \begin{pmatrix}
1 & 0 \\ \phi^{\hspace{0.5pt}2}(x) & 1
\end{pmatrix}\cdot\begin{pmatrix}
e^{\phi^{\hspace{0.5pt}3}(x)} & 0 \\ 0 & e^{-\phi^{\hspace{0.5pt}3}(x)}
\end{pmatrix}\cdot\begin{pmatrix}
1 & \phi^{\hspace{0.5pt}1}(x) \\ 0 & 1
\end{pmatrix}\,.
\end{equation}

From the canonical fields $\bigl( \phi^{\hspace{0.5pt}i}(x), \pi_i(x) \bigr)_{i=1,\dots,3}$, we will build Takiff currents $J^{[0]}(x)$ and $J^{[1]}(x)$ of multiplicity $m=2$. Here, we consider only these two currents and thus omit the additional index $r\in\lbrace 1,\dots,N\rbrace$ used above to label the different ``groups'' of Takiff currents (in the language of AGMs, this corresponds to a model with only one site, which has multiplicity 2). Quite explicitly, we define the currents $J^{[p]}(x)=J^{[p]}_a(x)\,\td^a$ by their components
\begin{equation*}
J^{[0]}_1(x) = \pi_1(x)\,, \qquad J^{[0]}_2(x) = \pi_3(x) - 2 \phi^{\hspace{0.5pt}1}(x)\,\pi_1(x)\,, 
\end{equation*}
\begin{equation}\label{Eq:RealSl2}
J^{[0]}_3(x) = e^{-2\phi^{\hspace{0.5pt}3}(x)}\, \pi_2(x) + \phi^{\hspace{0.5pt}1}(x) \, \pi_3(x) - \phi^{\hspace{0.5pt}1}(x)^2\,\pi_1(x) \,, \vspace{2pt}
\end{equation}
\begin{equation*}
J^{[1]}_1(x) = -\ell^{[1]}\, e^{2\phi^{\hspace{0.5pt}3}(x)}\,\p_x\phi^{\hspace{0.5pt}2}(x)\,, \qquad J^{[1]}_2(x) = 2\ell^{[1]}\bigl( \, e^{2\phi^{\hspace{0.5pt}3}(x)}\,\phi^{\hspace{0.5pt}1}(x)\,\p_x\phi^{\hspace{0.5pt}2}(x) - \p_x\phi^{\hspace{0.5pt}3}(x) \bigr)\,,
\end{equation*}
\begin{equation*}
J^{[1]}_3(x) = \ell^{[1]}\bigl( e^{2\phi^{\hspace{0.5pt}3}(x)}\,\phi^{\hspace{0.5pt}1}(x)^2\,\p_x\phi^{\hspace{0.5pt}2}(x) - 2\,\phi^{\hspace{0.5pt}1}(x)\,\p_x\phi^{\hspace{0.5pt}3}(x) - \p_x\phi^{\hspace{0.5pt}1}(x) \bigr)\,,
\end{equation*}
where $\ell^{[1]}\in\R$ is a real parameter. It is clear that this is a particular case of the general form \eqref{Eq:RealTakiff}. \\

Our claim is that the fields \eqref{Eq:RealSl2} define a Takiff realisation with Lie algebra $\g=\sl(2,\R)$, multiplicity $m=2$ and levels $\ell^{[0]}=0$ and $\ell^{[1]}$. Concretely, this means that they satisfy the Poisson bracket
\begin{subequations}\label{Eq:PbTakiff2}
\begin{eqnarray}
\bigl\lbrace J^{[0]}_a(x), J^{[0]}_b(y) \bigr\rbrace &=&
\f{ab}c \, J^{[0]}_c(x)\, \delta(x-y) \,, \label{Eq:PbJ00}\\
\bigl\lbrace J^{[0]}_a(x), J^{[1]}_b(y) \bigr\rbrace &=&
\f{ab}c \, J^{[1]}_c(x)\, \delta(x-y) - \ell^{[1]}\,\kappa_{ab}\,\p_x\delta(x-y) \,, \\
\bigl\lbrace J^{[1]}_a(x), J^{[1]}_b(y) \bigr\rbrace &=&
0\,,
\end{eqnarray}
\end{subequations}
where $\f{ab}c$ are the structure constants of $\sl(2,\R)$ in the basis \eqref{Eq:sl2} and $\kappa_{ab}$ the bilinear form entries \eqref{Eq:KappaSl2}. This can be checked directly from the canonical bracket \eqref{Eq:PbCan2} and its consequence $\lbrace \pi_i(x), \p_y\phi^{\,j}(y) \rbrace = -\delta_{i}^{\;j}\,\p_x\delta(x-y)$. Let us give an example. Using the Leibniz rule, we find
\begin{align*}
\bigl\lbrace J^{[0]}_1(x), J^{[0]}_2(y) \bigr\rbrace 
& = -2\,\bigl\lbrace \pi_1(x), \phi^{\hspace{0.5pt}1}(y)   \bigr\rbrace \, \pi_1(y) = -2\,\pi_1(x) \, \delta(x-y) = -2\,J^{[0]}_1(x)\,\delta(x-y)\,.
\end{align*}
This agrees with \eqref{Eq:PbJ00} for $(a,b)=(1,2)$, using $[\td_{1}, \td_{2}]=-2\td_{1}$, \textit{i.e.} $\f{12}c = -2\delta^{c}_{\;\,1}$. The other brackets in \eqref{Eq:PbTakiff2} can be computed in a similar way and are left as an exercise to the reader. In fact, in the exercise 12 below, we consider a slightly more general realisation, for which $J^{[0]}_2(x)$ and $J^{[0]}_3(x)$ also contain derivatives $\p_x\phi^{\hspace{0.5pt}i}(x)$: the effect of these additional terms is to turn on the level $\ell^{[0]}$.

Let us note that the fields $J^{[p]}_a(x)$ defining the realisation \eqref{Eq:RealSl2} are all real combinations of the canonical fields. In other words, the currents $J^{[p]}(x)$ are valued in the real Lie algebra $\g=\sl(2,\R)$. This realisation thus satisfies the reality conditions considered in the previous paragraph.\\

To summarise, we showed here that there is a $\sl(2,\R)$--Takiff realisation of multiplicity 2 in terms of the canonical fields in $\Ac^{\text{can}}_{T^\ast SL(2,\R)}$. This is in fact a special case of a more general result: indeed, for any simple Lie group $G$ with Lie algebra $\g$, there exists a $\g$--Takiff realisation of multiplicity 2 in terms of the canonical fields in $\Ac^{\text{can}}_{T^\ast G}$. We will come back to this general case in Subsection \ref{SubSec:PCM}: in fact, this will be the starting point for the construction of the Principal Chiral Model as an AGM.\vspace{10pt}

\begin{tcolorbox} \textit{\underline{Exercise 12:} Takiff realisation in $\Ac^{\text{can}}_{T^\ast SL(2,\R)}$.} \raisebox{0.5pt}{\Large$\;\;\star\star$} \vspace{5pt}\\
Consider the currents
\begin{equation*}
J^{[0]}_1(x) = \pi_1(x)\,, \qquad J^{[0]}_2(x) = \pi_3(x) - \ell^{[0]}\,\p_x\phi^{\hspace{0.5pt}3}(x) - 2 \phi^{\hspace{0.5pt}1}(x)\,\pi_1(x)\,, 
\end{equation*}
\begin{equation}\label{Eq:RealSl2k}
J^{[0]}_3(x) = e^{-2\phi^{\hspace{0.5pt}3}(x)} \pi_2(x) + \phi^{\hspace{0.5pt}1}(x) \bigl( \pi_3(x) - \ell^{[0]}\,\p_x\phi^{\hspace{0.5pt}3}(x) \bigr) - \phi^{\hspace{0.5pt}1}(x)^2\,\pi_1(x)  - \ell^{[0]}\,\p_x\phi^{\hspace{0.5pt}1}(x)\,, \vspace{2pt}
\end{equation}
\begin{equation*}
J^{[1]}_1(x) = -\ell^{[1]}\, e^{-2\phi^{\hspace{0.5pt}3}(x)}\,\p_x\phi^{\hspace{0.5pt}2}(x)\,, \qquad J^{[1]}_2(x) = 2\ell^{[1]}\bigl( \, e^{-2\phi^{\hspace{0.5pt}3}(x)}\phi^{\hspace{0.5pt}1}(x)\,\p_x\phi^{\hspace{0.5pt}2}(x) - \p_x\phi^{\hspace{0.5pt}2}(x) \bigr)\,,
\end{equation*}
\begin{equation*}
J^{[1]}_3(x) = \ell^{[1]}\bigl( e^{-2\phi^{\hspace{0.5pt}3}(x)}\phi^{\hspace{0.5pt}1}(x)^2\,\p_x\phi^{\hspace{0.5pt}2}(x) - \phi^{\hspace{0.5pt}1}(x)\,\p_x\phi^{\hspace{0.5pt}2}(x) - \,\p_x\phi^{\hspace{0.5pt}1}(x) \bigr)\,,\vspace{2pt}
\end{equation*}
built from the canonical fields $\bigl( \phi^{\hspace{0.5pt}i}(x), \pi_i(x) \bigr)_{i=1,\dots,3}$ of $\Ac^{\text{can}}_{T^\ast SL(2,\R)}$ and two real parameters $\ell^{[0]},\ell^{[1]}\in\R$. Show that they define a Takiff realisation with Lie algebra $\g=\sl(2,\R)$, multiplicity $m=2$ and levels $\ell^{[0]}$ and $\ell^{[1]}$.\\[4pt]
At this level of complexity, the computation of the Poisson brackets of  the fields $J^{[p]}_a$ is still doable by hand, although already a bit tedious. We note that it is also possible to obtain this result using symbolic computations software, such as Mathematica. These methods become quite useful for similar computations involving a higher number of fields. Solving this exercise with the help of such a software is thus a worthful practice as well.
\end{tcolorbox}~\vspace{-12pt}

\paragraph{Completeness of Takiff realisations.} We end this subsection by discussing a slightly technical but useful notion, that of the completeness of a Takiff realisation. To lighten the lectures, we only sketch the main aspects here and give a more detailed discussion in Appendix \ref{App:Comp}. Consider the realisation \eqref{Eq:RealTakiff}, expressing the currents $J^{[p]}_r(x)$ as composite fields in terms of the canonical fields $\bigl( \phi^{\hspace{0.5pt}i}(x), \pi_i(x) \bigr)_{i=1,\dots,d}$ in $\Acm$. We say that the realisation is \textit{complete} if
\begin{equation}
M\,\dim\g=2d\,,
\end{equation}
with $M=\sum_{r=1}^N m_r$ as defined in equation \eqref{Eq:M}. This means that there are as many current components $J^{[p]}_{r,a}(x)$ as there are canonical fields $\bigl( \phi^{\hspace{0.5pt}i}(x), \pi_i(x) \bigr)$ in $\Acm$. We note that this condition forces either $M$ or $\dim\g$ to be even: from now on, we will always suppose that $M$ is even.

As argued in Appendix \ref{App:Comp}, this completeness property in fact ensures that the canonical fields $\bigl( \phi^{\hspace{0.5pt}i}(x), \pi_i(x) \bigr)$ can be entirely and uniquely reconstructed from the data of the currents $J^{[p]}_r(x)$ and the initial values $\phi^{\hspace{0.5pt}i}(0)$ of the coordinate fields.\footnote{Adding these initial conditions is necessary since the currents involve the spatial derivatives $\p_x\phi^{\hspace{0.5pt}i}(x)$: schematically, reconstructing the fields $\phi^{\hspace{0.5pt}i}(x)$ from the currents thus requires an integration.} Schematically, this means that the functional map expressing $J^{[p]}_r(x)$ in terms of $\bigl( \phi^{\hspace{0.5pt}i}(x), \pi_i(x) \bigr)$ is invertible, up to a finite number of degrees of freedom, namely the initial values $\phi^{\hspace{0.5pt}i}(0)$. We note however that the inverse of this map can generally be a quite complicated and non-local functional.

We refer to Appendix \ref{App:Comp} for more details on this notion of completeness and for exercises. We finally note that the $\sl(2,\R)$--Takiff realisations \eqref{Eq:RealSl2} and \eqref{Eq:RealSl2k} considered in the previous paragraph are complete, since they involve $6$ current components written in terms of $6$ canonical fields (in the above notations, these cases correspond to $M=2$, $\dim\g=3$ and $d=3$). 

\subsection[From AGMs to integrable $\s$-models]{From AGMs to integrable $\bm\s$-models}

\paragraph{Image of the AGM in the Takiff realisation.} Let us consider an AGM with $N$ sites $\s_r$, having multiplicities $m_r \in \Z_{\geq 1}$ such that $M=\sum_{r=1}^N m_r$ is an even number. This model is defined in terms of $M$ Takiff currents $\Jc_r^{[p]}(x)$, generating its Poisson algebra $\Ac$. The main result of the AGM formalism is the construction of an integrable structure $\Zc$ in $\Ac$, generated by an infinite number of Poisson-commuting observables. The latter take the form of non-local charges $\Tc_V(\lambda)$ as well as higher-degree local charges $\Q_i^{(p)}$, built from the currents $\Jc_r^{[p]}(x)$ and depending on the various parameters $(\s_r,\ze_i,\ell_\infty)$ of the model (see Section \ref{Sec:ClassicalAGM} for details).

Suppose now that we are given a complete realisation $J_r^{[p]}(x)$ of these currents in a canonical Poisson algebra $\Acm$, as in the previous subsection. By definition, the realised currents $J_r^{[p]}(x)$ satisfy the same Takiff bracket as $\Jc_r^{[p]}(x)$. The image $\Zc_{T^\ast \Mc}$ of the AGM integrable structure in this realisation is then by construction a Poisson-commutative subalgebra of $\Acm$. Concretely, generators of $\Zc_{T^\ast \Mc}$ are obtained by considering the charges $\Tc_V(\lambda)$ and $\Q_i^{(p)}$ of the AGM and by replacing the abstract Takiff currents $\Jc_r^{[p]}(x)$ with the realised ones $J_r^{[p]}(x)$. We denote these charges by $\Tc_{V,\,T^\ast \Mc}(\lambda)$ and $\Q_{i,\,T^\ast \Mc}^{(p)}$. By construction, these are elements of $\Acm$, built from the canonical fields $\bigl( \phi^{\hspace{0.5pt}i}(x), \pi_i(x) \bigr)$, which are all Poisson-commuting with respect to the canonical bracket.

\paragraph{Momentum and Hamiltonian.} Recall from Subsection \ref{SubSec:MomHam} that the momentum $\Pc$ of the AGM can be characterised as the sum of the quadratic local charges $\Q_i$. Let us then consider its image
\begin{equation}
\Pc_{\TM} = \sum_{i=1}^M \Q_{i,\,\TM}
\end{equation}
in the realisation. In Appendix \ref{App:Comp}, we argue that due to the completeness of the realisation, this quantity is equal to the momentum
\begin{equation}
\Pc_{\TM} = \int_0^{2\pi} \pi_i(x)\,\p_x\phi^{\hspace{0.5pt}i}(x)\,\dd x
\end{equation}
of the canonical Poisson algebra $\Acm$. This fact ensures that all the results of Section \ref{Sec:SpaceTime} on the space-time symmetries of the AGM transfer to the realisation. In particular, we define a relativistic 2-dimensional field theory in $\Acm$ by choosing the Hamiltonian $\Hc_{\TM}$ as the image in the realisation of the AGM Hamiltonian \eqref{Eq:H}, \textit{i.e.}
\begin{equation}\label{Eq:HReal0}
\Hc_{\TM} = \sum_{i=1}^M \epsilon_i\,\Q_{i,\,\TM}\,,
\end{equation}
where the coefficients $\epsilon_i$ are equal to $\pm 1$ for $i\in\Ic_\pm$, to ensure the relativistic invariance of the theory. In this section, we will always suppose that there are as many $\epsilon_i$'s equal to $+1$ as ones equal to $-1$, \textit{i.e.} that the sets $\Ic_\pm$ have equal size $|\Ic_+|=|\Ic_-|=M/2$.\footnote{Recall that we supposed $M$ is even, so that this is always possible. Without this technical assumption, the Hamiltonian field theory we would obtain would be quite degenerate and in particular would not possess a Lagrangian formulation.} The main difference between the field theory considered in Section \ref{Sec:SpaceTime} and the one considered here is that the latter is defined in terms of the canonical fields $\bigl( \phi^{\hspace{0.5pt}i}(x), \pi_i(x) \bigr)$ in $\TM$, and not the abstract Takiff currents $\Jc_r^{[p]}(x)$. By construction, this theory is relativistic and integrable, with conserved Poisson-commuting charges $\Tc_{V,\,T^\ast \Mc}(\lambda)$ and $\Q_{i,\,T^\ast \Mc}^{(p)}$: we say that it is a \textit{relativistic realisation of the AGM} we started with.\\

Re-expressing the local charges $\Q_{i,\,\TM}$ in terms of the currents $J_r^{[p]}(x)$ and using the definition \eqref{Eq:RealTakiff} of the latter, one can write the Hamiltonian $\Hc_{\TM}$ in terms of the canonical fields $\bigl( \phi^{\hspace{0.5pt}i}(x), \pi_i(x) \bigr)$. Doing so, we find an expression of the form
\begin{equation}\label{Eq:HReal}
\Hc_{\TM} = \frac{1}{2} \int_0^{2\pi} \bigl(  G^{\hspace{0.5pt}ij}\,\pi_i\,\pi_j + S_{ij}\,\p_x\phi^{\hspace{0.5pt}i}\,\p_x\phi^{\,j} - 2\,B^{\hspace{0.5pt}i}_{\;\,j}\,\pi_i\,\p_x\phi^{\,j}  \bigr)\,\dd x\,,
\end{equation}
where $G^{\hspace{0.5pt}ij}$, $S_{ij}$ and $B^{\hspace{0.5pt}i}_{\;\,j}$ are tensors, symmetric for the first two, which are (potentially complicated) functions of the parameters $(\s_r,\ze_i,\ell_\infty,\epsilon_i)$ of the AGM and of the coordinate fields $\phi^{\,k}$ (but not of their derivatives or the momenta) -- more precisely, these tensors are built as specific combinations of the coefficients $A_{r,p,a}^{\hspace{19pt}i}$ and $B_{r,p,a\,;\,i}$ appearing in equation \eqref{Eq:RealTakiff}. This is in fact the general structure for the Hamiltonian arising from a Takiff realisation of the form \eqref{Eq:RealTakiff}, without any constraints on the coefficients $\epsilon_i$ appearing in equation \eqref{Eq:HReal0}. As recalled above, we restrict those to be equal to $+1$ or $-1$ to ensure the relativistic invariance of the theory: this induces some constraints on the tensors $G^{\hspace{0.5pt}ij}$, $S_{ij}$ and $B^{\hspace{0.5pt}i}_{\;\,j}$ appearing in $\Hc_{\TM}$. Namely, we have
\begin{equation}\label{Eq:GBH}
B^{\hspace{0.5pt}i}_{\;\,j} = G^{\hspace{0.5pt}ik}\, B_{kj} \qquad \text{ and } \qquad S_{ij} = G_{ij} + G^{kl}\,B_{ki}\,B_{lj}\,,
\end{equation}
where $B_{ij}=-B_{ji}$ is skew-symmetric and $G_{ij}$ is the inverse of $G^{\hspace{0.5pt}ij}$ (such that $G^{\hspace{0.5pt}ik}\,G_{kj}=\delta^{\hspace{0.5pt}i}_{\,\,j}$). The Hamiltonian is thus entirely characterised by the data of two tensors
\begin{equation}
G_{ij}( \Phi ) = G_{ji}( \Phi )  \qquad \text{ and } \qquad B_{ij}( \Phi ) = -B_{ji}( \Phi )
\end{equation}
on the manifold $\Mc$, the first one being symmetric and invertible and the second one being skew-symmetric.\footnote{As explained in Subsection \ref{SubSec:Real}, we are working in a local coordinate chart $\bigl(\phi^{\hspace{0.5pt}i}\bigr)_{i=1,\dots,d}$ on $\Mc$. Considering other charts and how they are patched together, one eventually finds, after various manipulations, that $G_{ij}$ and $B_{ij}$ indeed behave as covariant 2-tensors on $\Mc$.}

\paragraph{Time derivative of the coordinates.} The Hamiltonian \eqref{Eq:HReal} defines the dynamics $\p_t = \lbrace \Hc_{\TM}, \cdot \rbrace$ of the observables in the canonical Poisson algebra $\Acm$. Therefore, we will now see elements of $\Acm$ as time-dependent: in particular, we interpret the canonical fields $\bigl( \phi^{\hspace{0.5pt}i}(x,t), \pi_i(x,t) \bigr)$ as functions on the 2-dimensional worldsheet $\Sigma$. Starting from the canonical Poisson bracket and the expression \eqref{Eq:HReal} of the Hamitonian, it is straightforward to compute the time derivative of the coordinate fields $\phi^{\hspace{0.5pt}i}(x,t)$, yielding
\begin{equation}\label{Eq:dtPhi}
\p_t\phi^{\hspace{0.5pt}i} = G^{\hspace{0.5pt}ij}\,\pi_j - B^{\hspace{0.5pt}i}_{\;\,j}\,\p_x\phi^{\,j} = G^{\hspace{0.5pt}ij}\,\bigl( \pi_j - B_{jk}\,\p_x\phi^{\,k}\bigr)\,,
\end{equation}
where we have used equation \eqref{Eq:GBH} in the last equality. This relation is easily inverted to express the momenta as
\begin{equation}\label{Eq:Pi}
\pi_i = G_{ij}\,\p_t \phi^{\,j} + B_{ij} \, \p_x \phi^{\,j}\,.
\end{equation}

\paragraph{Lagrangian formulation and $\bm\s$-model action.} So far we have built an integrable field theory in the Hamiltonian formulation, expressed in terms of the canonical fields $\bigl( \phi^{\hspace{0.5pt}i}(x,t), \pi_i(x,t) \bigr)$. It is a standard result that such a theory also possesses a Lagrangian formulation. The first step in deriving this formulation is to eliminate the momenta $\pi_i(x,t)$ in terms of the coordinate fields $\phi^{\hspace{0.5pt}i}(x,t)$ and their derivatives: we did this in equation \eqref{Eq:Pi} above. The second one is to determine the \textit{action functional}, whose density is given by the inverse Legendre transform of the Hamiltonian, \textit{i.e.}\vspace{-3pt}
\begin{equation}\label{Eq:Action}
S\bigl[ \Phi \bigr] = \int_{\R} \left( \int_0^{2\pi} \pi_i\,\p_t \phi^{\hspace{0.5pt}i}\,\dd x - \Hc_{\TM} \right) \dd t\,.\vspace{-3pt}
\end{equation}
In this equation, all the momenta $\pi_i$ should be replaced by their Lagrangian expression \eqref{Eq:Pi}: we then obtain a functional depending only on the coordinate fields $\phi^{\hspace{0.5pt}i}(x,t)$ and their derivatives $\p_x\phi^{\hspace{0.5pt}i}(x,t)$ and $\p_t\phi^{\hspace{0.5pt}i}(x,t)$. In the present case, the explicit computation of this functional will be done in the exercise 14 below. In the end, using light-cone derivatives $\p_\pm = \frac{1}{2}(\p_t \pm \p_x)$, we find\vspace{-3pt}
\begin{equation}\label{Eq:Sigma}
S\bigl[ \Phi \bigr] = \iint_{\Sigma}\bigl( G_{ij}(\Phi) + B_{ij}(\Phi) \bigr)\,\p_-\phi^{\,i}\,\p_+\phi^{\,j}\, \dd x^+\,\dd x^- \, ,
\end{equation}
in terms of the two tensors $G_{ij}$ and $B_{ij}$ introduced above. This is the action of a \textit{non-linear $\s$-model}, with target space $\Mc$, metric $G_{ij}$ and B-field $B_{ij}$. Thus, relativistic AGMs with complete Takiff realisations of the form \eqref{Eq:RealTakiff} automatically give rise to integrable $\s$-models. We can then construct many such theories by varying the AGM we start with and the choice of the realisation (see Subsection \ref{SubSec:PCM} for a simple example and Subsection \ref{SubSec:Pano} for a brief overview of the panorama of models obtained this way). This is the main message of this section. Let us stress that the question of determining which choice of metric $G_{ij}$ and $B$-field $B_{ij}$ makes the $\s$-model \eqref{Eq:Sigma} integrable or not is in general a quite complicated one, which often requires some guesswork. In contrast, the AGM formalism provides a systematic approach to this question, by producing $\s$-models which are by construction automatically integrable.\vspace{10pt}

\begin{tcolorbox} \textit{\underline{Exercise 14:} The $\s$-model formulation.} \raisebox{0.5pt}{\Large$\;\;\star\star$} \vspace{5pt}\\
Consider the Hamiltonian $\Hc_{\TM}$, expressed as in \eqref{Eq:HReal} in terms of the canonical fields $\bigl( \phi^{\hspace{0.5pt}i}, \pi_i \bigr)$.\vspace{4pt}\\
1. Compute the time evolution of $\phi^{\hspace{0.5pt}i}$ with respect to the dynamics $\p_t=\lbrace \Hc_{\TM},\cdot\rbrace$ and check the equations \eqref{Eq:dtPhi} and \eqref{Eq:Pi}.\vspace{4pt}\\
2. Perform the inverse Legendre transform of the model and compute the action functional \eqref{Eq:Action}. Check that it takes the $\s$-model form \eqref{Eq:Sigma}.\vspace{4pt}\\
3. Determine the equations of motion arising from the action \eqref{Eq:Sigma} by considering an infinitesimal variation $\delta\phi^{\hspace{0.5pt}i}$ of the fields. Compare with the Hamiltonian equations of motion, by computing $\p_t\pi_i = \lbrace \Hc_{\TM}, \pi_i \rbrace$ and combining it with the result of question 1 for $\p_t\phi^{\hspace{0.5pt}i}$, in order to obtain a second order partial differential equation on $\phi^{\hspace{0.5pt}i}$.\vspace{3pt}\\
\textit{Hint:} You should find\vspace{3pt}
\begin{equation}\label{Eq:EoMSigma}
\p_+\p_-\phi^{\hspace{0.5pt}i} + \bigl( \Gamma^{\hspace{0.5pt}i}_{\;jk} + H^{\hspace{0.5pt}i}_{\;\,jk} \bigr) \p_+ \phi^{\,j} \p_- \phi^k = 0\,,
\end{equation}
where
\begin{equation*}
\Gamma^{\hspace{0.5pt}i}_{\;jk} = \frac{1}{2} G^{il} \bigl( \p_j G_{lk} + \p_k G_{jl} - \p_l G_{jk} \bigr) \qquad \text{ and } \qquad H^{\hspace{0.5pt}i}_{\;\,jk} = \frac{1}{2} G^{il} \bigl( \p_j B_{kl} + \p_k B_{lj} + \p_l B_{jk} \bigr)
\end{equation*}
are respectively the Christoffel symbols of the metric and the torsion tensor of the B-field.
\end{tcolorbox}~\vspace{-12pt}

\paragraph{Lax connection.} As mentioned in the previous paragraph, the $\s$-model \eqref{Eq:Sigma} obtained through the AGM realisation is automatically integrable. In particular, this integrability is reflected in the existence of a Lax connection, whose flatness is equivalent to the equations of motion \eqref{Eq:EoMSigma}. We end this subsection by a brief description of the form of this connection.

In the initial AGM we started with, it is given by equation \eqref{Eq:Lpm} and is built as a well-chosen linear combination of the Takiff currents $\Jc_r^{[p]}$, with coefficients depending rationally on the spectral parameter $\lambda$. By a slight abuse of notation, we will still denote the image of this connection in the realisation as $\Lc_\mu(\lambda)$. It is obtained by replacing the abstract Takiff currents $\Jc_r^{[p]}$ by their realisations $J_r^{[p]}$, defined as in equation \eqref{Eq:RealTakiff}. Using also the Lagrangian expression \eqref{Eq:Pi} of $\pi_i$, it is clear that the Lax connection is linear in the space-time derivatives $\p_t\phi^{\hspace{0.5pt}i}$ and $\p_x\phi^{\hspace{0.5pt}i}$ of the coordinate fields. In fact, the relativistic invariance of the model forces the light-cone component $\Lc_\pm(\lambda)$ to depend only on the combination $\p_\pm\phi^{\hspace{0.5pt}i}$ of these derivatives. In the end, we thus get a Lax connection of the form
\begin{equation}
\Lc_\pm(\lambda) = f_{\pm,i}\bigl(\lambda,\Phi\bigr) \,\p_\pm \phi^{\hspace{0.5pt}i}\,,
\end{equation}
where a sum over $i\in\lbrace 1,\dots,d\rbrace$ is implied. The coefficients $f_{\pm,i}$  depend non-linearly on the coordinate fields $\phi^{\,j}(x)$ (but not their derivatives) and are rational functions of the spectral parameter $\lambda$, with poles at the points $\bm{\ze}_\pm = \bigl( \ze_k)_{k\in\Ic_\pm}$. This is the typical form expected for the Lax connection of an integrable $\s$-model. Moreover, since the realisation is complete, one checks that the zero curvature equation for $\Lc_\pm(\lambda)$ capture all of the equations of motion \eqref{Eq:EoMSigma} of the $\s$-model, as typically required for an integrable field theory. Finally, we note that the spatial component $\Lc_+(\lambda)-\Lc_-(\lambda)$ of the Lax connection automatically satisfies a Maillet bracket, by construction.

\subsection{Example: the Principal Chiral Model}
\label{SubSec:PCM}

We now illustrate the general philosophy of this section by discussing a simple example of integrable $\s$-model obtained as a realisation of AGM, the so-called Principal Chiral Model (PCM).

\paragraph{The underlying AGM.} We start by specifying the AGM we will consider. It is characterised by its twist function $\vp(\lambda)$, which we take to be
\begin{equation}\label{Eq:TwistPCM}
\vp(\lambda) = \hay \,\frac{1-\lambda^2}{\lambda^2}\,,
\end{equation}
where $\hay$ is a positive parameter. In the complex plane, $\vp(\lambda)$ has a double pole at $0$: we are thus considering an AGM with a single site of multiplicity $2$, so that $N=1$ in the above notation. For simplicity, we will then omit the index $r\in\lbrace 1,\dots,N\rbrace$ labelling the sites. Performing the partial fraction decomposition of $\vp(\lambda)$, we read off the corresponding levels
\begin{equation}
\ell^{[0]} = 0\,, \qquad \ell^{[1]}=\hay\,, \qquad \ell_\infty = \hay\,.
\end{equation}
In terms of the Takiff currents $\Jc^{[0]}$ and $\Jc^{[1]}$ attached to the site, the Gaudin Lax matrix reads
\begin{equation}\label{Eq:GaudinPCM}
\Gamma(\lambda) = \frac{\Jc^{[0]}}{\lambda} + \frac{\Jc^{[1]}}{\lambda^2}\,.
\end{equation}

The zeroes of the twist function are $\bm\ze=(+1,-1)$. Recall that these zeroes are associated with the quadratic charges \eqref{Eq:Qi}. Here, we will denote them as $\Q_+$ and $\Q_-$. A direct computation shows that
\begin{equation}
\Q_\pm = \pm \int_0^{2\pi} \frac{\psb{\Jc^{[0]}\pm\Jc^{[1]}}{\Jc^{[0]}\pm\Jc^{[1]}}}{4\hay}\,\dd x\,.
\end{equation}
Following Section \ref{Sec:SpaceTime}, the momentum of the theory is given by
\begin{equation}\label{Eq:MomG}
\Pc = \Q_+ + \Q_- = \frac{1}{\hay}\int_0^{2\pi} \psb{\Jc^{[0]}}{\Jc^{[1]}}\,\dd x\,.
\end{equation}
Moreover, we define the Hamiltonian as a linear combination of $\Q_\pm$ with coefficients equal to $+1$ or $-1$ (to ensure relativistic invariance): this leaves four possible combinations. Two of them give $\pm\Pc$ and thus lead to quite degenerate models, where the time variable $t$ coincides with the spatial one $x$ or its opposite. Among the two other possibilities, which are opposite one of another, we choose the one which is positive when the underlying Lie algbera $\g$ is compact, namely
\begin{equation}\label{Eq:HamG}
\Hc = \Q_+ - \Q_- = \frac{1}{2\hay} \int_0^{2\pi} \Bigl( \psb{\Jc^{[0]}}{\Jc^{[0]}} + \psb{\Jc^{[1]}}{\Jc^{[1]}} \Bigr)\,\dd x\,.
\end{equation}

\paragraph{The Takiff realisation.} The next step to obtain an integrable $\s$-model from this AGM is to specify a realisation of the Takiff currents $\Jc^{[0]}$ and $\Jc^{[1]}$, with levels $\ell^{[0]} = 0$ and $\ell^{[1]}=\hay$. In fact, we have already encountered such a realisation in the case where $\g=\sl(2,\R)$, given by equation \eqref{Eq:RealSl2}. This was a realisation in terms of canonical fields in $T^\ast SL(2,\R)$: it can in fact be generalised for any real simple Lie algebra $\g$, working with canonical fields in $T^\ast G$, where $G$ is a Lie group with $\text{Lie}(G)=\g$.

To understand how this general realisation is built, let us analyse closer the expression \eqref{Eq:RealSl2} of the $\sl(2,\R)$ example, in particular the current $J^{[1]}(x)=J^{[1]}_a(x)\,\td^a$. We note that it contains spatial derivatives $\p_x\phi^{\hspace{0.5pt}i}(x)$ of the coordinate fields but not the momenta $\pi_i(x)$ (which are encoded in the other current $J^{[0]}(x)$). Using the expression \eqref{Eq:gSL2} of the $SL(2,\R)$-valued field $g(x)$, one in fact checks that $J^{[1]}(x)$ can be simply rewritten as $\ell^{[1]}\,g(x)^{-1}\p_x g(x)$. The main interest of this formulation is that it is now independent of the choice of coordinates $\phi^{\hspace{0.5pt}i}$ used to parametrise the group-valued field $g(x)$ and is easily generalisable to an arbitrary group $G$. The other current $J^{[0]}(x)$ can in fact also be rewritten in a coordinate-independent way, but this will take a bit more work.\\

Let us now consider the case of an arbitrary simple real Lie group $G$ of dimension $d$. For simplicity, we will suppose that $G$ can be represented as a matrix group, as was the case for $SL(2,\R)$. We let $g(x)$ be a $G$-valued field, which we then see as a $x$-dependent matrix. Following our observation above, let us consider the current
\begin{equation}
j(x) = g(x)^{-1}\p_x g(x)\,,
\end{equation}
which is valued in the Lie algebra $\g$. Choosing a coordinate chart on the group manifold $G$, we can parametrise the matrix $g(x)$ in terms of $d$ coordinate fields $\bigl( \phi^{\hspace{0.5pt}i}(x) \bigr)_{i=1,\dots,d}$, similarly to what we did in equation \eqref{Eq:gSL2} for $G=SL(2,\R)$. The above current can then be rewritten as
\begin{equation}\label{Eq:j}
j(x) = L^a_{\;\,i}(x)\, \p_x\phi^{\hspace{0.5pt}i}(x)\,\td_a\,, \qquad \text{ with } \qquad L^a_{\;\,i} = \psb{\td^a}{g^{-1}\p_i g}
\end{equation}
and where $\p_i g$ denotes the partial derivative of $g$ with respect to the coordinate $\phi^{\hspace{0.5pt}i}$. By construction, $j$ is independent of the choice of coordinates on $G$ and of basis of $\g$.

One shows that the matrix $\bigl( L^a_{\;\,i} \bigr)^{a=1,\dots,d}_{i=1,\dots,d}$ is invertible: we let $\bigl( \Lt^{\hspace{1pt}i}_{\;\;a} \bigr)^{i=1,\dots,d}_{a=1,\dots,d}$ be its inverse, which then satisfies $L^a_{\;\,i}\,\Lt^{\hspace{1pt}i}_{\;\;b}=\delta^a_{\;\,b}$ and $\Lt^{\hspace{1pt}i}_{\;\;a}\,L^a_{\;\,j}=\delta^{\hspace{0.5pt}i}_{\;\,j}$. We now introduce the momenta $\pi_i(x)$ conjugate to the coordinates $\phi^{\hspace{0.5pt}i}(x)$ and gather them into a $\g$-valued quantity
\begin{equation}\label{Eq:X}
X(x) = \Lt^{\hspace{1pt}i}_{\;\;a}(x)\,\pi_i(x)\,\td^a\,.
\end{equation}
One actually checks that $X$ is independent of the choice of coordinates on $G$ and of basis of $\g$. For instance, performing a change of the coordinates $\bigl(\phi^{\hspace{0.5pt}i}\bigr)_{i=1,\dots,d}$ on $G$ modifies the matrix $L^a_{\;\,i}$ defined in equation \eqref{Eq:j} as well as the momenta $\bigl(\pi_i\bigr)_{i=1,\dots,d}$ conjugated to these coordinates, in such a way that the combination \eqref{Eq:X} stays the same. In the $SL(2,\R)$ case, with $g(x)$ given by \eqref{Eq:gSL2}, one in fact finds that $X$ coincides with the current $J^{[0]}$ considered in the realisation \eqref{Eq:RealSl2}.\\

We now have a $G$-valued field $g(x)$ and two $\g$-valued fields $j(x)$ and $X(x)$, all expressed in a coordinate-independent way in terms of the canonical fields $\bigl( \phi^{\hspace{0.5pt}i}(x), \pi_i(x) \bigr)$ of $\Acg$. We now want to derive their Poisson bracket with respect to the canonical Poisson structure \eqref{Eq:PbCan2}. To lighten the discussion, we discuss this computation in the Appendix \ref{App:PbG} and the exercise 16 therein and only summarise the results here. Using tensorial notations, we find
\begin{equation*}
\bigl\lbrace X(x)\ti{1}, g(y)\ti{2} \bigr\rbrace = g(y)\ti{2}\,C\ti{12}\,\delta(x-y)\,, \qquad \bigl\lbrace X(x)\ti{1}, X(y)\ti{2} \bigr\rbrace = \bigl[ C\ti{12}, X(y)\ti{1} \bigr]\,\delta(x-y)\,,
\end{equation*}
\begin{equation}\label{Eq:PbXg}
\bigl\lbrace X(x)\ti{1}, j(y)\ti{2} \bigr\rbrace = \bigl[ C\ti{12}, j(y)\ti{1} \bigr]\,\delta(x-y) - C\ti{12}\,\p_x\delta(x-y)\,,
\end{equation}
while $g$ and $j$ have vanishing Poisson brackets with themselves and with each other (since they depend only on the coordinate fields and their derivatives, not on the momenta). From there, it is clear that
\begin{equation}\label{Eq:RealG}
J^{[0]} = X \qquad \text{ and } \qquad J^{[1]} = \hay\,j
\end{equation}
are $\g$-Takiff currents of multiplicity 2 with levels $\ell^{[0]}=0$ and $\ell^{[1]}=\hay$, as wanted. Note that this realisation relates $2\dim\g$ current components with $2\dim\g$ canonical fields and thus is complete. \vspace{10pt}

\begin{tcolorbox} \textit{\underline{Exercise 15:} From $SL(2,\R)$ to $G$.} \raisebox{0.5pt}{\Large$\;\;\star\star$} \vspace{5pt}\\
Check that the Takiff realisation \eqref{Eq:RealG} becomes the one \eqref{Eq:RealSl2} when $G=SL(2,\R)$ and $\ell^{[1]}=\hay$.\\[4pt]
\textit{Hint:} Start with the parametrisation of $g(x) \in SL(2,\R)$ given by equation \eqref{Eq:gSL2}, which defines our choice of coordinates $\bigl(\phi^{\hspace{0.5pt}i}(x) \bigr)_{i=1,\dots,3}$, as well as our choice of basis $\lbrace \td_a \rbrace_{a=1,\dots,3}$ of $\sl(2,\R)$ made in equation \eqref{Eq:sl2}. Compute the corresponding matrices $L^a_{\;\,i}$ and $\Lt^{\hspace{1pt}i}_{\;\;a}$ and use this result to derive the components $J^{[0]}_a$ and $J^{[1]}_a$, in order to compare with equation \eqref{Eq:RealSl2}. Note that the basis $\lbrace \td_a \rbrace$ is not orthonormal and one should thus be careful with the position of the Lie algebra indices. The use of a symbolic computations software is advised.
\end{tcolorbox}~\vspace{-12pt}

\paragraph{Momentum and Hamiltonian.} Let us consider the image $\Pc_{T^\ast G}$ of the momentum \eqref{Eq:MomG} under the realisation \eqref{Eq:RealG}. It reads
\begin{equation}\label{Eq:PG}
\Pc_{T^\ast G} = \int_0^{2\pi} \psb{X(x)}{j(x)}\,\dd x = \int_0^{2\pi} \pi_i(x)\,\p_x\phi^{\hspace{0.5pt}i}(x)\,\dd x \,,
\end{equation}
where the last equality is found by reinserting the expressions \eqref{Eq:j} and \eqref{Eq:X} of $j$ and $X$ and using the fact that $\Lt^{\hspace{1pt}i}_{\;\;a}$ is the inverse of $L^a_{\;\,i}$. As expected, we recognise in this last expression the momentum of the canonical Poisson algebra $\Acg$.\vspace{10pt}

\begin{tcolorbox} \textit{\underline{Exercise 17:} Momentum of $\Acg$.} \raisebox{0.5pt}{\Large$\;\;\star$} \vspace{5pt}\\
Check that $\Pc_{T^\ast G}$ is the momentum of $\Acg$ from its coordinate-independent expression in terms of $X$ and $j$ (first equality in \eqref{Eq:PG}).\\[4pt]
\textit{Hint:} Show that it generates the spatial derivatives on $g$ and $X$ using the brackets \eqref{Eq:PbXg}.
\end{tcolorbox}~\vspace{-5pt}

Similarly, we can compute the image $\Hc_{T^\ast G}$ of the Hamiltonian \eqref{Eq:HamG} in the realisation \eqref{Eq:RealG}. This gives
\begin{equation}
\Hc_{T^\ast G} = \int_0^{2\pi} \Bigl( \frac{1}{2\hay}\,\psb{X(x)}{X(x)} + \frac{\hay}{2}\,\psb{j(x)}{j(x)} \Bigr)\,\dd x\,.
\end{equation}
We will not need its expression in terms of the canonical fields $\bigl(\phi^{\hspace{0.5pt}i},\pi_i\bigr)$ and will work with the coordinate-independent fields $g$, $X$ and $j$. This defines the dynamics $\p_t = \lbrace \Hc_{T^\ast G},\cdot\rbrace$ of observables in $\Acg$. For instance, using the brackets \eqref{Eq:PbXg}, we get
\begin{equation*}
\hay\,\p_t g(y) = \int_0^{2\pi} \bigl\langle \bigl\lbrace X(x)\ti{1}, g(y)\ti{2} \bigr\rbrace, X(x)\ti{1} \bigr\rangle\ti{1} \,\dd x =   \int_0^{2\pi} \bigl\langle g(y)\ti{2}\, C\ti{12}, X(x)\ti{1} \bigr\rangle\ti{1}\, \delta(x-y)\, \dd x = g(y)\,X(y)\,,
\end{equation*} 
where we used the completeness relation \eqref{Eq:CasComp}. Thus, we have
\begin{equation}\label{Eq:XLag}
X = \hay\,g^{-1}\p_t g\,.
\end{equation}
This encodes in a coordinate-independent way the relation between the momenta and the time derivatives of the coordinate fields.

\paragraph{Action.} Let us now pass to the Lagrangian formulation. The action functional is obtained by performing the inverse Legendre transform \eqref{Eq:Action}. In the present case, it will be useful to rewrite the first term in this equation in the following coordinate-independent way:
\begin{equation}
\pi_i\,\p_t\phi^{\hspace{0.5pt}i} = \psb{X}{g^{-1}\p_t g}\,.
\end{equation}
It is then straightforward to compute the action, replacing $X$ by its Lagrangian expression \eqref{Eq:XLag}:
\begin{equation}
S[g] = \frac{\hay}{2}\,\iint_\Sigma\Bigl( \psb{g^{-1}\p_t g}{g^{-1}\p_t g} - \psb{g^{-1}\p_x g}{g^{-1}\p_x g} \Bigr)\, \dd t\,\dd x\,.
\end{equation}
Using light-cone derivatives, we finally get
\begin{equation}\label{Eq:PCM}
S[g] = \hay\,\iint_\Sigma\psb{g^{-1}\p_+ g}{g^{-1}\p_- g}\, \dd x^+\,\dd x^-\,.
\end{equation}
We recognise in this expression the action of the PCM, which is a well-known integrable $\s$-model with target space the Lie group $G$.

\paragraph{Lax connection and charges.} Since we obtained the PCM as a realisation of an AGM, we also automatically get a Lax connection $\Lc_\pm(\lambda)$ encoding its dynamics. In terms of the field $g$, it reads
\begin{equation}\label{Eq:LaxPCM}
\Lc_\pm(\lambda) = \frac{g^{-1}\p_\pm g}{1 \mp \lambda}\,.
\end{equation}
This coincides with the Lax connection originally proposed for the PCM in~\cite{Zakharov:1973pp}. The spatial component of this connection satisfies a Maillet bracket with twist function \eqref{Eq:TwistPCM}, which coincides with the one initially found in~\cite{Maillet:1985ec}. One can extract an infinite family of Poisson-commuting non-local charges from the trace of the monodromy of this component. \vspace{10pt}

\begin{tcolorbox} \textit{\underline{Exercise 18:} Lax connection of the PCM.} \raisebox{0.5pt}{\Large$\;\;\star$} \vspace{5pt}\\
1. Check that the Lax connection obtained from the AGM construction gives \eqref{Eq:LaxPCM} in the present case.\\[4pt]
2. Derive the equations of motion of the PCM \eqref{Eq:PCM}. Show that it can be recast as the zero curvature equation of the Lax connection \eqref{Eq:LaxPCM}, using also the Maurer-Cartan identity
\begin{equation}
\p_+ \bigl(g^{-1}\p_- g \bigr) - \p_+ \bigl(g^{-1}\p_- g \bigr) + \bigl[ g^{-1}\p_+ g, g^{-1}\p_- g \bigr] = 0\,,
\end{equation}
which holds off-shell.
\end{tcolorbox}~\vspace{-5pt}

In addition to the non-local charges extracted from the Lax connection, the model also admits two infinite families of higher degree local charges $\Q_{\pm,T^\ast G}^{(p)}$, associated with the zeroes $\pm 1$ of the twist function (and starting with the light-cone Hamiltonians at degree 2). For the PCM, these charges were initially found in the work~\cite{Evans:1999mj} (which was the main inspiration behind the general construction of local charges in AGMs~\cite{Lacroix:2017isl} reviewed in Subsection \ref{SubSec:Local}).

\paragraph{Adding a Wess-Zumino term.} The Takiff realisation \eqref{Eq:RealG} used to obtain the PCM is the generalisation for any group $G$ of the realisation \eqref{Eq:RealSl2}, which corresponds to $G=SL(2,\R)$. Recall from Exercise 12 that the latter possesses a deformation \eqref{Eq:RealSl2k}, which has for effect to turn on the level $\ell^{[0]}$. Such a deformation in fact exists for any group $G$: in terms of the resulting integrable $\s$-model, it amounts to adding the so-called Wess-Zumino term to the PCM. Let us briefly discuss this.

From the point of view of the underlying AGM, let us now consider the twist function
\begin{equation}\label{Eq:TwistPCM+WZ}
\vp(\lambda) = \hay\,\frac{1-\lambda^2}{(\lambda+\nu)^2} = \frac{\hay\,(1-\nu^2)}{(\lambda+\nu)^2} + \frac{2\nu\hay}{\lambda+\nu} - \hay \,,
\end{equation}
where $\nu\in\R$ is a real parameter. When $\nu=0$, we recover the twist function \eqref{Eq:TwistPCM} of the PCM alone. It is clear from the above partial fraction decomposition that this AGM has a site of multiplicity $2$ at $-\nu$ and levels $\ell^{[0]}=2\nu\hay$, $\ell^{[1]}=\hay\,(1-\nu^2)$ and $\ell_\infty=\hay$: in particular, we see that the main consequence of deforming the twist function by the parameter $\nu$ is to turn on the level $\ell^{[0]}$. We discuss this deformed AGM, the corresponding Takiff realisation and the resulting $\s$-model in Appendix \ref{App:WZ}: here, we just report the final result. In particular, the action and Lax connection of the model are
\begin{equation}\label{Eq:PCM+WZ}
S[g] = \hay\left( \iint_\Sigma\psb{g^{-1}\p_+ g}{g^{-1}\p_- g}\, \dd x^+\,\dd x^-\, + \nu\,\W g \right)\,, \qquad\quad \Lc_\pm(\lambda) = \frac{1 \pm \nu}{1 \mp \lambda}g^{-1}\p_\pm g\,,
\end{equation}
where $\W g$ is the Wess-Zumino term of $g$, which is defined in Appendix \ref{App:WZ}, equation \eqref{Eq:WZ2}. We note that this model is conformal (at the quantum level) when $\nu=\pm 1$.

\subsection[Panorama of integrable \texorpdfstring{$\s$}{sigma}-models obtained from AGMs: an overview]{Panorama of integrable \texorpdfstring{$\bm\s$}{sigma}-models obtained from AGMs: an overview}
\label{SubSec:Pano}

In the previous subsection, we have shown that the PCM (potentially with Wess-Zumino term) can be seen as a realisation of a well-chosen AGM. As explained earlier, the integrable structure of the PCM is then recovered from this construction since integrability is a built-in property of AGM realisations. We end this section with an overview of the panorama of other integrable $\s$-models which can be obtained in this way.

\paragraph{Duals and deformations of the PCM.} As a realisation of AGM, the PCM is characterised by the twist function \eqref{Eq:TwistPCM} and a pair of Takiff currents \eqref{Eq:RealG}, realised in terms of canonical fields in $T^\ast G$, where $G$ is the Lie group that serves as the target space of the PCM. There exists another realisation~\cite{Delduc:2019bcl} of the same current algebra, now in terms of canonical fields in $T^\ast\g$, where $\g=\text{Lie}(G)$. Considering the image of the AGM in this realisation, we obtain another integrable $\s$-model, whose target space is the Lie algebra $\g$. This is the so-called non-abelian T-dual of the PCM~\cite{Fridling:1983ha,Fradkin:1984ai}. The notion of non-abelian T-duality was initially proposed in the context of string theory and gives rise to some intricate relations between seemingly different $\s$-models. At the classical level, this relation takes the form of a non-local canonical (\textit{i.e.} Poisson-preserving) equivalence between the two models. In terms of AGMs, this is reflected in the fact that the PCM and its non-abelian T-dual are obtained as two different realisations of the same underlying AGM: they thus share the same formulation once written in terms of Takiff currents.

The twist function \eqref{Eq:TwistPCM} of the PCM possesses a double pole at infinity and a double pole in the complex plane.
One can consider a continuous deformation of the underlying AGM by splitting this finite double pole into a pair of simple poles with opposite residues.\footnote{A complete treatment of this deformation requires a careful discussion of reality conditions, which for conciseness we will not cover here.} This changes the Poisson algebra of the model, which is now described by a pair of Kac-Moody currents with opposite levels (rather than a pair of Takiff currents). However, one can still find a realisation~\cite{Delduc:2013fga} of this current algebra in terms of canonical fields in $T^\ast G$, which is a deformation of the Takiff realisation \eqref{Eq:RealG} giving rise to the PCM. From this setup, we obtain an integrable deformation of the PCM on $G$, which coincides with the so-called Yang-Baxter model or $\eta$-model, initially introduced by Klim\v{c}\'{i}k in~\cite{Klimcik:2002zj,Klimcik:2008eq}. There exists another realisation of the same current algebra, which gives rise to another type of integrable $\s$-model, introduced by Sfetsos in~\cite{Sfetsos:2013wia} and called the $\lambda$-model. The latter is not a deformation of the PCM but rather of its non-abelian T-dual. The $\eta$-model and the $\lambda$-model are said to be Poisson-Lie T-dual~\cite{Klimcik:1995ux,Klimcik:1995dy} one from another: this is a worldsheet duality generalising the notion of non-abelian T-duality mentioned above, which also takes the form of a non-local canonical equivalence~\cite{Sfetsos:1997pi} between classical $\s$-models. In the present language, this is again reflected in the fact that these two theories arise as different realisations of the same underlying AGM.

The $\eta$-model admits a further integrable deformation, which corresponds to turning on the Wess-Zumino term~\cite{Delduc:2014uaa,Hoare:2020mpv}. From the point of view of AGMs, this is obtained by relaxing the assumption made above that the two Kac-Moody currents have opposite levels. Finding an appropriate realisation~\cite{Delduc:2014uaa,Hoare:2020mpv} of this current algebra, one gets the $\eta$-model with Wess-Zumino term.

The PCM is characterised by its $G\times G$ isometries, acting on the field $g$ by left and right multiplications $g \mapsto h_L \,g\,h_R$ (it is clear that this transformation is a symmetry of the action \eqref{Eq:PCM} for constant $h_{L,R}$ in $G$). The Yang-Baxter deformation has for effect to break the left isometry and more precisely to deform it into a Poisson-Lie symmetry. It is possible to further deform the right isometry, obtaining a two-parameter integrable deformation called the bi-Yang-Baxter model, without~\cite{Klimcik:2014bta} or with~\cite{Delduc:2017fib} Wess-Zumino term. This model is obtained from AGMs by splitting the double pole at infinity in the twist function of the Yang-Baxter model into a pair of simple poles.\footnote{Technically, this is not yet established in the literature for the bi-Yang-Baxter model with Wess-Zumino term: however, it is strongly expected. Let us also note that, as mentioned in Subsection \ref{SubSec:Gen}, the fact that there is no double pole at infinity in $\vp(\lambda)\,\dd \lambda$ for these theories means that we are dealing with AGMs possessing a gauge symmetry.} In the case without Wess-Zumino term, the same AGM also gives rise to two other realisations, the so-called generalised $\lambda$-model~\cite{Sfetsos:2015nya} and the bi-$\lambda$-model~\cite{Sfetsos:2017sep}, related by various Poisson-Lie T-dualities.

For a pedagological introduction on integrable deformations of the PCM, their symmetries and their dualities, we refer the reader to the lecture notes~\cite{Hoare:2021dix}, which discuss in more details most of the concepts and models discussed above.\footnote{Another example of integrable deformations that can be recovered from AGMs and that we shall not develop further here are the so-called homogeneous Yang-Baxter deformations~\cite{Kawaguchi:2014qwa}.}

\paragraph{Coupling.} Beside deforming, another way to construct different integrable $\s$-models form AGMs is to consider twist functions with more involved analytical structure. For instance, instead of a twist function with one double pole in the complex plane, which corresponds to the PCM with Wess-Zumino term, one can consider a function with an arbitrary number $N$ of such poles. The corresponding AGM is thus described in terms of $N$ pairs of Takiff currents. Realising each of these pairs in independent copies of $\Acg$, we obtain an integrable $\s$-model on the product space $G^N$, which takes the form of $N$ coupled PCMs with Wess-Zumino terms, interacting in a non-trivial but very fine-tuned way, which ensures the integrability of the model. This coupled integrable theory was built in~\cite{Delduc:2019bcl} and constitutes the first example of an integrable $\s$-model which was not known before its construction from AGMs.

This theory can be further deformed while preserving integrability, by splitting the double poles into pairs of simple poles and attaching to these the realisations leading to $\eta$- or $\lambda$-deformations. This gives rise to a rich panorama of integrable coupled $\s$-models, possessing various properties of Poisson-Lie symmetries and T-dualities (for a non-exhaustive list of references, see for instance~\cite{Georgiou:2018gpe,Georgiou:2019plp,Bassi:2019aaf}).

\paragraph{Dihedral AGMs and coset $\bm\s$-models.} Another direction in the panorama of integrable $\s$-models is obtained by considering dihedral AGMs~\cite{Vicedo:2017cge}. These are generalisations of the ones considered in these lectures: as briefly sketched in Subsection \ref{SubSec:Gen}, they possess additional equivariance properties under a Lie-group automorphism $\Omega\in\text{Aut}(G^\C)$ of finite order $T$. In particular, these models possess an $H$-gauge symmetry, where $H$ is the group of real fixed points of $\Omega$. The simplest examples of integrable $\s$-models obtained as realisations of dihedral AGMs are the ones on so-called $\Z_T$-cosets $G/H$~\cite{Young:2005jv} (also named symmetric spaces for $T=2$). One can obtain $\eta$- and $\lambda$-deformations of these models by splitting double poles of the twist function into pairs of simple poles (see for instance the review~\cite{Hoare:2021dix} and references therein). As another direction, one can consider dihedral AGMs with more double poles and construct the coset version of the aforementioned coupled theories, resulting in integrable $\s$-models on $G^N/H_{\text{diag}}$~\cite{Arutyunov:2020sdo}.

\paragraph{Higher order poles and $\bm{\mathcal{E}}$-models.} In addition to coupling, gauging and deforming, the AGM formalism can also be used to construct more exotic examples of new integrable $\s$-models, for instance by considering higher order poles in the twist function (indeed, the cases discussed above are built from double and simple poles only). These models have been far less studied so far and should lead to a rich variety of new target spaces (see~\cite{Lacroix:2020flf,Klimcik:2021bqm,Liniado:2023uoo} for some first results and examples\footnote{Note that these references use the formalism of 4-dimensional Chern-Simons theory and/or $\mathcal{E}$-models rather than that of affine Gaudin models. As we will explain below, these are all intimately related.}).

The exploration of this general panorama can be aided by the deep relation between AGMs and a class of theories called $\mathcal{E}$-models, which were introduced by Klim\v{c}\'{i}k and \v{S}evera in~\cite{Klimcik:1995ux,Klimcik:1995dy,Klimcik:1996np}. These are $\s$-models defined in terms of current algebras and whose Lie-algebraic formulation makes their Poisson-Lie symmetries and dualities manifest. Such features are reminiscent of AGMs and their $\s$-models realisations, as described earlier. We note however that $\mathcal{E}$-models are not always integrable, although it has been shown that most integrable $\s$-models can be recast as such (see for instance~\cite{Severa:2017kcs,Lacroix:2020flf,Klimcik:2021bqm,Liniado:2023uoo} and references therein). In that context, AGMs can be seen as a very large class of $\mathcal{E}$-models which have the additional property of being integrable by construction. The rich existing literature on $\mathcal{E}$-models then offers promising perspectives for the study of the $\s$-model realisations of these AGMs and of their properties, including for instance their geometries, symmetries, dualities and renormalisation.

\paragraph{4-dimensional Chern-Simons theory.} We end this overview by mentioning the intimate relation between the formalism of AGMs and that of 4-dimensional Chern-Simons theory (with so-called disorder defects), which has been proposed in~\cite{Costello:2019tri} as another general framework to generate integrable $\s$-models. This is a gauge theory defined on a 4-dimensional manifold $\Sigma \times \CP$, where $\Sigma$ will eventually serve as the worldsheet of the generated integrable $\s$-model and $\CP$ will be identified with the Riemann sphere to which belongs its spectral parameter. As above, we will use $(x,t)$ to denote coordinates on the worldsheet $\Sigma$ and $\lambda$ to denote the spectral parameter, which defines a complex coordinate on $\CP$, with conjugate $\bar{\lambda}$. The field content of the theory is given by a gauge field $A_\mu(x,t,\lambda,\bar\lambda)$, which has components in the directions $\mu = x,t,\bar\lambda$ only and is valued in a complexified Lie algebra $\g^\C$. The action of the model then takes the form
\begin{equation}\label{Eq:4dCS}
S[A] = \iiiint_{\Sigma\times\CP} \vp(\lambda)\,\dd \lambda\wedge \text{CS}(A)\,,
\end{equation}
where $\text{CS}(A)$ is a 3-form built from the gauge field (called the Chern-Simons form and whose expression we will not need) and where $\vp(\lambda)\,\dd\lambda$ is any meromorphic 1-form on $\CP$ (which is part of the data defining the theory). The main idea proposed in~\cite{Costello:2019tri} is that the equations of motion of the theory can be naturally recast as the zero curvature equation of a connection $\bigl(\p_x + \Lc_x(\lambda,x,t),\p_t + \Lc_t(\lambda,x,t) \bigr)$ on $\Sigma$, extracted from the gauge field $A$ and depending meromorphically on $\lambda$. This is exactly the characteristic property of an integrable 2-dimensional field theory on $\Sigma$, with Lax connection $\Lc_\mu(\lambda)$ and spectral parameter $\lambda\in\CP$. The action of this 2-dimensional theory can be computed by performing explicitly the integration over $\CP$ in the above 4-dimensional functional. This way, one obtains the action  of a 2-dimensional $\s$-model, which by construction possesses a Lax connection and whose fields are related to the poles of the meromorphic function $\vp(\lambda)$. We refer to~\cite{Costello:2019tri} for details (for a review, see also the lectures~\cite{Lacroix:2021iit} and the references therein).

The remaining information needed to complete the proof of the integrability of the theory is to show that the Lax matrix $\Lc_x(\lambda,x,t)$ obeys a Maillet bracket. This was done in~\cite{Vicedo:2019dej}: moreover, is was shown in this reference that the $\Rc$-matrix appearing in this Maillet bracket is of the form \eqref{Eq:RTwist}, with the twist function given by the choice of 1-form $\vp(\lambda)\,\dd\lambda$ appearing in the action \eqref{Eq:4dCS} (as the notation suggested). Based on this result, one concludes that the integrable $\s$-model obtained by the 4-dimensional Chern-Simons theory can equivalently be seen as a realisation of the AGM with twist function $\vp(\lambda)$.\footnote{The choice of realisation of the current algebra underlying this AGM is related to a choice of boundary conditions in the 4-dimensional Chern-Simons theory.} This shows the deep relation between these two general approaches (see also~\cite{Levin:2022dnq} for a more geometric statement, in the language of affine Higgs bundles~\cite{Levin:2001nm}).

\section{Conclusion and perspectives}
\label{Sec:Conclusion}

\subsection{Summary and classical perspectives}

In these lectures, we reviewed the construction of classical affine Gaudin models, their integrable structures and their $\s$-model realisations. This approach has been used in the literature to explore the panorama of classical integrable $\s$-models in a systematic way, by reinterpreting known examples in this unifying framework but also by using it to build new ones (see Subsection \ref{SubSec:Pano} for an overview). Moreover, it offers a general Lie-algebraic formulation to study some of the classical properties of these models, as for instance their conserved charges, their symmetries and their dualities.

There are various natural perspectives concerning these subjects, in particular aiming towards the classification of integrable $\s$-models, or at least of large families thereof. A natural question is for instance to find all possible realisations of Takiff currents, which would then allow to construct all integrable $\s$-models arising from AGMs. Another direction to explore is to study generalisations of AGMs which describe even wider classes of integrable field theories, including for example models possessing a spectral parameter in a higher genus Riemann surface, which have been so far less explored in the context of $\s$-models (see however~\cite{Cherednik:1981df,Levin:2001nm,Costello:2019tri,Derryberry:2021rne,Bykov:2021dbk,Lacroix:2023qlz} for first results).

\subsection{Towards quantisation: the Renormalisation Group flow}

Another interesting perspective of the formalism of AGMs is its potential applications to the quantisation of integrable $\s$-models. In this paragraph, we develop a first aspect in this direction, which is the description of the Renormalisation Group (RG) flow of these theories. It is a long-standing and well-checked conjecture~\cite{Fateev:1992tk} that classically integrable $\s$-models are renormalisable at 1-loop, \textit{i.e.} that all counter-terms arising from UV divergences in 1-loop computations can be reabsorbed into a dependence of the parameters $\lbrace \gamma_i \rbrace$ of the classical model with respect to the renormalisation energy scale $\mu$. In that case, the form of this dependence is controlled by the RG-flow equations $\mu\frac{\dd\;}{\dd\mu}\gamma_i = \beta_i(\gamma)$, where the $\beta_i(\gamma)$'s are functions of the parameters called the $\beta$-functions. The existence of this RG-flow is a manifestation of the breaking of the scale-invariance of these theories at the quantum level.

For integrable $\s$-models arising as realisations of AGMs, there have been various progresses in recent years towards the proof of this conjecture and the determination of the corresponding $\beta$-functions. Such a model depends on various parameters $\lbrace \gamma_i \rbrace$, which can be of two different types. The first are the parameters defining the underlying AGM (for instance the levels $\ell_r^{[p]}$ and the positions $\s_r$ of the sites), which are then all contained in the data of the twist function $\vp(\lambda)$. In addition to those, one can also have extra parameters entering the definition of the currents in the Takiff realisations (in other words, the same Takiff current algebra can possess several realisations depending on external continuous parameters, which will then appear in the action of the corresponding $\s$-models). It was conjectured in~\cite{Delduc:2020vxy} that the 1-loop RG-flow of such a model satisfies the following properties. First of all, only the parameters of the first kind are running under this flow (the additional parameters entering the realisation are then essentially RG-invariants, independent of the scale $\mu$). Secondly, the 1-loop flow of the AGM parameters can be encoded in a very compact way in terms of the twist function $\vp(\lambda)$, which we recall contains all of these parameters. More precisely, one has
\begin{equation}\label{Eq:RG-flow}
\mu\frac{\dd\;}{\dd\mu} \vp(\lambda) = -4\pi\hbar \, \hv\,\frac{\dd\;}{\dd \lambda} \Bigl( \vp(\lambda) f(\lambda) \Bigr) + O(\hbar^2)\,, \qquad \text{ with } \quad f(\lambda) = \sum_{i=1}^M \frac{\epsilon_i}{\vp'(\ze_i)} \frac{1}{\lambda-\ze_i} + a + b\,\lambda\,.
\end{equation}
Here, recall that $\hv$ is the dual Coxeter number of the Lie algebra $\g$ (see footnote \ref{FootForm}), that the $\ze_i$'s ($i=1,\dots,M$) are the zeroes of the twist function $\vp(\lambda)$ and that the $\epsilon_i$'s are their associated coefficients, equal to $+1$ or $-1$, in the definition \eqref{Eq:H} of the Hamiltonian. This data determines the function $f(\lambda)$ appearing in the RG-flow \eqref{Eq:RG-flow}, together with the coefficients $a$ and $b$, which encode the freedom of performing translations and dilations of the spectral parameter $\lambda$ along the RG-flow and which can thus be taken to be arbitrary\footnote{More precisely, $a$ and $b$ are arbitrary if we did not fix this freedom of dilating and translating $\lambda$. If one considers a specific choice of $\vp(\lambda)$ where this freedom was fixed (for instance by setting two of its poles/zeroes to specific positions in the complex plane), then $a$ and $b$ will take well-chosen values which ensure that this choice is preserved.} (recall that such transformations do not change the AGM) . The RG-flow \eqref{Eq:RG-flow} was conjectured in~\cite{Delduc:2020vxy}, based on various examples, for the same class of AGMs as considered in these lectures, for which $\vp(\lambda)\,\dd\lambda$ has a double pole at $\lambda=\infty$. It was later proven in~\cite{Hassler:2020xyj,Hassler:2023xwn}, using the relation between AGMs and the formalism of $\mathcal{E}$-models. Its generalisation for an arbitrary pole structure around $\lambda=\infty$ was conjectured in~\cite{Kotousov:2022azm} and checked in some examples. The proof of this general case has not been established yet in the literature.

All the discussion above concerned the RG-flow of integrable $\s$-models at 1-loop. Its study at higher-loop is an active and current subject of research. In particular, it is expected that the higher-loop renormalisation of these models in general requires adding fine-tuned quantum corrections to their geometry (see for instance~\cite{Hassler:2020xyj,Hoare:2019ark,Georgiou:2019nbz,Levine:2021fof,Alfimov:2021sir} for a non-exhaustive list of references). The interpretation of these quantum corrections in the language of AGMs is an interesting perspective for future investigations. Moreover, it is a quite natural and still open question whether the higher-loop RG-flow of these models can also be recast in terms of their twist function, as done in equation \eqref{Eq:RG-flow} at 1-loop.

\subsection{Quantum integrable structure at conformal points (and beyond)}
\label{SubSec:RG}

In addition to the study of the RG-flow discussed in the previous subsection, another important question to address in the quantisation of integrable $\s$-models and AGMs is that of their first-principle quantum integrability. Recall that, at the classical level, the main result of the AGM formalism is the construction of an integrable structure $\Zc$, composed by an infinite family of Poisson-commuting conserved charges inside of the algebra of observables $\Ac$ of the theory. By ``establishing the first-principle quantum integrability of this model'', we mean here the following two steps:\vspace{-3pt}
\begin{enumerate}[(i)]\setlength\itemsep{0.2pt}
\item determine the algebra of quantum observables $\widehat{\Ac}$ of the theory, which is then non-commutative and reduces to the Poisson algebra $\Ac$ in the classical limit ;
\item construct a quantum integrable structure $\widehat{\Zc}$, \textit{i.e.} an infinite family of commuting operators in $\widehat{\Ac}$, which reduces to $\Zc$ in the classical limit.\vspace{-3pt}
\end{enumerate}
For a generic integrable $\s$-model, both of these points are famously difficult tasks. For certain classes of integrable field theories, one approach which is sometimes considered to study quantum integrability is the so-called Quantum Inverse Scattering Method, which aims at quantising the monodromy matrix. In the case of integrable $\s$-models, this approach is made difficult by the presence of derivatives of the Dirac-distribution $\p_x\delta(x-y)$ in the Poisson bracket \eqref{Eq:Maillet} of the Lax matrix. This issue is called the problem of non-ultralocality~\cite{Maillet:1985fn,Maillet:1985ek} and has been an obstacle for the first-principle quantisation of many integrable $\s$-models for now more than 30 years. Due to these difficulties, an alternative path is to suppose the quantum integrability of these models and use other methods to study their quantum properties, such as for instance the S-matrix bootstrap, the Thermodynamic Bethe Ansatz and the Quantum Spectral Curve, which led to a rich variety of powerful results.\\

The first-principle quantum integrability of these models is however still an important and open question. One programme that was proposed to approach it is to start with the quantum integrable structures of their conformal points, following the philosophy proposed in~\cite{Zamolodchikov:1989hfa,Bazhanov:1994ft} for other types of theories. The main idea is to follow the RG-flow of these quantum models towards conformal fixed points and focus on the latter. Indeed, their additional conformal symmetry allows a rigorous non-perturbative quantisation of these fixed points, using the standard techniques of 2-dimensional CFTs. In particular, a key role in this approach is played by the chiral (left- or right-moving) fields of the model, which form its algebra of extended conformal symmetry. The latter always contains the Virasoro algebra formed by the energy-momentum tensor and generating the conformal transformations but can can also include larger structures, such as current algebras and $\Wc$-algebras. Mathematically, it is is described by a well-understood object called a Vertex Operator Algebra.

A key observation at this point is the following: for many integrable 2-dimensional field theories, taking the conformal limit has for effect to split the integrable structure of the model into two parts, one built in terms of left-moving fields only and the other one in terms of right-moving fields. As explained above, the operatorial approach of 2-dimensional CFT provides us with a well controlled description of these chiral fields at the quantum level, in the form of a Vertex Operator Algebra. This offers a rigorous answer to the step (i) in the proof of first-principle quantum integrability (see above). We can now pass to the step (ii), namely the construction of a quantum integrable structure in this Vertex Operator Algebra, formed by infinitely many commuting operators built from the chiral fields. This is a hard but well-defined mathematical question. For $\s$-models, such conformal integrable structures were studied in various examples: see~\cite{Bazhanov:2013cua,Bazhanov:2017nzh,Litvinov:2018bou,Bazhanov:2018xzh,Litvinov:2019rlv,Alfimov:2020jpy,Kotousov:2022azm} for a non-exhaustive list of references.

Of particular interest for these lectures are the applications of AGMs to this question. The starting point is the observation that for many integrable $\s$-models obtained as realisations of AGMs, the two chiral halves of their conformal limits are described by current algebras or quotients thereof\footnote{For an illustration and further discussion of this phenomenon, see for instance~\cite{Kotousov:2022azm,Lacroix:2023ybi}.}, which are quantisations of the Kac-Moody/Takiff Poisson algebras considered throughout the lectures. In that context, the study of quantum integrable structures amounts to the construction of local and non-local commuting operators in terms of these quantum chiral currents, generalising the classical constructions presented in Section \ref{Sec:ClassicalAGM}. These ``chiral AGMs'' and their quantum integrable structures were the main motivations and subjects of the foundational work~\cite{Feigin:2007mr} of Feigin and Frenkel, which gave some first results and conjectures in this direction. This was further studied and extended in~\cite{Frenkel:2016gxg,Lacroix:2018fhf,Lacroix:2018itd,Gaiotto:2020fdr,Gaiotto:2020dhf,Kotousov:2021vih,Wu:2021jir,Kotousov:2022azm,Franzini:2022duf,Masoero:2023lam} and forms an open and active domain of research. One might hope that this approach opens the way for a systematic first-principle quantisation of many conformal integrable $\s$-models.

Let us note that beyond the construction of the integrable structure of these AGMs, a natural follow-up question is the determination of its spectrum, \textit{i.e.} the diagonalisation of these commuting operators in the Hilbert space of the theory and the computation of their eigenvalues. In the work~\cite{Feigin:2007mr}, it was conjectured that this question can be efficiently answered using the so-called ODE/IQFT (or ODE/IM) correspondence\cite{Dorey:1998pt,Bazhanov:1998wj,Bazhanov:2003ni}, which encodes these eigenvalues into well-chosen Ordinary Differential Equations.\footnote{In the approach of~\cite{Feigin:2007mr}, this correspondence is in fact the manifestation of a deeper mathematical structure, namely a generalisation of the so-called Geometric Langlands Correspondence.} For conciseness, we will not enter into more details on these aspects here.\\

Let us end this conclusion by mentioning an interesting but complicated perspective, namely the construction of quantum integrable structures beyond the conformal limits. In particular, a natural approach to this question would be to start with the integrable structures described above for the conformal points and to consider their deformations under non-marginal perturbations, breaking the conformal symmetry of the model and ``recoupling'' together its two chiralities. At the moment, this is a vastly open direction for future developments.
 
\section*{Acknowledgements}

I am grateful to Sibylle Driezen, Gleb Kotousov, Marc Magro and Benoît Vicedo for useful discussions during the preparation of these notes and to Ben Hoare for valuable comments on the draft. This work is supported by Dr. Max R\"ossler, the Walter Haefner Foundation and the ETH Z\"urich Foundation.\\

These notes were prepared for the Young Researchers Integrability School and Workshop (YRISW) held in Durham from 17 to 21 July 2023, see \href{https://indico.cern.ch/e/YRISW23}{{\ttfamily https://indico.cern.ch/e/YRISW23}} for the website. I wish to thank the organisers of the school (Patrick Dorey, Ben Hoare, Ana Retore, Fiona Seibold and Alessandro Sfondrini) for the opportunity to give the lectures, and for all the help and support before and during the school. I also thank the students for their engagement during the lectures and for the interesting questions. The YRISW 2023 was realised with the financial support of Durham University, the University of Padova, Imperial College London, GATIS+, UKRI and the European Union.
 
\newpage

\appendix

\section{Higher-degree local charges}
\label{App:Local}

This appendix contains various complements and technical computations related to the higher-degree local charges considered in the Subsection \ref{SubSec:Local} of the main text.

\subsection{EHMM-tensors}
\label{App:EHMM-Tensors}

In this subsection, we will collect some useful informations on the EHMM-tensors $\kappa_{p+1}$, which appear in the definition \eqref{Eq:Qip} of the local charges $\Q_i^{(p)}$. Recall that
\begin{equation}
\begin{array}{lccc}
\kappa_{p+1} : & \g^\C \oplus \dots \oplus \g^\C       & \longrightarrow & \C \\
               & (X_1,\dots,X_{p+1}) & \longmapsto & \kappa_{p+1}^{a_1\,\cdots\, a_{p+1}}\,X_{1,a_1} \cdots X_{p+1,a_{p+1}}
\end{array}
\end{equation}
is a symmetric $(p+1)$--multinear form on $\g^\C$. Here, we used a basis decomposition $X_k = X_{k,a}\,\td^a$ and thus described this multilinear form through a tensor $\kappa_{p+1}^{a_1\,\cdots\, a_{p+1}}$, depending symmetrically on $p+1$ indices $a_k \in \lbrace 1,\dots,\dim\g\rbrace$.

\paragraph{Properties of $\bm{\kappa_{p+1}}$.} As explained in the main text, the EHMM-tensors satisfy certain fundamental properties \eqref{Eq:adKappa} and \eqref{Eq:KappaKappa}. There, we formulated these properties in components, using a basis of $\g^\C$. In the rest of this appendix, we will work with a basis-independent formulation. In this spirit, we thus start by rewritting \eqref{Eq:adKappa} and \eqref{Eq:KappaKappa} in terms of the abstract multilinear forms $\kappa_{p+1}$. For instance, the first of these identities simply translates the ad-invariance of these forms:
\begin{equation}\label{Eq:adKappa2}
\kappa_{p+1} \bigl([X_1,Y],\dots,X_{p+1}\bigr) + \,\cdots\, + \kappa_{p+1} \bigl(X_1,\dots,[X_{p+1},Y] \bigr) = 0\,,
\end{equation}
for all $X_1,\dots,X_{p+1},Y\in\g^\C$.\\

To re-express the second property \eqref{Eq:KappaKappa}, we have to introduce some notations. For $p,q\in E$, we define the $(p+q)$-linear form
\begin{equation}\label{Eq:Omega}
\omega^{(p,q)}\bigl(X_1,\dots,X_{p+q}\bigr) = \kappa_{p+1}\bigl(X_{1},\dots,X_p,\td_a \bigr)\, \kappa_{q+1} \bigl( X_{p+1}, \dots, X_{p+q}, \td^a \bigr)\,,
\end{equation}
where a sum over $a\in\lbrace 1,\dots,\dim\g\rbrace$ is implied. In this definition, we used the basis $\lbrace \td_a \rbrace_{a=1,\dots,\dim\g}$ of $\g$ and its dual basis $\lbrace \td^a \rbrace_{a=1,\dots,\dim\g}$: however, one easily shows that $\omega^{(p,q)}$ is in fact basis-independent. It is clear that $\omega^{(p,q)}$ is symmetric with respect to its first $p$ entries and with respect to its last $q$ entries. It is however not invariant under permutations mixing these two sets. Let us then introduce 
\begin{equation}\label{Eq:OmSym}
\oms^{(p,q)}\bigl(X_1,\dots,X_{p+q}\bigr) = \omega^{(p,q)}\bigl(X_{(1},\dots,X_{p+q-1)},X_{p+q}\bigr)\,,
\end{equation}
where the round brackets $(1,  \dots, p+q-1)$ mean that we symmetrise over the $p+q-1$ entries $(X_1,\dots,X_{p+q-1})$. The identity \eqref{Eq:KappaKappa} then implies that
\begin{equation}\label{Eq:SymOmega}
\oms^{(p,q)}\bigl(X_{1},\dots,X_{p+q}\bigr) = \omega^{(p,q)}\bigl(X_{(1},\dots,X_{p+q)}\bigr)\,,
\end{equation}
\textit{i.e.} that $\oms^{(p,q)}$ is completely symmetric, although it was initially introduced as a symmetrisation over the first $p+q-1$ entries only. We will need this property in the next subsection, to prove the Poisson-commutativity of the local charges.

\paragraph{Construction.} The properties \eqref{Eq:adKappa2} and \eqref{Eq:SymOmega} -- or \eqref{Eq:adKappa} and \eqref{Eq:KappaKappa} in components -- define strong conditions on the tensors $\kappa_{p+1}$. Solutions to these constraints were found by Evans, Hassan, MacKay and Mountain in~\cite{Evans:1999mj,Evans:2000hx}, hence the name of EHMM-tensors. More precisely, these solutions were found for the cases where $\g^\C$ is a simple algebra of classical type A, B, C or D (in the classification of simple Lie algebras): however, similar tensors are expected to also exist for $\g^\C$ of exceptional type.

The EHMM-tensors $\kappa_{p+1}$ are labelled by the integer $p\in E$, taking value in a specific set\footnote{Technically, $E$ should rather be seen as a multiset, in the sense that some integers $p$ can appear several times in $E$. When that is the case, it means that there are multiple EHMM-tensors with the same degree $p+1$. However, this occurs only in very specific cases, namely the algebras $D_{2k}$, where some of the exponents have multiplicity 2. Due to this rarity and to avoid introducing heavier notations, we will generally treat $E$ as a set.} of integers $E\subset \Z_{\geq 1}$, called the affine exponents of $\g$. The latter can be found as follows. We denote by $\rk$ the rank of $\g$, \textit{i.e.} the dimension of a Cartan subalgebra of $\g^\C$, and by $\Cox$ the Coxeter number of $\g$, which is defined as the height of the highest-root of $\g^\C$ plus one. Following results of Kostant, we consider a set of $\rk$ integers $E_{\text{fin}}=\lbrace d_1,\dots,d_{\rk} \rbrace$, called the \textit{finite exponents} of $\g$. These integers appear naturally in the so-called principal decomposition of $\g^\C$ and satisfy:
\begin{equation}
1 = d_1 \leq d_2 \leq \dots \leq d_{\rk} = \Cox - 1\,. 
\end{equation}
In particular, the first finite exponent is always $d_1=1$ and the last one $d_{\rk} = \Cox - 1$. These integers can be equivalently characterised by the fact that the elementary ad-invariant polynomials on $\g$ have degrees $d+1$, $d\in E_{\text{fin}}$. In terms of these finite exponents, the affine ones are defined by repetition modulo the Coxeter number $\Cox$, \textit{i.e.}
\begin{equation}
E = \bigl\lbrace d + a\,\Cox\,,\; d\in E_{\text{fin}},\; a\in \Z_{\geq 0} \bigr\rbrace\, = E_{\text{fin}} + (\Cox)\,\Z_{\geq 0}.
\end{equation}
We gather the list of ranks, Coxeter numbers and exponents of all classical Lie algebras in Table \ref{Tab:Exp}.\vspace{7pt}

\begin{table}[H]
\begin{center}
\begin{tabular}{ccccccccc}
Type & ~~ & Algebra & ~~ & Rank & ~~ & Coxeter & ~~ & Exponents \\
\hline \hline
A & & $\sl(k,\C)$              & & $k-1$ & & $k$  & & $1,2,\dots,k-1$ \\
B & & $\mathfrak{so}(2k+1,\C)$ & & $k$   & & $2k$ & & $1,3,\dots,2k-1$ \\
C & & $\mathfrak{sp}(2k,\C)$   & & $k$   & & $2k$ & & $1,3,\dots,2k-1$\\
D & & $\mathfrak{so}(2k+2,\C)$ & & $k+1$   & & $2k$ & & $1,3,\dots,2k-1\,;\,k$
\end{tabular}
\caption{Rank, Coxeter number and finite exponents of classical Lie algebras.\label{Tab:Exp}}
\end{center}\vspace{-8pt}
\end{table}

The simplest elements in the classification are the ones of type A. They can be represented as the algebra $\g^\C = \sl(k,\C)$ of traceless $k\times k$ matrices, with $k=\rk+1$ in $\Z_{\geq 2}$. According to Table \ref{Tab:Exp}, the affine exponents of this algebra are given by all positive integers except the multiples of $k$, \textit{i.e.} $E=\lbrace 1,\dots,k-1,k+1,\dots,2k-1,2k+1,\dots\rbrace$. Let us quickly describe some of the corresponding EHMM-tensors $\kappa_{p+1}$. Since the latter are symmetric, it is enough to give the associated homogeneous polynomials $X \mapsto \kappa_{p+1}(X,\dots,X)$, which we still denote by $\kappa_{p+1}$ by a slight abuse of notations. The first few of these polynomials read
\begin{equation}
\kappa_2(X) = \Tr\bigl(X^2\bigr)\,,\qquad\quad \kappa_3(X) = \Tr\bigl(X^3)\,,
\end{equation}
\begin{equation*}
\kappa_4(X) = \Tr\bigl(X^4\bigr) - \frac{3}{2k}\, \Tr\bigl(X^2\bigr)\null^2\,,\qquad\quad \kappa_5(X) = \Tr\bigl(X^5\bigr) - \frac{10}{3k}\,\Tr\bigl(X^2\bigr)\,\Tr\bigl(X^3\bigr)\,,
\end{equation*}
\begin{equation*}
\kappa_6(X) = \Tr\bigl(X^6\bigr) - \frac{5}{3k}\,\Tr\bigl(X^3\bigr)\null^2 - \frac{15}{4k} \,\Tr\bigl(X^4\bigr)\,\Tr\bigl(X^2\bigr) + \frac{25}{8k^2} \Tr\bigl(X^2\bigr)\null^3\,.\vspace{4pt}
\end{equation*}
It is clear that the traces $X\mapsto \Tr(X^{p+1})$ define ad-invariant polynomials on $\g^\C$. The corresponding tensors thus satisfy the property \eqref{Eq:adKappa}. However, they do not obey the main identity \eqref{Eq:KappaKappa} of EHMM-tensors. This additional constraint requires to add fine-tuned corrections to these traces built from combinations of lower degree ones, as for instance the term proportional to $\Tr(X^2)^2$ in $\kappa_4(X)$. In~\cite{Evans:1999mj,Evans:2000hx}, a general construction of these corrections was proposed, based on certain generating functions: for conciseness, we will not develop this construction further here. We finally note that these corrections are such that the tensor $\kappa_{ak+1}$ vanishes for all $a\in\Z_{\geq 1}$: for instance, one checks that matrices $X\in\sl(3,\C)$ satisfy $\Tr(X^4) = \frac{1}{2} \Tr(X^2)\null^2$, so that $\kappa_4(X)=0$ for $k=3$. This reflects the fact that the multiples of the Coxeter number $\Cox=k$ are skipped in the set of affine exponents.\\

The case of types B, C and D is at first sight simpler but in fact ends up being quite richer. For these algebras, one can also build invariant polynomials by looking at traces $X\mapsto \Tr(X^{2p})$ in the fundamental representation (restricting to even powers, the odd ones being in fact vanishing). It was argued in~\cite{Evans:1999mj} that these trace polynomials directly define EHMM-tensors satisfying \eqref{Eq:KappaKappa}, without having to add correction terms built from combinations of lower degree traces. However, one in fact finds that such terms can be added in a specific way without breaking \eqref{Eq:KappaKappa}. More precisely, we obtain a one-parameter family of EHMM-tensors $\kappa_{p+1}$, encoding the freedom of adding these well-chosen corrections.\footnote{For algebras of type B, C and D, the local charges $\Q_i^{(p)}$ built in Subsection \ref{SubSec:Local} thus depend on the choice of this parameter in the definition of the EHMM-tensors. By construction, each choice yields a family of Poisson-commuting charges: however, we stress the fact that the charges corresponding to different choices do not have vanishing Poisson bracket one with another. This freedom should thus not be understood as a way of constructing a larger integrable structure but rather as the existence of an additional free parameter in its definition. This additional parameter only appears for charges of degree at least 4 and thus does not affect the definition of the quadratic charges and the Hamiltonian \eqref{Eq:H}. Its physical interpretation is not obvious. It is possible that some other considerations (for instance the quantisation of these local charges) single out a specific choice for this parameter.} Similarly to the type A case, these tensors can be obtained using well-chosen generating functions~\cite{Evans:1999mj}. Let us finally mention that for an algebra of type $D_{k+1}$, where $\g^\C=\mathfrak{so}(2k+2,\C)$, there exist ``exotic'' EHMM-tensors of degrees $(2a+1)k+1$, $a\in\Z_{\geq 0}$, which are not built from traces of powers in the fundamental representation, but rather from what is called the Pfaffian invariant. We refer to~\cite{Evans:1999mj} for more details about this.

\subsection{Poisson brackets of the local charges}
\label{App:LocalPb}

In this second subsection of the appendix, we detail the proof of the main result \eqref{Eq:PbQp}, which was only sketched in the main text.

\paragraph{Brackets with $\bm{\Wc^{(p+1)}}$.} Recall that, for $p \in E$, we defined
\begin{equation}
\Wc^{(p+1)}(\lambda,x) = \frac{1}{p+1}\; \kappa_{p+1} \bigl( \Gamma(\lambda,x), \dots, \Gamma(\lambda,x) \bigr)\,.
\end{equation}
If $\Oc$ is any observable, using the Leibniz rule and the symmetry of $\kappa_{p+1}$, we find that
\begin{equation}
\bigl\lbrace \Oc, \Wc^{(p+1)}(\lambda,x) \bigr\rbrace = \kappa_{p+1} \bigl( \bigl\lbrace \Oc , \Gamma(\lambda,x) \bigr\rbrace, \Gamma(\lambda,x), \dots, \Gamma(\lambda,x) \bigr)\,.
\end{equation}
Defining the linear form
\begin{equation}\label{Eq:Beta}
\begin{array}{lccc}
\beta_X^{(p)} : & \g^\C & \longrightarrow & \C \\
                &   Y   &   \longmapsto   & \kappa_{p+1}(Y,X,\dots,X)
\end{array} 
\end{equation}
for $X\in\g^\C$, we then rewrite the above bracket as
\begin{equation}\label{Eq:PbWO}
\bigl\lbrace \Oc,\Wc^{(p+1)}(\lambda,x) \bigr\rbrace = \beta_{\Gamma(\lambda,x)}^{(p)} \bigl( \bigl\lbrace  \Oc, \Gamma(\lambda,x) \bigr\rbrace \bigr)\,.
\end{equation}

\paragraph{Bracket of $\bm{\Wc^{(p+1)}}$ with the Gaudin Lax matrix.} Using equation \eqref{Eq:PbWO} and tensorial notations, one gets
\begin{equation}
\bigl\lbrace \Gamma(\lambda,x), \Wc^{(p+1)}(\mu,y) \bigr\rbrace = \beta_{\Gamma(\mu,y)}^{(p)}\,\ti{2} \Bigl( \bigl\lbrace \Gamma(\lambda,x)\ti{1}, \Gamma(\mu,y)\ti{2} \bigr\rbrace \Bigr)\,,
\end{equation}
where the index $\bm{\underline{2}}$ means that the linear form $\beta_{\Gamma(\mu,y)}^{(p)}$ acts on the second tensor factor (in particular, the end result is then $\g^\C$-valued, as expected). We can now use our key formula \eqref{Eq:PbGaudin} for $\lbrace \Gamma(\lambda,x)\ti{1}, \Gamma(\mu,y)\ti{2} \rbrace$. The term containing $\Gamma(\mu)$ in \eqref{Eq:PbGaudin} will give a contribution proportional to\footnote{Here we used the Casimir identity \eqref{Eq:CasId} to move $\Gamma(\mu)$ on the second tensor factor and the presence of a Dirac-distribution $\delta(x-y)$ to set $y=x$ in all fields.}
\begin{equation}\label{Eq:abc}
\beta_{\Gamma(\mu,x)}^{(p)}\,\ti{2}\,\bigl[ C\ti{12},  \Gamma(\mu,x)\ti{2} \bigr]\,.
\end{equation}
For any $X,Y\in\g^\C$, we now note that
\begin{equation}\label{Eq:BetaInv}
\beta^{(p)}_X\bigl( [Y,X] \bigr) = \kappa_{p+1} \bigl( X,\dots,X,[Y,X] \bigr) = 0
\end{equation}
by ad-invariance \eqref{Eq:adKappa2} of $\kappa_{p+1}$. Thus the contribution \eqref{Eq:abc} vanishes. We are then left with
\begin{align}\label{Eq:PbWpG}
\bigl\lbrace \Gamma(\lambda,x), \Wc^{(p+1)}(\mu,y) \bigr\rbrace = -\frac{1}{\lambda-\mu}\,\left[ \beta_{\Gamma(\mu,x)}^{(p)}\,\ti{2} \,C\ti{12},\Gamma(\lambda,x)\ti{1} \right]\, \delta(x-y) \hspace{30pt} \\
+\, \beta_{\Gamma(\mu,y)}^{(p)}\,\ti{2} \,C\ti{12}\,\frac{\vp(\lambda)-\vp(\mu)}{\lambda-\mu} \, \p_x\delta(x-y)\,, \notag
\end{align}
where we exchanged the order in which the operations $\beta_{\Gamma(\mu,x)}^{(p)}\,\ti{2}$ and $\bigl[ \cdot, \Gamma(\lambda,x)\ti{1} \bigr]$ are applied, which we are allowed to do since they act on different tensor spaces.

\paragraph{Poisson bracket of $\bm{\Wc^{(p+1)}}$ and $\bm{\Wc^{(q+1)}}$.} We now apply the equation \eqref{Eq:PbWO} again, to get
\begin{equation*}
\bigl\lbrace \Wc^{(p+1)}(\lambda,x), \Wc^{(q+1)}(\mu,y) \bigr\rbrace = \beta_{\Gamma(\lambda,x)}^{(p)} \Bigl( \bigl\lbrace \Gamma(\lambda,x), \Wc^{(q+1)}(\mu,y) \bigr\rbrace \Bigr)\,.
\end{equation*}
We can thus reinsert our previous result \eqref{Eq:PbWpG} for $\bigl\lbrace \Gamma(\lambda,x), \Wc^{(q+1)}(\mu,y) \bigr\rbrace$. The first term in the right-hand side of this equation gives a contribution proportional to
\begin{equation*}
\beta_{\Gamma(\lambda,x)}^{(p)}\,\ti{1} \left[ \beta_{\Gamma(\mu,x)}^{(q)}\,\ti{2} \,C\ti{12},\Gamma(\lambda,x)\ti{1} \right] = 0\,,
\end{equation*}
where we used the identity \eqref{Eq:BetaInv} to get zero. Essentially, we have shown that the commutator term in the Poisson bracket \eqref{Eq:PbGaudin} of the Gaudin Lax matrix does not create any contribution in the bracket between densities $\Wc^{(p+1)}$, due to the ad-invariance of the EHMM-tensors $\kappa_{p+1}$. We are thus left with
\begin{equation}\label{Eq:PbWWp}
\bigl\lbrace \Wc^{(p+1)}(\lambda,x), \Wc^{(q+1)}(\mu,y) \bigr\rbrace = \beta_{\Gamma(\lambda,x)}^{(p)}\,\ti{1}\,\beta_{\Gamma(\mu,y)}^{(q)}\,\ti{2} \,C\ti{12}\,\frac{\vp(\lambda)-\vp(\mu)}{\lambda-\mu} \, \p_x\delta(x-y)\,.
\end{equation}

\paragraph{Analysis of the bracket of densities.} Let us now analyse the bracket \eqref{Eq:PbWWp}. As a first observation, we note that if we evaluate it at different zeroes of the twist function, \textit{i.e.} at $\lambda=\ze_i$ and $\mu=\ze_j$ with $i\neq j$, the term involving twist functions vanishes. Thus, we proved
\begin{equation}\label{Eq:PbWWpij}
\bigl\lbrace \Wc^{(p+1)}(\ze_i,x), \Wc^{(q+1)}(\ze_j,y) \bigr\rbrace = 0\, \qquad \text{ if } \; i\neq j\,.
\end{equation}
This means that the densities of the local charges $\Q_i^{(p)}$ and $\Q_j^{(q)}$ Poisson-commute for $i\neq j$.\\

In contrast, if we evaluate the bracket \eqref{Eq:PbWWp} at the same zero, \textit{i.e.} at $\lambda=\mu=\ze_i$, the term involving the twist functions gives a non-vanishing contribution $\vp'(\ze_i)$. To get the Poisson-commutation of charges associated to the same zero, we thus need to develop another strategy. We start by noting that the definitions \eqref{Eq:Cas}, \eqref{Eq:Omega} and \eqref{Eq:Beta} imply
\begin{equation}
\beta^{(p)}_X\null\ti{1} \, \beta^{(q)}_Y\null\ti{2}\,C\ti{12} = \omega^{(p,q)}(X^{\times p},Y^{\times q})\,,
\end{equation}
for any $X,Y\in\g^\C$, where $(X^{\times p},Y^{\times q})$ is our notation for $p$ entries $X$ and $q$ entries $Y$. We thus rewrite the bracket \eqref{Eq:PbWWp} as
\begin{equation*}
\bigl\lbrace \Wc^{(p+1)}(\lambda,x), \Wc^{(q+1)}(\mu,y) \bigr\rbrace = \,\frac{\vp(\lambda)-\vp(\mu)}{\lambda-\mu}\, \omega^{(p,q)}\bigl(\Gamma(\lambda,x)^{\times p}, \Gamma(\mu,y)^{\times q} \bigr)  \, \p_x\delta(x-y)\,.
\end{equation*}
We now apply the identity \eqref{Eq:DerDelta} to transform all $y$'s in $x$'s. To compute the result, we need
\begin{equation}
\p_y\, \omega^{(p,q)}\bigl(\Gamma(\lambda,x)^{\times p}, \Gamma(\mu,y)^{\times q} \bigr) = q\,\omega^{(p,q)}\bigl(\Gamma(\lambda,x)^{\times p}, \Gamma(\mu,y)^{\times (q-1)}, \p_y \Gamma(\mu,y) \bigr)\,,
\end{equation}
which follows from the Leibniz rule and the symmetry of $\omega^{(p,q)}$ with respect to its last $q$ entries. We then find
\begin{align*}
\bigl\lbrace \Wc^{(p+1)}(\lambda,x), \Wc^{(q+1)}(\mu,y) \bigr\rbrace &= q\,\frac{\vp(\lambda)-\vp(\mu)}{\lambda-\mu}\, \omega^{(p,q)}\bigl(\Gamma(\lambda,x)^{\times p}, \Gamma(\mu,x)^{\times (q-1)} , \p_x \Gamma(\mu,x) \bigr) \, \delta(x-y) \\
& \hspace{30pt} + \, \Ac_3^{(p,q)}(\lambda,\mu,x)\,\p_x\delta(x-y)\,,
\end{align*}
where we introduced
\begin{equation*}
\Ac_3^{(p,q)}(\lambda,\mu,x) = \frac{\vp(\lambda)-\vp(\mu)}{\lambda-\mu}\, \omega^{(p,q)}\bigl(\Gamma(\lambda,x)^{\times p}, \Gamma(\mu,x)^{\times q} \bigr) \,.
\end{equation*}

The reason why we do not get zero in the evaluation $\lambda=\mu=\ze_i$ is because the vanishing of $\vp(\lambda)-\vp(\mu)$ is counterbalanced by the diverging term $1/(\lambda-\mu)$. Our strategy will be to cancel this divergence in another way, by adding some terms by hand. More precisely, let us define
\begin{equation*}
\Ac^{(p,q)}_1(\lambda,\mu) = q\, \frac{\omega^{(p,q)}\bigl(\Gamma(\lambda)^{\times p}, \Gamma(\mu)^{\times (q-1)} , \p_x \Gamma(\mu) \bigr) - \omega^{(p,q)}\bigl(\Gamma(\mu)^{\times p}, \Gamma(\mu)^{\times (q-1)} , \p_x \Gamma(\mu) \bigr)}{\lambda-\mu}\,.
\end{equation*}
By construction, this is of the form $\frac{f(\lambda,\mu)-f(\mu,\mu)}{\lambda-\mu}$ and thus stays finite when $\lambda=\mu$. Making $\Ac_1^{(p,q)}(\lambda,\mu,x)$ appearing by hand in the above bracket, we find 
\begin{align*}
\bigl\lbrace \Wc^{(p+1)}(\lambda,x), \Wc^{(q+1)}(\mu,y) \bigr\rbrace &= q \,\frac{\vp(\lambda)-\vp(\mu)}{\lambda-\mu}\, \omega^{(p,q)}\bigl(\Gamma(\mu,x)^{\times (p+q-1)}, \p_x \Gamma(\mu,x) \bigr) \, \delta(x-y) \\
& \hspace{30pt} + \bigl( \vp(\lambda)-\vp(\mu) \bigr) \Ac^{(p,q)}_1(\lambda,\mu,x)\,\delta(x-y) + \, \Ac_3^{(p,q)}(\lambda,\mu,x)\,\p_x\delta(x-y)\,,
\end{align*}
where we used $\omega^{(p,q)}(X^{\times p},X^{\times (q-1)},Y)=\omega^{(p,q)}(X^{\times (p+q-1)},Y)$.\newpage

We are almost done. The next observation to make is that the $\omega^{(p,q)}$ in the first line can be replaced by a $\oms^{(p,q)}$, as defined in \eqref{Eq:OmSym}, since we can symmetrise over the identical $p+q-1$ entries $\Gamma(\mu,x)$. Now recall that the EHMM-tensors satisfy one crucial identity \eqref{Eq:SymOmega}, which can be phrased as the fact that $\oms^{(p,q)}$ is completely symmetric under all its $p+q$ entries. Using this and the Leibniz rule, we get
\begin{equation}
\oms^{(p,q)}\bigl(\Gamma(\mu,x)^{\times (p+q-1)}, \p_x \Gamma(\mu,x) \bigr) = \frac{1}{p+q} \p_x\,\omega^{(p,q)}\bigl(\Gamma(\mu,x)^{\times (p+q)} \bigr)\,.
\end{equation}
Introducing
\begin{equation}
\Ac_2^{(p,q)}(\lambda,\mu,x) = \frac{q}{p+q} \,\frac{\vp(\lambda)-\vp(\mu)}{\lambda-\mu}\,\omega^{(p,q)}\bigl(\Gamma(\mu,x)^{\times (p+q)} \bigr)\,,
\end{equation}
we thus finally get
\begin{align}\label{Eq:PbWpWqApp}
\bigl\lbrace \Wc^{(p+1)}(\lambda,x), \Wc^{(q+1)}(\mu,y) \bigr\rbrace &= \Bigl( \vp(\lambda) \, \Ac_1^{(p,q)}(\lambda,\mu,x)  -\vp(\mu)\, \Ac_1^{(p,q)}(\lambda,\mu,x)  + \p_x\Ac_2^{(p,q)}(\lambda,\mu,x) \Bigr)\, \delta(x-y) \notag \\
& \hspace{30pt} + \, \Ac_3^{(p,q)}(\lambda,\mu,x)\,\p_x\delta(x-y)\,. 
\end{align}
This ensures that $\bigl\lbrace \Q_i^{(p)},\Q^{(q)}_i \bigr\rbrace$ vanishes, as wanted. Indeed, up to global factors, this bracket is obtained by evaluating \eqref{Eq:PbWpWqApp} at $\lambda=\mu=\ze_i$ and integrating over $x$ and $y$. The contribution of the first-two terms on the right-hand side of \eqref{Eq:PbWpWqApp} vanishes due to the evaluations at $\lambda=\mu=\ze_i$, using the fact that we constructed $\Ac_1^{(p,q)}(\lambda,\mu,x)$ to be regular at $\lambda=\mu$. The remaining terms are spatial derivatives and thus do not contribute once integrated over $x$ and $y$.

\paragraph{Zero curvature equations.} Let us now consider the bracket
\begin{equation}
\bigl\lbrace q_i^{(p)}(y)\,, \Lc_x(\lambda,x) \bigr\rbrace = - \frac{\bigl\lbrace \Gamma(\lambda,x), \Wc^{(p+1)}(\ze_i,y) \bigr\rbrace}{\vp(\lambda)\,\vp'(\ze_i)^{(p+1)/2}}\,,
\end{equation}
where we used equations \eqref{Eq:Lax} and \eqref{Eq:Density}. We can compute it from our earlier result \eqref{Eq:PbWpG}, evaluated at $\mu=\ze_i$. Using $\vp(\ze_i)=0$ and introducing the $\g^\C$-valued field
\begin{equation}
\Kc_i^{(p)}(\lambda,x) = - \frac{1}{\vp'(\ze_i)^{(p+1)/2}} \frac{\beta_{\Gamma(\ze_i,x)}^{(p)}\,\ti{2} \,C\ti{12}}{\lambda-\ze_i}\,,
\end{equation}
we get
\begin{equation*}
\bigl\lbrace q_i^{(p)}(y), \Lc_x(\lambda,x) \bigr\rbrace = -\bigl[ \Kc_i^{(p)}(\lambda,x), \Lc_x(\lambda,x)\bigr]\, \delta(x-y) + \Kc_i^{(p)}(\lambda,y) \, \p_x\delta(x-y)\,.
\end{equation*}
We finally integrate over $y$ to obtain $\Q_i^{(p)}$ on the left-hand side, using the Dirac-distribution on the right-hand side. We then find
\begin{equation}
\bigl\lbrace \Q_i^{(p)}, \Lc_x(\lambda,x) \bigr\rbrace - \p_x \Kc_i^{(p)}(\lambda,x) + \bigl[\Kc_i^{(p)}(\lambda,x), \Lc_x(\lambda,x) \bigr] = 0 \,.
\end{equation}
This is the equation \eqref{Eq:ZCEip} in the main text. As explained there, it ensures the Poisson-commutation of $\Q_i^{(p)}$ with the non-local charges in $\Tc_V(\lambda)$.

\section{Completeness of Takiff realisations}
\label{App:Comp}

\paragraph{Preliminary.} Consider a Takiff realisation in a canonical Poisson algebra $\Acm$, with currents $J^{[p]}_r(x) = J_{r,\,a}^{[p]}(x)\,\td^a$ built as in \eqref{Eq:RealTakiff}. We gather the indices
\begin{equation*}
r\in\lbrace 1,\dots,N\rbrace\,, \qquad p\in\lbrace 0,\dots,m_r-1 \rbrace\qquad \text{ and } \qquad a\in\lbrace 1,\dots,\dim\g\bigr\rbrace
\end{equation*}
labelling the components of these currents into a multi-index $A$, which ranges from $1$ to $M\dim\g$, with $M=\sum_{r=1}^N m_r$. With this notation, we then see the collection of the current components as a $(M\dim\g)$--tuple\footnote{As explained in Subsection \ref{SubSec:Real}, we suppose that the Takiff realisation satisfies some reality conditions, namely that the fields $J_{r,\,a}^{[p]}(x)$ are either real or come in pairs $\bigl(J_{r,\,a}^{[p]}(x),J_{\bar r,\,a}^{[p]}(x)\bigr)$ of complex conjugates. In the latter case, we more precisely define the corresponding multi-indices $A$ to label the real and imaginary parts of the pair $\bigl(J_{r,\,a}^{[p]}(x),J_{\bar r,\,a}^{[p]}(x)\bigr)$, rather than the two complex fields forming it. In the end, all the components in $\bigl( J_A(x) \bigr)_{A=1,\dots,M\dim\g}$ are thus real fields.} 
\begin{equation}
\bigl( J_A(x) \bigr)_{A=1,\dots,\,M\dim\g}\,.
\end{equation}
Similarly, we gather the momentum fields and the derivatives of the coordinate fields in a $2d$-tuple
\begin{equation}
\bigl( \Psi_I(x) \bigr)_{I=1,\dots,\,2d} = \bigl(  \pi_i(x), \p_x\phi^{\hspace{0.5pt}i}(x) \bigr)_{i=1,\dots,d}\,,
\end{equation}
where we recall that $d=\dim \Mc$. We can thus rewrite the realisation \eqref{Eq:RealTakiff} as a linear relation
\begin{equation}\label{Eq:RealE}
J_A(x) = E_A^{\;\;\,I}\bigl( \Phi(x) \bigr)\,\Psi_I(x)\,,
\end{equation}
where a summation over $I\in\lbrace 1,\dots,\,2d \rbrace$ is implied and where the coefficients $\bigl( E_A^{\;\;\,I} \bigr)_{I=1,\dots,\,2d}^{A=1,\dots,\,M\dim\g}$ are a repackaging of the ones $A_{r,p,a}^{\hspace{19pt}i}$ and $B_{r,p,a\,;\,i}$ of equation \eqref{Eq:RealTakiff}, using the new labelling introduced above.\\

In these notations, the Poisson bracket \eqref{Eq:PbTakiffReal} of the current components take the form
\begin{equation}\label{Eq:PbJA}
\bigl\lbrace J_A(x), J_B(y) \bigr\rbrace = \F{AB}C\,J_C(x)\,\delta(x-y) - K_{AB}\,\p_x\delta(x-y)\,,
\end{equation} 
where the coefficients $\F{AB}C$ are built from the structure constants $\f{ab}c$ of the initial Lie algebra $\g$ and the coefficients $K_{AB}$ are built from the bilinear form $\kappa_{ab}$ of $\g$ and the levels $\ell_r^{[p]}$. We will admit here that the matrix $\mathbb{K}=(K_{AB})_{A,B=1,\dots,\,M\dim\g}$ is invertible due to the assumption made earlier that the levels with ``highest Takiff order'' $\ell_r^{[m_r-1]}$ are non-vanishing: this can be proven by writing down the explicit expression of $K_{AB}$.

On the other hand, we note that the canonical bracket \eqref{Eq:PbCan2} implies
\begin{equation}
\bigl\lbrace \Psi_I(x), \Psi_J(y) \bigr\rbrace = -D_{IJ}\,\p_x(x-y)\,, \qquad \text{ where } \qquad \mathbb{D}=\bigl( D_{IJ} \bigr)_{I,J=1,\dots,\,2d} = \begin{pmatrix}
0 & \Id \\ \Id & 0
\end{pmatrix}\,.
\end{equation}
Comparing the $\p_x\delta(x-y)$--term of the equation \eqref{Eq:PbJA} with what we obtain from the above bracket and the expression \eqref{Eq:RealE} of $J_A(x)$ in terms of $\Psi_I(x)$, one finds that
\begin{equation}\label{Eq:EE}
E_A^{\;\;\,I}\bigl( \Phi(x) \bigr)\, E_B^{\;\;\,J}\bigl( \Phi(x) \bigr)\,D_{IJ} = K_{AB}\,.
\end{equation}
Note that this holds for all values of the coordinate field $\Phi(x)$ on which $E_A^{\;\;\,I}$ and $E_B^{\;\;\,J}$ depend. This already imposes some constraints on the form of the realisation \eqref{Eq:RealE} (the remaining ones are more complicated and come from the analysis of the $\delta(x-y)$--term of the Poisson bracket). In particular, we note that the compatibility of this equation with the invertibility of $(K_{AB})_{A,B=1,\dots,\,M\dim\g}$ requires
\begin{equation}
M\dim\g \leq 2d\,.
\end{equation}

\paragraph{Completeness.} As defined in the main text, we will say that the realisation is \textit{complete} if
\begin{equation}
M\dim\g=2d\,,
\end{equation}
\textit{i.e.} if there are as much (real) current components as there are canonical fields. In that case, the matrix $\mathbb{E}=\bigl( E_A^{\;\;\,I} \bigr)^{I=1,\dots,\,2d}_{A=1,\dots,\,M\dim\g}$ is square and has the same dimension $2d\times 2d$ as $\mathbb{K}$ and $\mathbb{D}$. We note that the relation \eqref{Eq:EE} can be rewritten in matrix terms as $\mathbb{E}\cdot\mathbb{D}\cdot\mathbb{E}^{\mathrm{T}}=\mathbb{K}$, where $\mathbb{E}^{\mathrm{T}}$ denotes the transpose of $\mathbb{E}$. For a complete realisation, since all the matrices involved are square and $\mathbb{K}$ is invertible, this means that $\mathbb{E}=\bigl( E_A^{\;\;\,I} \bigr)_{A,I=1,\dots,\,2d}$ is also invertible.

As a consequence, there exist coefficients $\bigl( \widetilde{E}^{\,i,A}, \; \widetilde{E}_{\,i}^{\;\,A} \bigr)_{i=1,\dots,\,d}^{A=1,\dots,\,2d}$ built from the coordinate fields $\phi^{\,j}(x)$ such that
\begin{equation}\label{Eq:InverseRealisation}
\p_x \phi^{\hspace{0.5pt}i}(x) = \widetilde{E}^{\,i,A} \bigl( \phi^{\,j}(x) \bigr)\,J_A(x)\,, \qquad\quad \pi_i(x) = \widetilde{E}_{\,i}^{\;\,A} \bigl( \phi^{\,j}(x) \bigr)\,J_A(x)\,.
\end{equation}
Let us imagine that we are given a fixed configuration of the current components $J_A(x)$. We can see the first part of \eqref{Eq:InverseRealisation} as a system of $d$ coupled ODEs on the functions $\bigl( \phi^{\hspace{0.5pt}i}(x) \bigr)_{i=1,\dots,d}$, depending on some external functions $J_A(x)$. Formally, this system admits a unique solution for every choice of initial condition $\Phi_0=\bigl( \phi^{\hspace{0.5pt}i}(0) \bigr)_{i=1,\dots,d}$. From this solution, one can then use the second equation in \eqref{Eq:InverseRealisation} to obtain the momenta $\bigl( \pi_i(x) \bigr)_{i=1,\dots,d}$ in terms of the given configurations of $J_A(x)$ and the initial values $\Phi_0$.

To summarise, if the realisation is complete, one can fully and uniquely reconstruct the canonical fields $\bigl( \phi^{\hspace{0.5pt}i}(x), \pi_i(x) \bigr)_{i=1,\dots,d}$ from the currents $J^{[p]}_r(x)$ and the initial values $\Phi_0=\bigl( \phi^{\hspace{0.5pt}i}(0) \bigr)_{i=1,\dots,d}$ of the coordinate fields. Note that this reconstruction is always possible in principle but can be generally quite involved in practice, as it requires solving a non-linear system of coupled ODEs to reconstruct the functions $\phi^{\hspace{0.5pt}i}(x)$: the expression of $\bigl( \phi^{\hspace{0.5pt}i}(x), \pi_i(x) \bigr)$ in terms of $J^{[p]}_r(x)$ and $\Phi_0$ is thus a complicated non-local functional map. As mentioned in Subsection \ref{SubSec:Real}, the realisation \eqref{Eq:RealTakiff} can be seen as a Poisson map from $\Ac$ to $\Acm$, sending $\Jc_r^{[p]}(x)$ to the composite fields $J_r^{[p]}(x)$ built from $\bigl( \phi^{\hspace{0.5pt}i}(x), \pi_i(x) \bigr)$. If a realisation is complete, we essentially proved that this map becomes invertible if we add a finite number of degrees of freedom to its domain $\Ac$, namely the initial conditions $\Phi_0=\bigl( \phi^{\hspace{0.5pt}i}(0) \bigr)_{i=1,\dots,d}$. We finally note that the notion of completeness introduced here is inspired by similar conditions considered in the work~\cite{Kotousov:2022azm}.

\paragraph{Momentum of a complete Takiff realisation.} Recall that the momentum $\Pc$ of the Poisson algebra $\Ac$ was built as a specific quadratic local charge in terms of the currents $\Jc_r^{[p]}(x)$ in Subsection \ref{SubSec:MomHam}. More precisely, $\Pc$ is given in terms of the Kac-Moody currents $\Jc_r(x)$ by equation \eqref{Eq:Pr} for the case without mulitplicities. The case with multiplicites is more involved and was discussed in Exercise 9, equation \eqref{Eq:PrTak}. This momentum is characterised by the fact that it generates the spatial derivative $\p_x = \lbrace \Pc,\cdot \rbrace$ on the currents $\Jc_r^{[p]}(x)$.

Let us now consider its image $\Pc_{T^\ast \Mc}$ under a complete realisation, where the abstract currents $\Jc_r^{[p]}(x)$ are replaced by $J_r^{[p]}(x)$, built from the canonical fields  $\bigl( \phi^{\hspace{0.5pt}i}(x), \pi_i(x) \bigr)$ in $\Acm$. Since the realisation is complete, the latter can be reconstructed fully from $J_r^{[p]}(x)$ and the initial values $\phi^{\hspace{0.5pt}i}(0)$: thus, the observable $\Pc_{T^\ast \Mc}$, which generate spatial translations on $J_r^{[p]}(x)$, will also generate such translations on the canonical fields $\bigl( \phi^{\hspace{0.5pt}i}(x), \pi_i(x) \bigr)$. We thus identify $\Pc_{T^\ast \Mc}$ as the momentum of the canonical Poisson algebra $\Acm$. We note that the latter also always has a simple expression in terms of the canonical fields, namely
\begin{equation}\label{Eq:PQ}
\Pc_{T^\ast \Mc} = \int_0^{2\pi} \pi_i(x)\,\p_x\phi^{\hspace{0.5pt}i}(x)\,\dd x\,.
\end{equation}
The fact that $\Pc_{T^\ast \Mc}$ takes this form in terms of the canonical fields can also be proven by a direct computation, which we will not present here for conciseness. We stress that this property only holds for complete realisations. Schematically, for a non-complete realisation, the currents $J_r^{[p]}(x)$ do not encode all the canonical fields $\bigl( \phi^{\hspace{0.5pt}i}(x), \pi_i(x) \bigr)$: the image of $\Pc$ in the realisation, which generates the spatial translations on $J_r^{[p]}(x)$, then does not generate the translations on all the fields in $\Acm$. We will see an illustration of this in the exercise 13 below. We finally note that the results discussed here, seen as consequences of the completeness of the realisation, are related to the notion of admissible realisations considered in~\cite{Delduc:2019bcl}. \vspace{10pt}

\begin{tcolorbox} \textit{\underline{Exercise 13:} Momentum of Takiff realisations.} \raisebox{0.5pt}{\Large$\;\;\star\star$} \vspace{5pt}\\
1. {\Large$\;\star\;$} Show that $\Pc_{T^\ast \Mc}$, as given by equation \eqref{Eq:PQ}, is the momentum of the canonical Poisson algebra $\Acm$.\vspace{4pt}\\
2. Consider the Poisson algebra $\Ac$ generated by Takiff currents $\Jc^{[0]}(x)$ and $\Jc^{[1]}(x)$, with multiplicity 2 and levels $\ell^{[0]}$ and $\ell^{[1]}$. Show that the quantity
\begin{equation}\label{Eq:PTak2}
\Pc = \frac{1}{\ell^{[1]}} \int_0^{2\pi} \left( \psb{\Jc^{[0]}(x)}{\Jc^{[1]}(x)} - \frac{\ell^{[0]}}{2\ell^{[1]}} \psb{\Jc^{[1]}(x)}{\Jc^{[1]}(x)} \right)\,\dd x
\end{equation}
is the momentum of $\Ac$. The curious reader is invited to compare this expression with the one coming from equation \eqref{Eq:PrTak}. \vspace{4pt}\\
3. For the case $\g=\sl(2,\R)$, check that the image of \eqref{Eq:PTak2} under the realisation \eqref{Eq:RealSl2k} coincides with the momentum \eqref{Eq:PQ} of the canonical Poisson algebra $\Ac^{\text{can}}_{T^\ast SL(2,\R)}$. This is expected since the realisation expresses $6$ current components in terms of $6$ canonical fields and thus is complete. The use of a symbolic computations software such as Mathematica is advised.\vspace{4pt}\\
4. Consider the current $J^{[0]}(x)$ of the realisation \eqref{Eq:RealSl2k} on its own, forgetting about $J^{[1]}(x)$: it is a Kac-Moody current of level $\ell^{[0]}$ and thus defines a realisation of multiplicity $m=1$ in $\Ac^{\text{can}}_{T^\ast SL(2,\R)}$. Argue that this realisation is non-complete. Moreover, prove that the image
\begin{equation*}
\frac{1}{2\ell^{[0]}} \int_0^{2\pi} \psb{J^{[0]}(x)}{J^{[0]}(x)} \,\dd x
\end{equation*}
of the momentum of this Kac-Moody current alone does not coincide with the canonical momentum \eqref{Eq:PQ}. By construction, it generates spatial translations on the particular combination $J^{[0]}(x)$ of the canonical fields in $\Ac^{\text{can}}_{T^\ast SL(2,\R)}$, but not on all of them.
\end{tcolorbox}
\newpage

\section[Poisson brackets in \texorpdfstring{$\Acg$}{ATG}]{Poisson brackets in \texorpdfstring{$\bm{\Acg}$}{ATG}}
\label{App:PbG}

\paragraph{Generalities.} In this appendix, we discuss the Poisson bracket of the fields $g$, $j$ and $X$ introduced in Subsection \ref{SubSec:PCM} with respect to the canonical Poisson structure of $\Acg$: we refer to this subsection for definitions and notations. We will need the following result: if $F(\phi^{\,j})$ is a function of the coordinates $(\phi^{\,j})_{j=1,\dots,d}$, one has
\begin{equation}\label{Eq:PbPiF}
\bigl\lbrace \pi_i(x), F\bigl(\phi^{\,j}(y)\bigr) \bigr\rbrace = \p_i F\bigl( \phi^{\,j}(x)\bigr) \, \delta(x-y)\,,
\end{equation}
where $\p_i F$ is the partial derivative of $F$ with respect to the coordinate $\phi^{\hspace{0.5pt}i}$.

\paragraph{Bracket of $\bm X$ with $\bm g$.} Applying equation \eqref{Eq:PbPiF}, we get
\begin{equation}
\bigl\lbrace \pi_i(x), g(y) \bigr\rbrace = \p_i g(x) \, \delta(x-y)= g(x)\,L^a_{\;\,i}(x)\, \td_a\,\delta(x-y)\,,
\end{equation}
where we used the definition \eqref{Eq:j} of $L^a_{\;\,i}$. Recall from equation \eqref{Eq:X} that $X(x) = \Lt^{\hspace{0.5pt}i}_{\;\,a}(x)\,\pi_i(x)\,\td^a$. Since $\Lt^{\hspace{0.5pt}i}_{\;\,a}(x)$ depends only on the coordinate fields $\phi^{\hspace{0.5pt}i}(x)$, it Poisson commutes with $g(y)$ and we get
\begin{equation*}
\bigl\lbrace X(x)\ti{1}, g(y)\ti{2} \bigr\rbrace = \Lt^{\hspace{0.5pt}i}_{\;\,a}(x)\,\td^a \otimes \,\bigl\lbrace \pi_i(x), g(y) \bigr\rbrace = L^b_{\;\,i}(x)\,\Lt^{\hspace{0.5pt}i}_{\;\,a}(x)\,\td^a \otimes g(x)\,\td_b \, \delta(x-y) \\
\end{equation*}
Using the fact that $L^b_{\;\,i}\,\Lt^{\hspace{0.5pt}i}_{\;\,a} = \delta^b_{\;\,a}$ by definition of $\Lt^{\hspace{0.5pt}i}_{\;\,a}$, we get
\begin{equation*}
\bigl\lbrace X(x)\ti{1}, g(y)\ti{2} \bigr\rbrace = g(x)\ti{2}\,(\td^a \otimes \td_a) \, \delta(x-y)\,.
\end{equation*}
By the definition \eqref{Eq:Cas} of the split Casimir, we finally obtain
\begin{equation}\label{Eq:PbXgApp}
\bigl\lbrace X(x)\ti{1}, g(y)\ti{2} \bigr\rbrace = g(y)\ti{2}\,C\ti{12} \, \delta(x-y)\,.
\end{equation}

\paragraph{Bracket of $\bm X$ with $\bm j$.} Using the Leibniz rule, we get
\begin{align*}
\bigl\lbrace X(x)\ti{1}, j(y)\ti{2} \bigr\rbrace
& = \bigl\lbrace X(x)\ti{1}, g(y)^{-1}\p_y g(y)\ti{2} \bigr\rbrace \\
& = g(y)^{-1}\ti{2}\, \bigl\lbrace X(x)\ti{1}, \p_y g(y)\ti{2} \bigr\rbrace + \bigl\lbrace X(x)\ti{1}, g(y)^{-1}\ti{2} \bigr\rbrace \, \p_y g(y)\ti{2} \\
& = g(y)^{-1}\ti{2}\,  \p_y \bigl\lbrace X(x)\ti{1}, g(y)\ti{2} \bigr\rbrace - g(y)^{-1}\ti{2}\, \bigl\lbrace X(x)\ti{1}, g(y)\ti{2} \bigr\rbrace \,g(y)^{-1}\ti{2}\, \p_y g(y)\ti{2}\,.
\end{align*}
Reinserting our previous result \eqref{Eq:PbXgApp}, we find
\begin{align*}
\bigl\lbrace X(x)\ti{1}, j(y)\ti{2} \bigr\rbrace &= C\ti{12}\,\p_y \delta(x-y) + g(y)^{-1}\ti{2}\,  \p_y g(y)\ti{2}\,  C\ti{12}\,\delta(x-y) - C\ti{12} \,g(y)^{-1}\ti{2}\, \p_y g(y)\ti{2}\,\delta(x-y)\\
&= - C\ti{12}\,\p_x\delta(x-y) + \bigl[ j(y)\ti{2}, C\ti{12} \bigr]\,\delta(x-y)\,.
\end{align*}
We finally use the Casimir identity \eqref{Eq:CasId} to get
\begin{equation}
\bigl\lbrace X(x)\ti{1}, j(y)\ti{2} \bigr\rbrace = \bigl[ C\ti{12}, j(x)\ti{1} \bigr]\,\delta(x-y) - C\ti{12}\,\p_x\delta(x-y)\,.
\end{equation}

\paragraph{Bracket of $\bm X$ with itself.} The last bracket to compute is the one of $X$ with itself. This is the most complicated one and we treat it in the exercise 16 below. In the end, we will find
\begin{equation}\label{Eq:PbXX}
\bigl\lbrace X(x)\ti{1}, X(y)\ti{2} \bigr\rbrace = \bigl[ C\ti{12}, X(x)\ti{1} \bigr]\,\delta(x-y)\,.
\end{equation}

\begin{tcolorbox} \textit{\underline{Exercise 16:} Poisson bracket of $X$ with itself.} \raisebox{0.5pt}{\Large$\;\;\star\star\star$} \vspace{5pt}\\
1. Prove the Maurer-Cartan equation
\begin{equation*}
\p_i\bigl( g^{-1}\p_j g \bigr) - \p_j\bigl( g^{-1}\p_i g \bigr) + \bigl[ g^{-1}\p_i g, g^{-1}\p_j g \bigr] = 0
\end{equation*}
and deduce from it the identity
\begin{equation*}
\p_i L^a_{\;\,j} - \p_j L^a_{\;\,i} + f_{bc}^{{\color{white}bc}a}\, L^b_{\;\,i}\,L^c_{\;\,j} = 0\,.
\end{equation*}
2. Show that
\begin{equation*}
\Lt^{j}_{\;\,a}\,\p_j \Lt^{\hspace{0.5pt}i}_{\;\,b} - \Lt^{j}_{\;\,b}\,\p_j \Lt^{\hspace{0.5pt}i}_{\;\,a} =  f_{ab}^{{\color{white}ab}c}\, \Lt^{\hspace{0.5pt}i}_{\;\,c} \,.
\end{equation*}
\textit{Hint:} the fact that $\Lt^{\hspace{0.5pt}i}_{\;\,a}\,L^a_{\;\,j}=\delta^{\hspace{0.5pt}i}_{\;\,j}$ implies $\p_j\Lt^{\hspace{0.5pt}i}_{\;\,a}=-\Lt^{\hspace{0.5pt}i}_{\;\,c}\,\Lt^k_{\;\,a}(\p_j L^c_{\;\,k})$.  \vspace{4pt}\\
3. Consider $X=X_a\,\td^a$, with $X_a=\Lt^{\hspace{0.5pt}i}_{\;\,a}\,\pi_i$ -- see the definition \eqref{Eq:X}. Use the equation \eqref{Eq:PbPiF} and the result of question 2 to prove
\begin{equation*}
\bigl\lbrace X_a(x), X_b(y) \bigr\rbrace = f_{ab}^{{\color{white}ab}c}\,X_c(x)\,\delta(x-y)\,,
\end{equation*}
which is our desired result \eqref{Eq:PbXX}, written in components.
\end{tcolorbox}~\vspace{-12pt}\\

\section{PCM with Wess-Zumino term}
\label{App:WZ}

\paragraph{The Wess-Zumino term.} In subsection \ref{SubSec:PCM}, we consider a model formulated in terms of a field $g$ valued in a Lie group $G$. Locally, this field is described by $d=\dim G$ coordinates $\bigl( \phi^{\hspace{0.5pt}i} \bigr)_{i=1,\dots,d}$. Recall that $\p_i$ denotes the derivative with respect to $\phi^{\hspace{0.5pt}i}$. We then define the following 3-tensor on $G$:
\begin{equation}
\omega_{ijk} = \psb{g^{-1}\p_i g}{\bigl[g^{-1}\p_j g,g^{-1}\p_k g\bigr]}\,.
\end{equation}
By the ad-invariance \eqref{Eq:AdInv} of $\psd$, it is completely skew-symmetric with respect to permutations of $(i,j,k)$ and thus can be interpreted as a 3-form on $G$. We admit that this 3-form is locally exact, in the sense that, locally, there exists a 2-form $\lambda_{ij}(\phi^k)=-\lambda_{ji}(\phi^k)$ such that\footnote{In general, this 2-form is not defined globally on $G$: this brings various topological and global subtleties in the treatment of the Wess-Zumino term, which we will not discuss here for simplicity.}
\begin{equation}\label{Eq:dLambda}
\omega_{ijk} = \p_i\lambda_{jk} + \p_j\lambda_{ki} + \p_k\lambda_{ij}\,.
\end{equation}
In the Lagrangian formulation, we see $g$ as a $G$-valued field $g(t,x)$ on the 2-dimensional worldsheet $\Sigma$. To define the Wess-Zumino term, we first consider a 3-dimensional manifold $\mathbb{B}$ whose boundary $\p\mathbb{B}$ coincides with $\Sigma$ and which we describe in terms of three coordinates $(\xi,t,x)$. We admit that there exists an extension of the 2-dimensional field $g$ to $\mathbb{B}$, which we still denote as $g$ by a slight-abuse of notation: more explicitly, this is a function $g(\xi,t,x)$ on $\mathbb{B}$, which coincides with the initial 2-dimensional field $g(t,x)$ on $\p\mathbb{B}=\Sigma$. We then define the Wess-Zumino term as
\begin{equation}\label{Eq:WZ2}
\W{g} = \iiint_{\mathbb{B}}\, \psb{g^{-1}\p_\xi g}{\bigl[g^{-1}\p_t g,g^{-1}\p_x g\bigr]} \,\dd \xi\, \dd t\,\dd x = \iiint_{\mathbb{B}} \omega_{ijk}\,\p_\xi\phi^{\hspace{0.5pt}i}\,\p_t\phi^{\,j}\,\p_x\phi^{j} \,\dd \xi\, \dd t\,\dd x\,.
\end{equation}
Since the 3-form $\omega_{ijk}$ is locally exact, we can use Stoke's theorem to rewrite this term as a 2-dimensional integral on $\p\mathbb{B}=\Sigma$:
\begin{equation}\label{Eq:WZ0}
\W{g} = \iint_{\Sigma}\, \lambda_{ij}\,\p_t\phi^{\hspace{0.5pt}i}\,\p_x\phi^{\,j}\, \dd t\,\dd x\,.
\end{equation}
This description uses a choice of coordinate chart $(\phi^{\hspace{0.5pt}i})_{i=1,\dots,d}$ on the group manifold $G$. It is natural to wonder if we can find a coordinate-independent formulation of the above 2-dimensional integral. For that, recall the coefficients $L^a_{\;\,i}=\ps{\td^a}{g^{-1}\p_i g}$ defined in equation \eqref{Eq:j} and their inverse $\Lt^{\hspace{0.5pt}i}_{\;\,a}$. We can then use the former to write
\begin{equation}
g^{-1} \p_t g =  L^a_{\;\,i}\,\p_t\phi^{\hspace{0.5pt}i}\,\td_a
\end{equation}
and the later to define a new $\g$-valued field
\begin{equation}\label{Eq:W}
W = \lambda_{ij}\,\Lt^{\hspace{0.5pt}i}_{\;\,a}\,\p_x\phi^{\,j}\,\td^a\,.
\end{equation}
We can then rewrite the Wess-Zumino term \eqref{Eq:WZ0} as
\begin{equation}\label{Eq:WZ}
\W g = \iint_{\Sigma}\; \psb{g^{-1}\p_t g}{W}\, \dd t\,\dd x\,.
\end{equation}

\paragraph{Takiff realisation.} Note that the current $W$ depends on the coordinate fields $\phi^{\hspace{0.5pt}i}$ and their spatial derivatives $\p_x\phi^{\hspace{0.5pt}i}$ but not on their time derivatives: we can thus consider it in the Hamiltonian formulation without having to re-express these time derivatives in terms of the momentum fields. In particular we can then consider its Poisson brackets with other Hamiltonian observables. Since $W$ does not contain the momenta, its Poisson brackets with $g$ and $j$ vanish. The non-trivial bracket to compute is the one with the current $X$, which encodes the momentum fields. We will admit here that this bracket satisfies
\begin{equation}\label{Eq:PbXW}
\bigl\lbrace X(x)\ti{1}\,,W(y)\ti{2} \bigr\rbrace + \bigl\lbrace W(x)\ti{1}\,,X(y)\ti{2} \bigr\rbrace = \bigl[ C\ti{12}, W(x)\ti{1} + j(x)\ti{1} \bigr]\,\delta(x-y)\,.
\end{equation}
This can be proven using the same type of techniques as in Appendix \ref{App:PbG} and the identity \eqref{Eq:dLambda}. From this bracket, one proves (see exercise 19 below) that
\begin{equation}\label{Eq:RealGk}
J^{[0]} = X - \nu\hay\,W + \nu\hay\,j \qquad \text{ and } \qquad J^{[1]} = \hay\left(1-\nu^2\right)\,j
\end{equation}
form a Takiff realisation of levels $\ell^{[0]}=2\nu\hay$ and $\ell^{[1]}=\hay(1-\nu^2)$. When $\nu=0$, this realisation coincides with the one \eqref{Eq:RealG} used to obtain the PCM. Moreover, for $G=SL(2,\R)$, it coincides with the one defined in equation \eqref{Eq:RealSl2k}.

\paragraph{The $\bm\s$-model.} Let us now describe the integrable $\s$-model built from this realisation. We start with an AGM with twist function \eqref{Eq:TwistPCM+WZ}. By construction, this AGM has one site of multiplicity 2 at $-\nu$, with levels $\ell^{[0]}=2\nu\hay$ and $\ell^{[1]}=\hay(1-\nu^2)$. We can thus realise it in terms of the currents \eqref{Eq:RealGk} introduced above, hence producing an integrable field theory in its Hamiltonian formulation. Passing to the Lagrangian one, we obtain the $\s$-model action and Lax connection \eqref{Eq:PCM+WZ} of the PCM with Wess-Zumino term. This is the subject of the exercise 19 below.

\begin{tcolorbox} \textit{\underline{Exercise 19:} PCM with Wess-Zumino term.} \raisebox{0.5pt}{\Large$\;\;\star\star\star$} \vspace{5pt}\\
1. Show that the currents \eqref{Eq:RealGk} define a Takiff realisation, using the brackets \eqref{Eq:PbXg} and \eqref{Eq:PbXW}.\\[4pt]
\textit{Bonus:} Check that this realisation coincides with the one \eqref{Eq:RealSl2k} for $G=SL(2,\R)$.\\[4pt]
2. Check that $\ps{j}{W}=0$, from the expressions \eqref{Eq:j} and \eqref{Eq:W} of $j$ and $W$.\\[4pt]
3. The twist function \eqref{Eq:TwistPCM+WZ} possesses the same zeroes $\pm 1$ as the one \eqref{Eq:TwistPCM} of the PCM alone. We associate quadratic charges $\Q_{\pm}$ to these zeroes following Subsection \ref{SubSec:Quad}. Compute the image $\Q_{\pm,T^\ast G}$ of these charges in the realisation \eqref{Eq:RealGk}, as well as the momentum and Hamiltonian\vspace{-2pt}
\begin{equation*}
\Pc_{T^\ast G} = \Q_{+,T^\ast G} + \Q_{-,T^\ast G} \qquad \text{ and } \qquad \Hc_{T^\ast G} = \Q_{+,T^\ast G} - \Q_{-,T^\ast G}\,.\vspace{-2pt}
\end{equation*}
Show that they take the form\vspace{-2pt}
\begin{equation*}
\Pc_{T^\ast G} = \int_0^{2\pi} \psb{X}{j}\,\dd x \qquad \text{ and } \qquad \Hc_{T^\ast G} = \int_0^{2\pi} \Bigl( \frac{1}{2\hay}\,\psb{Y}{Y} + \frac{\hay}{2}\,\psb{j}{j} \Bigr)\,\dd x\,,\vspace{-2pt}
\end{equation*}
where $Y=X-\nu\hay\,W$.\\[4pt]
4. Use the above expression of the Hamiltonian to compute the time derivative of the field $g$. Check that $Y = \hay\,g^{-1}\p_t g$.\\[4pt]
5. Perform the inverse Legendre transform and determine the action of the model, following the same approach as in the PCM case. Using equation \eqref{Eq:WZ}, show that it takes the form \eqref{Eq:PCM+WZ}.\\[4pt]
6. Compute the AGM Lax connection in terms of the currents $g^{-1}\p_\pm g$. Compare to \eqref{Eq:PCM+WZ}.
\end{tcolorbox}~\vspace{-5pt}

\providecommand{\href}[2]{#2}\begingroup\raggedright

\end{document}